\documentclass[pra, showpacs, twocolumn, a4paper, nofootinbib]{revtex4}
\usepackage{graphicx}
\usepackage{amsmath, amsfonts, amssymb, bm}
\usepackage{amsmath}
\usepackage{verbatim}
\usepackage{bm}

\usepackage{graphicx,psfrag,bm,times,amsmath,amssymb,CJK}

\newcommand{\ket}[1]{| #1 \rangle}
\newcommand{\bra}[1]{\langle #1 |}

\begin{document}

\title{Quantum entanglement and disentanglement of multi-atom systems}

\author{Zbigniew \surname{Ficek}}
\email{zficek@kacst.edu.sa}
\affiliation{The National Centre for Mathematics and Physics, KACST, P.O. Box 6086, Riyadh 11442, Saudi Arabia}

%\date{\today}

\begin{abstract}
We present a review of recent research on quantum entanglement, with special emphasis on entanglement between single atoms, processing of an encoded entanglement and its temporary evolution. Analysis based on the density matrix formalism are described. We give a simple description of the entangling procedure and explore the role of the environment in creation of entanglement and in disentanglement of atomic systems. A particular process we will focus on is spontaneous emission, usually recognized as an irreversible loss of information and entanglement encoded in the internal states of the system. We illustrate some certain circumstances where this irreversible process can in fact induce entanglement between separated systems. We also show how spontaneous emission reveals a competition between the Bell states of a two qubit system that leads to the recently discovered "sudden" features in the temporal evolution of entanglement. An another problem illustrated in details is a deterministic preparation of  atoms and atomic ensembles in long-lived stationary squeezed states and entangled cluster states. We then determine how to trigger the evolution of the stable entanglement and also address the issue of a steered evolution of entanglement between desired pairs of qubits that can be achieved simply by varying the parameters of a given system.
\end{abstract}

\pacs{03.67.Bg, 03.67.Mn, 42.50.Dv}

\maketitle

\section{Introduction}

The studies of quantum entanglement engineering between trapped atoms and controlled transmission of information from one group of atoms to another are of fundamental importance and vital to the development of quantum information technologies~\cite{nil,cha04}. In the simplest way, entanglement or non-separability can be described as a property of a system of objects that cannot be identified without reference to each other -- regardless of their spatial separations. In terms of quantum-mechanical states of a system of objects, entangled states are linear superpositions of the internal states of the system that cannot be separated into product states of the individual objects. Entanglement is recognized as entirely quantum-mechanical effect which is not only at the heart of the distinction between quantum and classical mechanics but also have played a crucial role in many discussions on fundamental issues of quantum mechanics~\cite{epr,re87,bel}. Apart from the importance to understand the fundamentals of quantum physics, entanglement provides a fascinating avenue of research in different areas of physics, mathematics and biology. It is now well established that entanglement is an essential ingredient for implementation of quantum communication and quantum computation, emerging branches of science and technology based on the laws of quantum mechanics. 

The universality of entanglement is that it can be created between different objects such as individual photons, atoms, nuclear spins and even between biological living cells. Entanglement can be used as a resource for transmission of quantum information over long distances. However, a transfer of the information requires ingredients for the control and transmission in the form of long-lived entangled states that are immune to environmental noise and decoherence. Theoretical and experimental studies have demonstrated that atoms are promising candidates to achieve of all these specific requirements. The transfer of information has been accomplished between light, which appears as a carrier of the information, and atoms whose stable ground states serve as a storage system~\cite{sj02,dr05,sk06}. The ability to store quantum information in long-lived atomic ground states opens possibilities for a long time controlled processing of information and communication of the information on~demand.

A key model for study of entanglement creation, storage and processing is a system composed of two-level atoms$-$quantum bits, called {\it qubits}. A single qubit, often called a fuel for quantum information and quantum computation, is modeled as a system composed of two energy levels that are directly or indirectly coupled to each other via eg. electric dipole transitions, or transitions through auxilary levels. If the qubit interacts with an another qubit, an initial excitation can be periodically exchanged between the qubits leading to a coherence or quantum correlations being created, and then the qubits become entangled. 

An important issue in the creation of entanglement between the qubits is the presence of unavoidable decoherence which is a source of an irreversible loss of the coherence. It is often regarded as the main obstacle in practical implementation of coherent effects and entanglement. A~typical source of decoherence is spontaneous emission resulting from the interaction of a system with an external environment. For a composed system, each part of the system can interact with own independent environments or all the parts can be coupled to a common environment. In both cases, an initial entanglement encoded into the system degradates during the evolution. Nevertheless, the degradation process can be much slower when the parts of the system are coupled to a common rather than separate environments and, contrary to our intuition, might even entangle initially unentangled qubits. This effect, called the environment induced entanglement has been studied for discrete (atomic)~\cite{bra02,kl02,yy03,ba02,ja02,ft03,vw09}, as well as for continuous variable systems~\cite{lg07,hb08}. A crucial parameter in the case of the atoms coupled to the same external environment is the collective damping~\cite{dic,ft02,fs05}, which results from an incoherent spontaneous exchange of photons between the qubits. The effect of the collective damping is to slow down the spontaneous emission from the effective collective states of the system~\cite{leh,ag74}. Under some circumstances, the spontaneous emission can be completely eliminated from some of the collective states. The states form a sub-space called decoherence-free subspace that contains long-lived entangled states that are immune to dissipation and decoherence~~\cite{ph99,br07,ke07}. A challenge then is to prepare a system of qubits in entangled states which belong only to the decoherence-free subspace.

The destructive effect of spontaneous emission on entanglement encoded into qubits can take different time scales. The decoherence time depends on the damping rate of the state in which the entanglement was initially encoded and usually the decay process induced by spontaneous emission occurs exponentially in time. However, some entangled states can have interesting "strange" decoherence properties that an initial entanglement can vanish abruptly in a finite time that is much shorter than the exponential decoherence time of spontaneous emission~\cite{eb04,eb07,zyc01}. This drastic non-asymptotic feature of entanglement has been termed as the "entanglement sudden death", and is characteristic of the dynamics of a special class of initial entangled states. The qubits may remain separable for the rest of the evolution or entanglement may revival after a finite time~\cite{ft06}. The phenomenon of entanglement sudden death presents a fascinating example of a dynamical process in which spontaneous emission affects entanglement and coherence in very different ways. A recent experiment of Almeida {\it et al.}~\cite{al07} with correlated $H$ and $V$-polarized photons has shown evidence of the sudden death of entanglement under the influence of independent environments. The required initial correlations  were created by the parametric down-conversion process.

Although the sudden death feature is concerned with the disentangled properties of spontaneous emission there is an  interesting "sudden" feature in the temporal creation of entanglement from initially independent qubits~\cite{ft08,lo08}. The phenomenon termed as sudden birth of entanglement, as it is opposite to the sudden death of entanglement arises dynamically during the spontaneous evolution of an initially separate qubits. The sudden birth of entanglement is now intensively studied as it would provide a resource for a controled creation of entanglement on demand in the presence of a dissipative environment~\cite{fic09}. 

Entanglement and coherence effects can be preserved from the destructive spontaneous emission by placing a system of atoms inside a cavity or by sending atoms across the cavity field. Cavity Quantum Electrodynamics (CQED), the study of the interaction of atoms with the cavity modified electromagnetic field, provides a different context to explore entanglement properties and its dynamics. Different types of cavities have been proposed and implemented raging from microwave~\cite{rb01}, and high-$Q$ optical cavities~\cite{yariv,hj}, to cavities engineered inside photonic band gaps materials~\cite{sj}.
When atoms are placed inside or are slowly passing through a cavity, they interact with a modified electromagnetic field that is limited in space to a finite solid angle centered around the cavity axis. As a result, their radiative properties change, so that spontaneous emission can be modified or even suppressed. The suppression arises from a reduction of the number of field modes to which atoms can be coupled. With the help of a cavity, entanglement of atomic systems can be tested by probing the atomic states of the atoms leaving the cavity, or by a measurement of the cavity photons~\cite{ph99,os01}. For example, the absence of fluorescence can be a signature for maximally entangled atom pairs which, in principle, can be used to create cluster states for one-way quantum computing~\cite{bv06}.

An another scheme based on interactions in CQED has been proposed that uses a combination of the cavity field with atomic Raman transitions between a pair of long-lived atomic ground states~\cite{gr06,ps06,de07,mp08,lk09}. The Raman transitions are mediated by two laser fields that are highly detuned from the transitions between the atomic ground and the excited states. The transitions occur through a virtual level without population of the atomic excited states, thereby avoiding spontaneous emission.  

In the CQED systems there is an another loss mechanism, the dissipation of the cavity mode though imperfections of the cavity mirrors. Although the cavity loss cannot be avoided in a similar fashion to spontaneous emission, it is possible to achieve conditions in which the dissipation of the cavity mode could be constructive rather than destructive for entanglement creation and protection from decoherence. In particular, the cavity damping can cause an atomic system to decay to decoherence-free entangled states.  For example, it is well known that cavity QED measurements on photons can be an effective tool for manipulating and observing the states of atom-cavity systems and have applications in quantum feedback~\cite{ww05}. Cavity QED has an arsenal of detection methods such as homodyne, intensity second-order correlation, intensity-field third-order correlation techniques which available to use and has been successfully applied for quantum dynamics of atoms interacting with a cavity field.

Bearing in mind all of those difficulties with the elimination of spontaneous emission, we may conclude that the best for entanglement prevention from decoherence is to encode it into states which belong to a group of states contained inside a decoherence-free subspace. However, this creates a problem of how to access the internal states of the subspace to trigger an evolution of the encoded information. It has been proposed to implement a dynamical technique of addressing individual atoms by well stabilized laser fields to achieve controlled manipulation of the internal states of the atoms and evolution of entanglement~\cite{cz94,ge96}. However, under realistic conditions the problem of addressing individual atoms poses a substantial experimental challenge. Thus, more hopeful systems from the experimental point of view have been proposed where the evolution of an initially stable entanglement could be triggered by a change of internal  parameter of the system~\cite{cr09}. The parameters that could be changed are, for example, the coupling constants between the atoms and the field modes. Alternatively, it could be done by changing frequencies of the cavity modes to detune them from the atomic resonance frequencies. These  ideas could then be applied to a controlled steering of the evolution of an initial entanglement to a desired pair of qubits including a situation where qubits are completely isolated, such as completely separated cavities each containing a single atom.

Apart from two qubit systems, various attempts have been made to produce entanglement in more complex systems. An extensive research have been done on generation of multipartite continuous variable entangled states of atomic ensembles form multi-mode squeezed light~\cite{bl05,jz03,at03,pf04}. The reason for using atomic ensembles is twofold. On the one hand, atomic ensembles are macroscopic systems that are easily created in the laboratory. On the other hand, the collective behavior of the atoms enables to achieve almost a perfect coupling of the ensembles to external squeezed fields without the need to achieve a strong field-single atom coupling. Particularly interesting are schemes involving ensembles of hot or cold atoms trapped inside a high-$Q$ ring cavity~\cite{krb03,na03,kl06}. The schemes are based on a suitable driving of four-level atoms with two external laser fields and coupling to a damped cavity mode that prepares the atoms in a pure entangled state. Similar schemes have been proposed to realize an effective Dicke model operating in the phase transition regime~\cite{de07}, to create a stationary subradiant state in an ultracold atomic gas~\cite{cb09}, and to prepare trapped and cooled ions in pure entangled vibrational states~\cite{lw06}. Simple procedures have been proposed to prepare ensembles of cold atoms in pure multi-mode entangled states and ensembles of hot atoms in pure entangled cluster states~\cite{lk09}.

In this paper we review the major processes for entanglement creation and processing of entangled states in atomic systems. The plan of this review is as follows. We begin in Sec.~2 with an introduction of the concept of quantum entanglement and some of its features. In Sec.~3, we give a brief description of the formalism we adopt to overview the research on entanglement and disentanglement. The formalism is based entirely on the density matrix representation of systems in mixed states. We then in Sec.~4 introduce measures of entanglement. We give a short review of measures of entanglement between two qubits, but we mostly focus on concurrence, the necessary and sufficient conditions for a two-qubit entanglement. In Sec.~5, we introduce the elementary models for study entanglement. We specialize to two-atom systems and pay particular attention to the dynamics of atoms interacting with a common environment. In a series of examples, we examine specific systems and processes for entanglement creation between two atoms. In Sec.~6, we employ the dynamics of the atoms to illustrate some "strange" temporary behaviors of the disentanglement process, the sudden death and sudden birth of entanglement. We demonstrate under what kind of conditions the entanglement could revival during the evolution. Then, we establish a connection between sudden birth and the presence of long-lived states of the two-atom system. In Secs.~7 and 8, we specialize to a cavity QED situation, where two atoms could be located inside a single standing-wave cavity or in two completely isolated cavities. The cases of a good and bad cavity are investigated and the role of an imperfect matching of the atoms to the cavity modes is discussed. 
Next, in Sec.~9, we tackle the problem of triggered evolution of entanglement. Our focus is on how one could trigger an evolution of a stabile or sometimes referred to as "locked" or "frozen" entangled state without the presence of any external fields.  We consider three practical schemes where the atoms are coupled to a single mode cavity field or are subject of the interaction with an environment whose the internal modes are in their vacuum states. In Sec.~10, we establish how one could steer the evolution of entanglement to a desired state. The role of the rotating-wave (RWA) approximation in the evolution of an initial single excitation entangled state is discussed in Sec.~11. In the RWA, the so-called counter rotation terms in the atom-field interaction Hamiltonian are ignored, and in this section we inquire how and under which circumstances, the time evolution of the initial entanglement could be modified by the counter-rotation terms. The remaining two sections, Secs.~12 and~13, are devoted to discuss the recently proposed procedures for deterministic creation of pure continuous variable entangled states and cluster states in atomic ensembles located inside a high-$Q$ ring cavity.

\section{Quantum entanglement: A simple picture}\label{sec2}

\noindent We devote this introductory section to an elementary discussion of the concept of quantum entanglement or simply non-separability of systems composed of qubits$-$ quantum bits. A qubit may be simply a single atom for example, or a quantum dot, or even a large,
complicated molecule with a single transition within its nondegenerate energy levels.
For the purpose of this review, we restrict ourselves to two-qubit systems only, which is the simplest system for study entanglement. We shall characterize entanglement in terms of pure and mixed states of a given system, and explore various properties of entanglement, which have a particular significance in describing unusual behaviors of entanglement and lead to further understanding of the notion of entanglement.

Our focus is on a system composed of two quantum subsystems, two qubits labeled by the suffices $A$ and $B$, and prepared in a pure quantum state described by the state vector~$\ket\Psi$. We adopt the following definition of entanglement. The qubits are said to be entangled or non-separable, if the state vector~$\ket\Psi$ cannot be factorized into the tensor product of the state vectors~$\ket{\Psi_{A}}$ and $\ket{\Psi_{B}}$ of the individual qubits 
\begin{eqnarray}
\ket \Psi \neq \ket {\Psi_{A}}\otimes\ket {\Psi_{B}} .\label{2.1}
\end{eqnarray}
Otherwise, when the factorization is possible, the qubits are separable. An example of entangled state of two qubits is a superposition state
\begin{eqnarray}
\ket\Psi = a\ket {e_{1}}\otimes\ket {g_{2}} +b\ket {g_{1}}\otimes\ket {e_{2}} ,\label{2.2}
\end{eqnarray}
where $\ket {e_{1}}\otimes\ket {g_{2}}$ and $\ket {g_{1}}\otimes\ket {e_{2}}$ are the basis states of the Hilbert space of the system and $\ket {e_{i}}$ and $\ket {g_{i}}$ represent respectively the excited and ground state of the $i$th qubit. The parameters $a$ and $b$ which, in general are complex numbers and  $|a|^{2}+|b|^{2}=1$, determine the degree of superposition of the two product states. 
When $a\neq b$, we say that the state (\ref{2.2}) is a {\it non-maximally} entangled state, and reduces to a {\it maximally} entangled state when~$a=b$. 

In order to develop our discussion to the entangled properties of a system consisting of two parts and prepared in a mixed state, we consider a pair of qubits prepared in a quantum state described by the density matrix~$\rho$. The reduced density matrix describing the properties of the $A(B)$ subsystem is obtained by tracing the total density matrix $\rho$ over the~$B(A)$ subsystem
\begin{eqnarray}
\rho_{A} = {\rm Tr}_{B}\rho ,\quad \rho_{B} ={\rm Tr}_{A}\rho .\label{2.3}
\end{eqnarray}
If the two-qubit density matrix $\rho$ can be written as the tensor product of the density matrices of the individual qubits
\begin{eqnarray}
\rho = \rho_{A}\otimes\rho_{B} ,\label{2.4}
\end{eqnarray}
then the qubits are separable. If such a factorization is not possible, then the qubits are entangled.

For further understanding of entanglement, consider some characteristic properties and fundamental features of entanglement. It is often asserted that entanglement is a quantum phenomenon. By this it is meant that entanglement is a property resulting from the superposition principle, the basic principle of quantum mechanics. 
The characteristic property of entanglement is the impossibility to distinguish between two objects prepared in a maximally entangled state. This is manifested in a simultaneous measurement of the state of the qubits. If the qubits are prepared in the state (\ref{2.2}) with $a=b$, then the probability of finding the qubit $A$ in the excited state and simultaneously the qubit $B$ in the ground state is equal to $1/2$, and the same results is found for the probability of finding the qubit~$A$ in the ground state and simultaneously the qubit $B$ in the excited state.
 
A fundamental feature of entanglement is that it can easily achieve the optimal value if a system of qubits is prepared in a pure state, and rapidly degrades when it is shared between different states or when the qubits are coupled to an another system of qubits or to a surrounding environment. It may result in a mixed state of the system. We can easy understand that property just by analyzing a simple example of entanglement sharing between two entangled states. Consider two Bell states, the prototypes of two-qubit maximally entangled states~\cite{dic,ft02}
\begin{eqnarray}
\ket{\Psi_{s}} = \frac{1}{\sqrt{2}}\left( \ket {e_{1}}\otimes\ket {g_{2}}
+\ket {g_{1}}\otimes\ket {e_{2}}\right)  ,\nonumber \\
\ket{\Psi_{a}} = \frac{1}{\sqrt{2}}\left( \ket {e_{1}}\otimes\ket {g_{2}}
-\ket {g_{1}}\otimes\ket {e_{2}}\right) .\label{2.5}
\end{eqnarray}
The states (\ref{2.5}) are symmetric and antisymmetric linear superpositions of the product states of individual atomic states which cannot be separated into product states of the individual atoms.

If the system of the two atoms is prepared in one of these states, say $\ket{\Psi_{s}}$, then the atoms are maximally entangled. If, on other hand, the system is prepared in an equal superposition of these two states then the resulting state, that is a superposition of these two states is 
\begin{eqnarray}
\ket\Upsilon = \frac{1}{\sqrt{2}}\left(\ket{\Psi_{s}} + \ket{\Psi_{a}}\right) =  \ket {e_{1}}\otimes\ket {g_{2}} \ ,\label{2.6}
\end{eqnarray}
which is a product state of the individual qubits. Thus, this simple example clearly shows that the process of sharing of an entanglement between different entangled states can diminish entanglement or even can result in the complete separability of the qubits.

Entanglement sharing may take place not only among states of the same system but also among combined states of one system and another, e.g. an another qubit or a set of qubits. The entanglement may result from the direct or indirect interaction between them. We will deal with this issue latter in the Sec.~5.

\section{Density operator representation}

\noindent Traditionally, the concept of entanglement in an atomic system is introduced and understood in terms of the atomic density operator. Here, we provide a general discussion of the use of the density operator  formalism in the description of entangled systems. We discuss general properties of the density operator and the basic conditions that the density operator must satisfy. Then, we introduce different orthogonal basis that are employed in the description of the density operator, using a two-atom system as an illustration. 

The possible states of an atomic system are specified by the atomic density (matrix) operator $\rho$. The operator~$\rho$ of any quantum system must satisfy the three basic conditions:
\begin{enumerate}
\item $\rho$ is Hermitian, $\rho = \rho^{\dagger}$, 
\item $\rho$ has unit trace, ${\rm Tr}\rho = 1$,
\item $\rho$ is positive. Thus for any state $\ket \Psi$: $\bra\Psi\rho\ket\Psi \geq 0$.
\end{enumerate}

It is useful to consider the eigenvalues $\lambda_{i}$ and normalized eigenvectors $\ket{\lambda_{i}}$ of $\rho$:
\begin{eqnarray}
\rho\ket{\lambda_{i}} =\lambda_{i}\ket{\lambda_{i}} ,\quad \rho = \sum_{i}\lambda_{i}\ket{\lambda_{i}}\bra{\lambda_{i}} .\label{3.1}
\end{eqnarray}
Then, the two basic conditions hold as follows
\begin{enumerate}
\item $0\leq \lambda_{i}\leq 1$,
\item $\sum_{i}\lambda_{i} = 1$.
\end{enumerate}
Other properties also follow
\begin{itemize}
\item The determinant of $\rho$ and all principle minors are positive.
\item Tr$\rho^{2}=\sum_{i}\lambda_{i}^{2}\leq 1$.
\item For any state $\ket\Psi$: $\bra\Psi \rho\ket\Psi \leq 1$.
\end{itemize}

A possible situation is to have all the eigenvalues of $\rho$ equal to zero, except one which is unity. In this case, the system is in a pure state characterized by
\begin{eqnarray}
\rho =\ket\lambda\bra\lambda ,\quad \rho = \rho^{2} \quad {\rm and}\quad {\rm Tr}\rho^{2} =1 .\label{3.2}
\end{eqnarray}

The subjects of entanglement generation and its dynamics can be studied in any complete set of basis states of a given system. For example, in the representation of an orthogonal basis, $\{\ket n\}$, the density matrix for any quantum system can be  defined by its matrix elements 
\begin{eqnarray}
\rho_{ij} = \bra i \rho\ket j ,\label{3.3}
\end{eqnarray}
where the states $\ket i$ and $\ket j$ satisfy the orthonormality and completeness relations, $\langle i\ket j =\delta_{ij}$ and $\sum_{i}\ket i\bra i =1$, respectively.

Usually, we choose the basis that provide equations of motions for the density matrix elements in a simple mathematical structure that allows for a particularly transparent physical interpretation. Moreover, in order to get a better inside into physics involved in a specific process, we often change the analysis to another orthogonal basis, which is done via a transformation
\begin{eqnarray}
\ket \alpha = \sum_{i} C_{i\alpha}\ket i ,\label{3.4}
\end{eqnarray}
where is transformation coefficients $C_{i\alpha}$ satisfy the normalization condition $\sum_{i}|C_{i\alpha}|^{2} = 1$.

In terms of the new basis states, the density matrix elements are given as
\begin{eqnarray}
\rho_{\alpha\beta} = \bra\alpha \rho\ket\beta = \sum_{ij} C^{\ast}_{i\alpha}\rho_{ij}C_{j\beta}  ,\label{3.5}
\end{eqnarray}
which shows that the new density matrix elements can be easily related to the old density matrix elements through the transformation coefficients.

It is natural to choose a set of orthogonal states as basis for the representation of the density operator. If we deal with a finite-dimensional system consisting of two two-level atoms, we often determine the density matrix of the system in the basis spanned by four product (separable) states
\begin{eqnarray}
&&\ket{\Psi_{1}} = |g_{1}\rangle\otimes|g_{2}\rangle ,\quad 
 \ket{\Psi_{2}} = \ket{g_{1}}\otimes\ket{e_{2}}  ,\nonumber \\
&& \ket{\Psi_{3}} = \ket{e_{1}}\otimes\ket{g_{2}} , \quad
\ket{\Psi_{4}} = |e_{1}\rangle\otimes|e_{2}\rangle .\label{3.6}
\end{eqnarray}
Of course, the form of the density matrix written in the basis (\ref{3.6}) will depend on a specific process and on the detailed dynamics involved. The density matrix can have a diagonal or non-diagonal form and the presence of any non- diagonal terms simply indicates the existence of coherence effects in the system. The coherence effects are essentially a manifestation of the correlation which may exist between the atoms.

Various type of coherences can be distinguished depending on which of the density matrix elements are nonzero. In general, the density matrix of a system of two atoms written in the basis (\ref{3.6}) is of the form
\begin{eqnarray}
  \rho = \left(
    \begin{array}{cccc}
      \rho_{11} & \rho_{12} & \rho_{13} & \rho_{14} \\
      \rho_{21} & \rho_{22} & \rho_{23} & \rho_{24} \\
      \rho_{31} & \rho_{32} & \rho_{33} & \rho_{34} \\
      \rho_{41} & \rho_{42} & \rho_{43} &\rho_{44}
    \end{array}\right) .\label{3.7}
\end{eqnarray}
In all cases we will refer to the atomic density matrix elements as follows. The {\it one-photon} atomic coherences are $\rho_{12}, \rho_{13}, \rho_{21}, \rho_{31}, \rho_{23}, \rho_{32}, \rho_{42}, \rho_{24}, \rho_{34}, \rho_{43}$, the {\it two-photon} atomic coherences are $\rho_{14}, \rho_{41}$, and the populations of the states are $ \rho_{11}, \rho_{22}, \rho_{33}, \rho_{44}$. Among the populations, we distinguish the population $\rho_{11}$ of the ground state with zero photons, excited one-photon states with populations $\rho_{22}, \rho_{33}$, and a two-photon excited state with the population $\rho_{44}$. The appropriateness of this terminology describing atomic density matrix elements follows from the examination of the number of photons involved in each state and transitions between the states. 

For majority of situations considered in this review, the density matrix of a two-atom system will occur in a simplified form, the so-called $X$-state form~\cite{yue06} 
\begin{eqnarray}
  \rho = \left(
    \begin{array}{cccc}
      \rho_{11} & 0 & 0 & \rho_{14} \\
      0 & \rho_{22} & \rho_{23} & 0 \\
      0 & \rho_{32} & \rho_{33} & 0 \\
      \rho_{41} & 0 & 0 &\rho_{44}
    \end{array}\right) ,\label{3.8}
\end{eqnarray}
where non-zero matrix elements occur only along the main diagonal and anti-diagonal. 
Physically, the $X$-state form corresponds to a situation where all one-photon coherences between the ground and one-photon states and between one- and two-photon states are zero. The $X$-state density matrix can be easily created by an appropriate initial preparation of a two-atom system, e.g. by the pumping of the atoms by correlated (squeezed) light beams as, for example, can be obtained from the output of a parametric amplifier.  
Also, one can find processes that not only retain the initial $X$-state form of the density matrix, but even could lead to the $X$-state form under the evolution.

A variety of situations considered here will include the coupling of atoms to a common environment (reservoir). In this case, the product states (\ref{3.6}) do not correspond in general to the eigenstates of the system. A different choice of basis states is found particularly useful to work with, the basis of collective states of the system, or the Dicke states, defined as~\cite{dic,ft02}
\begin{eqnarray}
&&\ket{\Psi_{1}} = |g_{1}\rangle\otimes|g_{2}\rangle ,\nonumber \\
&&\ket{\Psi_{s}} = \frac{1}{\sqrt{2}}\left(\ket{e_{1}}\otimes\ket{g_{2}} +\ket{g_{1}}\otimes\ket{e_{2}}\right) ,\nonumber \\
&&\ket{\Psi_{a}} = \frac{1}{\sqrt{2}}\left(\ket{e_{1}}\otimes\ket{g_{2}} -\ket{g_{1}}\otimes\ket{e_{2}}\right) ,\nonumber \\
&&\ket{\Psi_{4}} = |e_{1}\rangle\otimes|e_{2}\rangle  .\label{3.9}
\end{eqnarray}
We see that the collective basis contains two states, $\ket{\Psi_{s}}$ and $\ket{\Psi_{a}}$ that are linear symmetric and antisymmetric superpositions of the product states, respectively. The most important is that the states are in the form of maximally entangled states. As we shall see in Sec.~5, the entangled states are created by the interaction of the atoms through the dipole-dipole potential induced by the coupling of the atoms to the same environment. The density matrix, if diagonal in the collective basis indicates that the dynamics of the states are then independent of each other. Nevertheless, we will see that under some circumstances an entanglement, initially encoded in one of the entangled states may not evolve independently of the evolution of the remaining states. 

The Dicke states are found as a useful basis in situations where only one-photon coherences are present. When two-photon coherences are involved, it is often found convenient to work in the Bell state basis that is composed of four maximally entangled states
\begin{eqnarray}
&&\ket{\Psi_{s}} = \frac{1}{\sqrt{2}}\left(\ket{e_{1}}\otimes\ket{g_{2}} +\ket{g_{1}}\otimes\ket{e_{2}}\right) ,\nonumber \\
&&\ket{\Psi_{a}} = \frac{1}{\sqrt{2}}\left(\ket{e_{1}}\otimes\ket{g_{2}} -\ket{g_{1}}\otimes\ket{e_{2}}\right) ,\nonumber \\
&&\ket{\Phi_{s}} =  \frac{1}{\sqrt{2}}\left(\ket{g_{1}}\otimes\ket{g_{2}} +\ket{e_{1}}\otimes\ket{e_{2}}\right) ,\nonumber \\
&&\ket{\Phi_{a}} = \frac{1}{\sqrt{2}}\left(\ket{g_{1}}\otimes\ket{g_{2}} -\ket{e_{1}}\otimes\ket{e_{2}}\right) .\label{3.10}
\end{eqnarray}
We shall refer to the states $\ket{\Psi_{s}}$ and $\ket{\Psi_{a}}$ as {\it one-photon} Bell states, since they are characterized by states involving a single photon, and to the states $\ket{\Phi_{s}}$ and $\ket{\Phi_{a}}$ as {\it two-photon} Bell states, since they are characterized by states involving two photons. The states $\ket{\Psi_{s}}$ and $\ket{\Psi_{a}}$ are often called Bell states with {\it anti-correlated} spins, whereas $\ket{\Phi_{s}}$ and $\ket{\Phi_{a}}$ are called Bell states with {\it correlated} spins.
The advantage of working in the Bell state basis is the most evident in the case when two-photon coherences~$\rho_{14}$ and~$\rho_{41}$ are different from zero. In this case, the density matrix, when written in the basis formed by the Bell states may be represented in the diagonal form. It is of interest to mention that the transformation is also useful even if the resulting transformed matrix is still in a non-diagonal form. In this situation, one could find the matrix useful, for example,  for analyzes of the departure of the states of a given system from the Bell states.

In closing this section, let us conclude that a typical reason of making a transformation from one basis to another is in the simplification of the density matrix involved. A~convenient change of the basis may transfer the density matrix from a non-diagonal to diagonal form.

\section{Measures of entanglement}

\noindent In order to determine the amount of entanglement for a pure or a general mixed state of  quantum systems of an arbitrary dimension, it is essential to have an appropriate measure of entanglement. A "good" entanglement measure should vary from zero, for separable states, to unity for maximally entangled states~\cite{hh09}. It is not difficult to give a good entanglement measure for a system in a pure state~\cite{gi96,bb96,vp97,vpj97,bbp96,ck00}. The classic example of pure states for study entanglement are the Bell states that represent a class of maximally entangled states, and thus a good measure of entanglement should give unity for such states. 

A measure, which satisfies the requirements for being a good measure for entanglement is the pure state entropy of entanglement
\begin{eqnarray}
E\left(\Psi\right) = S(\rho_{A}) = S(\rho_{B}) ,\label{4.1}
\end{eqnarray}
where $\ket\Psi$ is a pure state of the system composed of two subsystems $A$ and $B$, 
\begin{eqnarray}
S(\rho_{i}) = - {\rm Tr}\left(\rho_{i}\log_{2}\rho_{i}\right) ,\quad i=A,B \label{4.2}
\end{eqnarray}
is the von Neumann entropy, and 
\begin{eqnarray}
\rho_{i} = {\rm Tr}_{j}\ket\Psi\bra\Psi ,\quad i\neq j =A,B \label{4.3}
\end{eqnarray}
are reduced density operators of the subsystems. 

The quantity $E(\Psi)$, which determines the amount of entanglement in a pure state $\ket\Psi$
ranges from zero for a product state to maximum $E(\Psi) =1$ for a Bell state.

As is well known, the entropy is a measure of the disorder in a system and obviously depends on the correlation properties. It is probably for this reason that the measure of entropy has played an important role in the development of the theory of entanglement measures.

The problem of defining a good measure of entanglement is more complex for a system in a mixed state. There are some difficulties with ordering the states according to various entanglement measures that different measures can give different orderings of pairs of mixed states. Therefore, there is a problem of the definition of the maximally entangled mixed state. 

A number of measures of entanglement of mixed states have been proposed among which concurrence and negativity satisfy the necessary conditions for being good measures of entanglement~\cite{woo,pe96,hh96}. Both measures are unity for Bell states and zero for product states. For intermediate values of concurrence or negativity, there is some entanglement in the system, and the value of either of the two measures can be treated as a degree of entanglement. Concurrence and negativity are now widely accepted measures of entanglement.

For the purpose of this review article, we adopt concurrence as the measure to quantify entanglement between qubits. The advantage of using concurrence is that it is relatively simple to calculate. It involves a standard procedure of multiplication of matrices. 
We begin with a brief outline of the approach for the calculation of concurrence. Next, we employ the concurrence to derive general expressions for entanglement of a class of physical systems determined by a common density matrix. 

For any system composed of two qubits and determined by the density operator $\rho$, the entanglement of formation is defined by
\begin{eqnarray}
\varepsilon\left( \rho \right) =\inf \sum\limits_{n}p_{n}\rho_{n}^{p} ,
\end{eqnarray}
where $0\leq p_{n}\leq 1$ is a probability distribution and the infimum is taken over all pure state decompositions of~$\rho _{n}^{p}$. The entanglement of formation for a mixed quantum system can be written in terms of the Shannon entropy and concurrence as
\begin{equation}
\varepsilon\left( \rho\right) =H\left( \frac{1}{2}+\frac{1}{2}\sqrt{1-{\cal C}^{2}\left(\rho \right) }\right) ,
\end{equation}
where $H(\alpha)$ is the Shannon entropy and ${\cal C}\left( \rho \right) $ is called concurrence.

The concurrence can  be calculated explicitly from $\rho$ as follows~\cite{woo}
\begin{eqnarray}
{\cal C}(\rho) = \max\left(0,\sqrt{\lambda_{1}}-\sqrt{\lambda_{2}}-\sqrt{\lambda_{3}} -\sqrt{\lambda_{4}}\,\right) ,\label{4.4}
\end{eqnarray}
where the quantities $\lambda_{i}$ are the the eigenvalues in decreasing order of the matrix
\begin{eqnarray}
  R=\rho\left(\sigma_{y}\otimes\sigma_{y}\right)\rho^{\ast}\left(\sigma_{y}\otimes\sigma_{y}\right) ,\label{4.5}
\end{eqnarray}
where $\rho^{\ast}$ denotes the complex conjugate of $\rho$ given in the basis of the product states (\ref{3.6}) and $\sigma_{y}$ is the Pauli matrix expressed in the same basis as
\begin{eqnarray}
  \sigma_{y} = \left(
    \begin{array}{cc}
      0 & -i \\
      i & 0
    \end{array}\right) .\label{4.6}
\end{eqnarray}
Concurrence varies from ${\cal C} =0$ for separable qubits to~${\cal C} =1$ for maximally entangled qubits, and the intermediate cases $0< {\cal C}<1$ characterize a partly entangled qubits. 

The concurrence is specified by the density matrix of a given system and thus can be determined from the knowledge of the density matrix elements. In a general case such as described by the density matrix (\ref{3.7}), the concurrence appears to have a very complicated form, not convenient in analytical discussion. The difficulty consists in obtaining the roots of the fourth order polynomial that determine the four eigenvalues of the density matrix. In the following, we consider some specific situations giving simple analytical expressions for the concurrence suitable for discussion and interpretation. A particularly simple expression for the concurrence is obtained when the density matrix of a given system, written in the vector space spanned by the product states (\ref{3.6}), has a block diagonal~form
\begin{eqnarray}
  \rho(t) = \left(
    \begin{array}{cccc}
      \rho_{11}(t) & 0 & 0 & 0 \\
      0 & \rho_{22}(t) & \rho_{23}(t) & 0\\
      0 & \rho_{32}(t) & \rho_{33}(t) & 0\\
      0 & 0 & 0 &\rho_{44}(t)
    \end{array}\right) ,\label{4.7}
\end{eqnarray}
in which all the coherences except $\rho_{23}(t)$ and $\rho_{32}(t)$ are equal to zero. In practice, it could correspond to a situation of two qubits coupled to a noisy reservoir with no initial and external coherences. In this simplified case, the concurrence can be easily computed as
\begin{eqnarray}
  {\cal C}(t) = 2 \max\left\{0,\, |\rho_{23}(t)| - \sqrt{\rho_{11}(t)\rho_{44}(t)}\right\} .\label{4.8} 
\end{eqnarray}
It is seen that a non-zero coherence between the $\ket{\Psi_{2}}$ and $\ket{\Psi_{3}}$ states is the necessary condition for entanglement, but not in general sufficient one since there is also a threshold term in the concurrence, as seen from Eq.~(\ref{4.5}), involving the populations $\rho_{11}(t)$ and $\rho_{44}(t)$. For some situations, the quantity $\sqrt{\rho_{11}(t)\rho_{44}(t)}$ will be different from zero, and $|\rho_{23}|$ may be positive but not large enough to enhance the concurrence ${\cal C}$ above the threshold for entanglement. 
Thus, the necessary and sufficient condition for entanglement between two qubits whose the dynamics are determined by the density matrix (\ref{4.7}) is
\begin{eqnarray}
   |\rho_{23}(t)|\neq 0 \quad {\rm and}\quad  |\rho_{23}(t)|>\sqrt{\rho_{11}(t)\rho_{44}(t)} .\label{4.9} 
\end{eqnarray}
We can see immediately that the threshold behavior depends on the population of the two-photon state $\ket{\Psi_{4}}$.  Thus, no threshold features can be observed if the entanglement creation and the evolution of the system involves a single photon only, which rules out the possibility to populate the state~$\ket{\Psi_{4}}$. Nevertheless, the threshold behavior still could be possible to observe. It will be seen by explicit calculations, which we will do very soon, that the non-Rotating Wave Approximation effects will give some interesting results on the threshold behavior of entanglement even if only a single photon is present initially in the system.

The threshold for entanglement is a sort of a "boundry" between classical and quantum behavior of a two-qubit system. Above the threshold, the behavior of the system is determined in terms of superposition states, the basic principe of quantum mechanics. Below the threshold, the qubits are separable and no superposition principle is required to determined their properties. 

A particularly interesting behavior of concurrence is found for a system of qubits determined by the density matrix of the $X$-state form, Eq.~(\ref{3.8}). Note that the $X$-state form arises naturally in a variety of practical situations that we will discuss in details in the following sections. First, we find that the square roots of the eigenvalues of the matrix $R$ are
\begin{eqnarray}
 && \sqrt{\lambda_{1,2}} = |\rho_{14}(t)|\pm \sqrt{\rho_{22}(t)\rho_{33}(t)} , \nonumber \\
  \nonumber\\ 
 &&\sqrt{\lambda_{3,4}} = |\rho_{23}(t)| \pm
  \sqrt{\rho_{11}(t)\rho_{44}(t)} ,\label{4.10} 
\end{eqnarray}
from which it is easily verified that for a particular value of the matrix elements there are two possibilities for the largest eigenvalue, either $\sqrt{\lambda_{1}}$ or $\sqrt{\lambda_{3}}$. The two possibilities result in the concurrence of the form 
\begin{eqnarray}
  {\cal C}(t) = 2\max\left\{0,\,{\cal C}_{1}(t),\,{\cal C}_{2}(t)\right\} ,\label{4.11} 
\end{eqnarray}
where
\begin{eqnarray}
{\cal C}_{1}(t) = |\rho_{14}(t)| -\sqrt{\rho_{22}(t)\rho_{33}(t)} ,\label{4.12}
\end{eqnarray}
and
\begin{eqnarray}
{\cal C}_{2}(t) = |\rho_{23}(t)| -\sqrt{\rho_{11}(t)\rho_{44}(t)} .\label{4.13} 
\end{eqnarray}
From this it is clear that the concurrence ${\cal C}(t)$ can always be regarded as being made up of the sum of nonnegative contributions, ${\cal C}_{1}(t)$ and ${\cal C}_{2}(t)$ associated with
two different coherences that can be generated in a two qubit system. From the forms of ${\cal C}_{1}(t)$ and ${\cal C}_{2}(t)$, it is obvious that ${\cal C}_{1}(t)$ provides a measure of an entanglement produced by the two-photon coherence $\rho_{14}$, whereas ${\cal C}_{2}(t)$
provides a measure of an entanglement produced by the one-photon coherence $\rho_{23}$. The contributions ${\cal C}_{1}(t)$ and ${\cal C}_{2}(t)$ are traditionally understood as the {\it criteria} for one and two-photon entanglement, respectively. They are often identified as weight functions for the distribution of entanglement between the two coherences.

Inspection of Eq.~(\ref{3.8}) shows that the two separate contributions to the concurrence are associated with two separate blocks the density matrix elements are grouped. We can distinguish an inner block composed of the one-photon populations, $\rho_{22}, \rho_{33}$ together with the coherences $\rho_{23}, \rho_{32}$, and an outer block composed of the populations $\rho_{11},\rho_{44}$ together with the two-photon coherences~$\rho_{14},\rho_{41}$. Further analysis show that the inner block could be related to the one-photon Bell states, whereas the outer to the two-photon Bell states. 
Thus, the threshold behaviors of the concurrence, seen in Eqs.~(\ref{4.12}) and (\ref{4.13}), can be interpreted as a competition between one and two-photon processes or equivalently between one and two-photon Bell states in the creation of entanglement of the $X$-state form in the two-qubit system. If an entanglement is created among the one-photon states, the amount of entanglement created is limited by a population of the two-photon states, and vice versa, if entanglement is created among the two-photon states, the entanglement is limited by a population of the one-photon states.

The previous discussion on the concurrence focused on the detailed analysis of the relations between the two different groups the density matrix elements could be divided, without specifying the state of the system. There are specific cases of interest, where two qubits are found in a pure entangled state, or in a mixed state involving only a single or two entangled states. It is worth discussing these specific cases because properties of the concurrence can be easily determined from the nature of the entangled states and their populations. For example, consider a system of two qubits prepared in a pure non-maximally entangled state of the form of the wave function~(\ref{2.2}). This state is associated with the one-photon coherence and could be created between the states of the inner block of the matrix (\ref{3.8}). Note that in the special case of $|a|=|b|=1/\sqrt{2}$, the state coincides with the Dicke state $\ket{\Psi_{s}}$. We can write the state (\ref{2.2}) in an equivalent form
\begin{eqnarray}
\ket\Psi = \frac{1}{\sqrt{1+|\alpha|^{2}}}\left(\ket {e_{1}}\otimes\ket {g_{2}} +\alpha\ket {g_{1}}\otimes\ket {e_{2}}\right) ,\label{4.14}
\end{eqnarray}
where $\alpha =b/a$. A pure-state density matrix formed from the state (\ref{4.14}) is of the form
\begin{eqnarray}
  \rho = \left(
    \begin{array}{cc}
      \frac{1}{1+|\alpha|^{2}} & \frac{\alpha^{\ast}}{\sqrt{1+|\alpha|^{2}}}  \\
      \frac{\alpha}{\sqrt{1+|\alpha|^{2}}} & \frac{|\alpha|^{2}}{1+|\alpha|^{2}}
    \end{array}\right) ,\label{4.15}
\end{eqnarray}
and leads to a particularly simple expression for the concurrence
\begin{eqnarray}
  {\cal C} =  \frac{2|\alpha|}{1+|\alpha|^{2}}  .\label{4.16} 
\end{eqnarray}
The concurrence involves only a cross term, the product of the off-diagonal elements of the density matrix. Note an interesting feature of the concurrence (\ref{4.16}), namely, the concurrence of the pure state is determined only by a single parameter $|\alpha|$, the ratio of the probability amplitudes of the product states. 

If, in addition to the entangled state, the Hilbert space of the system is expanded to include few separable states, then the expanded system could be found in a mixed state. Nevertheless, the concurrence still could be determined by the parameters characteristic of the entangled state. A consequence of the presence of the separable states could be in a possibility of the threshold behavior for entanglement. A familiar example of this situation is the Dicke model, which refers to two atoms coupled to a common reservoir and confined to a region much smaller than the resonance wavelength of the atomic transitions~\cite{dic,ft02}. The Hilbert space of the Dicke model is spanned by three collective state vectors, $\ket{\Psi_{1}}, \ket{\Psi_{s}}$ and $\ket{\Psi_{4}}$. In the absence of the coherences, the density matrix of the system, written in the basis of the three state vectors, takes a diagonal form
\begin{eqnarray}
  \rho = \left(
    \begin{array}{ccc}
      \rho_{11} & 0 & 0 \\
      0 & \rho_{ss} & 0\\
      0 & 0 &\rho_{44}
    \end{array}\right) ,\label{4.17}
\end{eqnarray}
and then the concurrence is given by
\begin{eqnarray}
  {\cal C} = \max\left\{0,\, \rho_{ss} - 2\sqrt{\rho_{11}\rho_{44}} \right\} .\label{4.18} 
\end{eqnarray}
We see that positive values of the concurrence are brought by a non-zero population of the entangled state and unlike the pure state situation, the concurrence is not determined by a single parameter. The involvement of the other density matrix elements can lead to a threshold behavior for entanglement. It happens whenever the term $2\sqrt{\rho_{11}\rho_{44}}$ overweights the population $\rho_{ss}$ of the state $\ket{\Psi_{s}}$.
It is interesting to note that the threshold could occur only if the population is shared between the entangled state and {\it both} of the separable states. If only one of the separable states is populated, no threshold for entanglement is observed. 

A further very instructive example is a situation when the Hilbert space of the system is spanned by state vectors among which two states are entangled and the other are separable. An example of a physical system corresponding to this situation is the case of two qubits coupled to the same environment. The Hilbert space of this system is composed of the four collective states~(\ref{3.9}), and the resulting density has the following form 
\begin{eqnarray}
  \rho = \left(
    \begin{array}{cccc}
      \rho_{11} & 0 & 0 & 0 \\
      0 & \rho_{ss} & \rho_{sa} & 0\\
      0 & \rho_{as} & \rho_{aa} & 0\\
      0 & 0 & 0 &\rho_{44}
    \end{array}\right) .\label{4.19}
\end{eqnarray}
With the density matrix (\ref{4.19}), the concurrence is of the form
\begin{eqnarray}
  {\cal C} = \max\left\{0,\,{\cal C}_{2}\right\} ,\label{4.20} 
\end{eqnarray}
with
\begin{eqnarray}
 {\cal C}_{2} = \sqrt{\left(\rho_{ss}-\rho_{aa}\right)^{2} - \left(\rho_{sa}
  -\rho_{as}\right)^{2}} - 2\sqrt{\rho_{11}\rho_{44}} .\label{4.21}
\end{eqnarray}
It is seen that the criterion for entanglement becomes very different when two or more entangled  states are present. One would expect that a larger number of entangled states involved in the description of a system should result in a better entanglement. This is not the case. According to Eq.~(\ref{4.21}), the concurrence depends crucially on the population difference rather than the sum of the populations  of the two entangled states involved. A further analysis of Eq.~(\ref{4.21}) results in  two important conclusions. First of all, any two qubis represented by two maximally entangled states, which are equally populated, are completely separable. This is true independent of whether the entangled states are correlated or not. 
Hence, two equally populated Dicke entangled states will not give rise to entanglement even though they may be strongly correlated. Secondly, a close look at Eq.~(\ref{4.21}) reveals that the correlations between the entangled states have a further destructive rather than constructive effect on entanglement. This is in contrast to the correlations between product states that always appear as constructive correlations for entanglement. 

Finally it may be mentioned, that threshold behavior of entanglement is also possible in this system, as indicated by the presence of the term $2\sqrt{\rho_{11}\rho_{44}}$ in Eq.~(\ref{4.21}). Thus, similar to the above example, a threshold of entanglement could be observed only if the redistribution of the population involves both of the separable states.

The examples presented in this section show that we can find a variety of physical systems which exhibit the same entangled properties. We have established the relation between the concurrence and the form of the density matrix of a physical system. We hope that the general situations considered here will lead to a better understanding of the detailed dynamics of entanglement discussed in the succeeding sections, where we specialize to specific situations of two atoms and atomic ensembles evolving under the influence of noisy environment and/or external fields.

\section{Elementary models for entanglement}

\noindent The theoretical study of multi-atom systems is generally complicated and the physics involved is often difficult to grasp in the necessarily complicated analysis. Although less important from the point of view of practical applications, studying simpler systems such as single two-level atoms is significant in terms of enabling analytic treatments to be undertaken and thus facilitates a better insight into the fundamentals of the atomic dynamics.
It is our purpose in this section to specialize the preceding general discussion on entanglement creation and processing to the simplest examples of potential atomic systems for entanglement. Four cases are studied: (1) Two atoms in free space and interacting with a common environment, (2)~two atoms interacting with a single-mode cavity field, (3)~two atoms located in separate cavities and interacting with local single-mode fields,~(4) atomic ensembles interacting with radiation modes of a high-$Q$ ring cavity. 

The systems considered here involve a set of identical atoms located at fixed positions $\vec{r}_{i}$, and separated by a distance $r_{ij} =|\vec{r}_{j} -\vec{r}_{i}|$ large compared to the atomic diameter, so that overlap between the atoms can be ignored. The atoms are modeled as two-level systems with ground states $\ket{g_{i}}$ and excited states $\ket{e_{i}}$, separated by transition frequencies $\omega_{0}$ and connected by a transition dipole moment $\vec{\mu}$. 
Both atoms are assumed to be damped with the same rates $\gamma-$the spontaneous emission rates arising from the coupling of the atoms to a common environment. The atoms can be prepared initially in an arbitrary state that could be a separable (product) or a superposition (entangled) state. The initial state may evolve in time under the action of the Hamiltonian of the system, or its evolution may be "frozen" for all times.

The dynamics of the atoms can be studied in a complete set of basis states of a given system. 
For the first scheme of two atoms interacting with a common environment, we will work in the basis of the Dicke states~(\ref{3.9}). For the second scheme of two atoms located inside a single-mode cavity field, it will prove convenient to study the dynamics of the atoms in the basis of four product states (\ref{3.7}). The third scheme involves two atoms located in two separate cavities. We shall consider two cases, where there is only a single or two excitations present in the system. Thus, in the first case, we will choose the Hilbert space of the system spanned by four product state vectors, defined as follows
\begin{eqnarray}
&&|\xi_{1}\rangle = \ket{e_{1}}\otimes\ket{g_{2}}\otimes{\ket 0}_{1}\otimes{\ket 0}_{2} ,\nonumber\\
&&|\xi_{2}\rangle = \ket{g_{1}}\otimes\ket{e_{2}}\otimes{\ket 0}_{1}\otimes{\ket 0}_{2}  ,\nonumber \\ 
&&|\xi_{3}\rangle =  \ket{g_{1}}\otimes\ket{g_{2}}\otimes{\ket 1}_{1}\otimes{\ket 0}_{2} ,\nonumber \\
&&|\xi_{4}\rangle = \ket{g_{1}}\otimes\ket{g_{2}}\otimes{\ket 0}_{1}\otimes{\ket 1}_{2}  .\label{5.1}
\end{eqnarray}
Here, for example, $ \ket{e_{1}}\otimes\ket{g_{2}}\otimes{\ket 1}_{1}\otimes{\ket 0}_{2}$ represents the state in which the atom $1$ is in the excited state, the atom $2$ is in the ground state, one photon is present in the mode of the cavity~$1$, and no photons are present in the mode of the cavity~$2$.

In the second case of this scheme, we will assume that there are two excitations present, one in each sub-system, and choose the Hilbert space of the system spanned by the vectors 
\begin{eqnarray}
&&|\chi_{1}\rangle  =  \ket{e_{1}}\otimes\ket{e_{2}}\otimes{\ket 0}_{1}\otimes{\ket 0}_{2} ,\nonumber\\
&&|\chi_{2}\rangle  = \ket{e_{1}}\otimes\ket{g_{2}}\otimes{\ket 0}_{1}\otimes{\ket 1}_{2} ,\nonumber\\
&&|\chi_{3}\rangle  =  \ket{g_{1}}\otimes\ket{e_{2}}\otimes{\ket 1}_{1}\otimes{\ket 0}_{2} ,\nonumber\\
&&|\chi_{4}\rangle  =  \ket{g_{1}}\otimes\ket{g_{2}}\otimes{\ket 1}_{1}\otimes{\ket 1}_{2} .
\label{5.2}
\end{eqnarray}
In this case we also include the ground state
\begin{eqnarray}
|\chi_{0}\rangle = \ket{g_{1}}\otimes\ket{g_{2}}\otimes{\ket 0}_{1}\otimes{\ket 0}_{2} ,\label{5.3}
\end{eqnarray}
for which there is no excitation present in the system. In practice, this state could result from a loss of the excitation due to spontaneous emission from the atoms or through the damping of the cavity modes. The possibility of the presence of this state will allow us to study the case where the initial state is in the form of a non-maximally entangled Bell state with correlated spins
\begin{eqnarray}
\ket{\chi_{sd}}= \cos\alpha |\chi_{1}\rangle +{\rm e}^{i\beta}\sin\alpha|\chi_{0}\rangle ,\label{5.4}
\end{eqnarray}
where $\alpha$ and $\beta$ are real numbers.

In the fourth scheme, which involves atomic ensembles coupled to radiation modes of a ring cavity, we discuss entanglement procedures in terms of the field (bosonic) representation of the atomic operators. In this representation, we first express each atomic ensemble in terms of collective atomic operators 
\begin{eqnarray}
J_{k_{m}n}^{\pm} = \sum\limits_{j=1}^{N} S^{\pm}_{jn}{\rm e}^{i\vec{k}_{m}\cdot\vec{r}_{jn}} ,\quad J_{zn} =  \sum\limits_{j=1}^N S^{z}_{jn} ,\label{5.5}
\end{eqnarray}
where $N$ is the number of atoms in the $n$th ensemble, $\vec{k}_{m}$ is the propagation vector of the $m$th field mode, $\vec{r}_{jn}$ is the position vector of the $j$th atom in the $n$th ensemble,~$S^{+}_{jn}$, $S^{-}_{jn}$, and $S^{z}_{jn}$ are the raising, lowering and the energy difference operators, respectively, of the $j$th atom in the $n$th ensemble.

Next, we reformulate the operators in terms of boson variables by applying the Holstein-Primakoff representation of the collective atomic operators~\cite{hp40}. The representation transforms the collective operators into harmonic oscillator annihilation and creation operators, $c_{n}$ and $c_{n}^{\dagger}$ of a single bosonic mode
\begin{eqnarray}
J_{k_{m}n}^{+} = c_{n}^{\dagger}\sqrt{N -c_n^\dag c_n} ,\quad J_{zn}=c_n^\dag c_n-N/2 .\label{5.6}
\end{eqnarray}
Because the ensembles are independent of each other, the bosonic operators satisfy the commutation relation
\begin{eqnarray}
 \left[c_{m},c_{n}^{\dagger}\right] = \delta_{mn} ,\label{5.7}
\end{eqnarray}
where the subscripts $n$ and $m$ enumerate the ensembles.
The commutation property of the modes will allow us to prepare each mode separately in a desired state and independent of the other modes. In other words, an arbitrary transformation performed on the operators of a given mode will not affect the remaining modes.

\subsection{Two atoms in free space}

\noindent The simplest model for the study of entanglement is a system composed of two two-level atoms interacting in free space with a quantized multimode electromagnetic field. The field appears as a vacuum reservoir (environment) to the atoms that causes initial atomic excitation and coherence to decay in time. The atoms may decay independently or their radiation field could  exert a strong dynamical influence on one another through the vacuum field modes. This would result in cooperative behavior of the system. In this section we concentrate on the problem of entanglement creation in the process of spontaneous emission from an excited initially unentangled state. We shall demonstrate under what kind of conditions the initially separated atoms become entangled during the spontaneous emission process. We explore the role of the collective states of the system in the entanglement creation via spontaneous emission. Our analysis reveals the necessity of including the antisymmetric state into the dynamics of the system if the spontaneous decay of an initial excitation is to generate an entanglement during the evolution of the system.

The first step is to define the approach we take to study the spontaneous dynamics of the system. We shall work in the density matrix formalism and adopt the master equation technique to study the time evolution of the density matrix of the system. For two two-level atoms interacting in free space with a vacuum field at zero temperature, the evolution of the density matrix of the system is given by the Lehmberg$-$Agarwal master equation~\cite{leh,ag74}
\begin{eqnarray}
      \frac{\partial \rho}{\partial t} &=& -i\omega_{0}\sum_{i=1}^{2} \left[S^{z}_{i},\rho\right]
      -i\!\sum_{i\neq j=1}^{2}\!\Omega_{ij}\!\left[S^{+}_{i}S^{-}_{j},\rho\right] \nonumber \\
      &-& \frac{1}{2}\!\sum_{i,j=1}^{2}\!\gamma _{ij}\!\left( \left[\rho S_{i}^{+},S_{j}^{-}\right]\!+\!\left[
      S_{i}^{+},S_{j}^{-}\rho\right]\right) ,\label{5.8}
\end{eqnarray}
where where $S_{i}^{+}$, $S_{i}^{-}$, and $S_{i}^{z}$ are the dipole raising, lowering, and population difference operators, respectively, of the $i$th atom, and $\gamma_{ii}\equiv \gamma$ are the spontaneous decay rates of the atoms, equal to the Einstein $A$ coefficient for spontaneous emission. The terms in the master equation that depend on $\gamma_{ij}$ and~$\Omega_{ij}\ (i\neq j)$ are the so-called {\it collective} terms because they result form the mutual exchange of photons between the atoms and thus determine the atomic interaction. The parameter~$\gamma_{ij}$ represents the collective damping which results from an incoherent exchange of photons between the atom. The collective damping  leads in general to a change in the lifetime of the collective states from the single-atom radiative lifetime. The parameter~$\Omega_{ij}$ represents the collective shift of the atomic levels and results from a coherent exchange of photons, the dipole-dipole interaction between the atoms. The effect of $\Omega_{12}$ on the atomic system is the shift of the energy of the single excitation collective states from the single-atom energy. The collective parameters are given by the expressions~\cite{ft02,fs05,leh,ag74,se08,step} 
\begin{eqnarray}
&&\gamma_{ij} = \frac{3}{2}\gamma\left\{ \left[1\!-\!\left( \hat{\mu}\cdot \hat{r}_{ij}\right)^{2} \right]
 \frac{\sin\!\left( kr_{ij}\right)}{kr_{ij}}\right.  \nonumber \\
&&\left. +\!\left[ 1\!-\!3\!\left( \hat{\mu}\cdot\hat{r}_{ij}\right)^{2} \right]\!\!\left[ \frac{\cos\!\left(kr_{ij}\right) }
{\left( kr_{ij}\right) ^{2}}\!-\!\frac{\sin\!\left(kr_{ij}\right) }{\left( kr_{ij}\right) ^{3}}\right]\! \right\} ,\label{5.9}
\end{eqnarray}
and
\begin{eqnarray}
&&\Omega_{ij} = \frac{3}{4}\gamma\left\{ -\left[1-\left( \hat{\mu}\cdot \hat{r}_{ij}\right)^{2} \right] 
\frac{\cos\!\left( kr_{ij}\right)}{kr_{ij}}\right.  \nonumber \\
&&\left. +\!\left[ 1\!-\!3\!\left(\hat{\mu}\cdot\hat{{r}}_{ij}\right)^{2} \right]\!\!\left[\!\frac{\sin\!\left(kr_{ij}\right) }
{\left(kr_{ij}\right) ^{2}}\!+\!\frac{\cos\!\left(kr_{ij}\right) }{\left( kr_{ij}\right) ^{3}}\right]\!\right\} ,\label{5.10}
\end{eqnarray}
where $\hat{\mu}$ is the unit vector along the dipole moments of the atoms, which we have assumed to be parallel $(\hat{\mu}=\hat{\mu}_{i}=\hat{\mu}_{j})$,  $\hat{r}_{ij}$ is the unit vector in the direction of $\vec{r}_{ij}$, $k =\omega_{0}/c$, and $r_{ij}$ is the distance between the atoms. 

The parameters (\ref{5.9}) and (\ref{5.10}) are oscillatory functions of~$kr_{ij}$ multiplied by inverse powers of $kr_{ij}$ ranging from~$(kr_{ij})^{-3}$ to $(kr_{ij})^{-1}$. In the two limiting cases of~$kr_{ij}\ll 1$ and $kr_{ij}\gg 1$, the parameters simplify to the following~forms
\begin{eqnarray}
\gamma_{ij} \Longrightarrow\left\{
\begin{array}{cc}
     \gamma\left[1-O\left(kr_{ij}\right)^{2}+\cdots\right] &\quad (kr_{ij}\ll 1)  \label{5.11} \\
O(kr_{ij})^{-1} &\quad (kr_{ij}\gg 1) ,
\end{array}
\right.
\end{eqnarray}
and 
\begin{eqnarray}
\Omega_{ij} \Longrightarrow\left\{
\begin{array}{cc}
     V_{ij} &\quad (kr_{ij}\ll 1)  \label{5.12} \\
O(kr_{ij})^{-1} &\quad (kr_{ij}\gg 1) ,
\end{array}
\right.
\end{eqnarray}
where
\begin{eqnarray}
V_{ij} = \frac{3\gamma}{4\left(kr_{ij}\right)^{3}}
\left[1-3\left( \hat{\mu}\cdot \hat{r}_{ij}\right)^{2}\right] .\label{5.13}
\end{eqnarray}
For small $kr_{ij}$, the collective damping $\gamma_{ij}$ becomes equal to $\gamma$, and $\Omega_{ij}$ reduces to the quasistatic dipole-dipole interaction potential. For large $kr_{ij}$, the collective effects become negligible and the system reduces to two independent atoms.  

We choose the collective states (\ref{3.9}), as a convenient basis for the representation of the density operator and the analysis of the spontaneous dynamics of the system. Consider first the evolution of the diagonal matrix elements, which correspond to the populations of the collective states. The equations of motion are found from the master equation~(\ref{5.8}), and are given by
\begin{eqnarray}
      &&\dot{\rho}_{44} = -2\gamma \rho_{44} ,\nonumber \\
     && \dot{\rho}_{ss} = -\left(\gamma +\gamma_{12}\right)\left(\rho_{ss} -\rho_{44}\right) ,\nonumber \\
     && \dot{\rho}_{aa} = -\left(\gamma -\gamma_{12}\right)\left(\rho_{aa} -\rho_{44}\right) .\label{5.14}
\end{eqnarray}
Equations (\ref{5.14}) are in the form of rate equations for the populations of the collective states. The equations give us an important information about the transitions rates between the collective  states~\cite{ftk81,ftk83,hf82}. One can see that the transitions rates to and from the symmetric and antisymmetric states are changed by the collective damping $\gamma_{12}$. The transitions to and from the symmetric state occur with an enhanced rate $\gamma +\gamma_{12}$, whereas the transitions to and from the antisymmetric state occur with a reduced rate $\gamma -\gamma_{12}$. The collective damping depends on the distance between the atoms. For small~$kr_{12}$, the state~$\ket{\Psi_{s}}$ becomes {\it superradiant} with a decay rate double that of the single atom $\gamma$, and the state $\ket{\Psi_{a}}$ becomes {\it subradiant}, with a decay rate of order $(kr_{12})\gamma$ which vanishes in the limit of small distances $kr_{12}\ll 1$. For large $kr_{12}$, the effects of the collective damping vanish as $(kr_{12})^{-1}$ and become negligible for $kr_{12}\gg 1$. In this limit,  the states decay with a single rate identical to that of a single atom. 
\begin{figure}[hbp]
\includegraphics[width=7cm,keepaspectratio,clip]{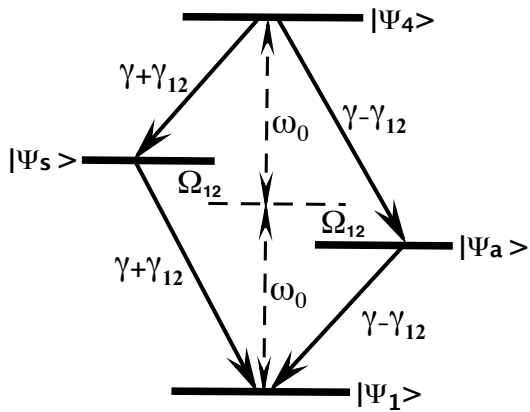}
\caption{The collective states of a two-atom system with possible one (solid arrows) and two-photon (dashed arrows) transitions. Transitions $\ket{\Psi_{4}}\rightarrow\ket{\Psi_{s}}\rightarrow\ket{\Psi_{1}}$ through the symmetric state occur with an enhanced rate $\gamma +\gamma_{12}$, while the transitions $\ket{\Psi_{4}}\rightarrow\ket{\Psi_{a}}\rightarrow\ket{\Psi_{1}}$ through the antisymmetric state occur with a reduced rate $\gamma-\gamma_{12}$. The intermediate states are shifted from their resonant positions by $\Omega_{12}$, the energy of the dipole-dipole interaction.}
\label{figc1}
\end{figure}

Using Eq.~(\ref{5.8}), we find that the off-diagonal matrix elements are decoupled from the diagonal elements and obey equations of motion that can be grouped into two independent sets: One containing two independent equations for the coherence $\rho_{as}$ between the symmetric and antisymmetric states, and the two-photon coherence~$\rho_{14}$ between the ground state $\ket{\Psi_{1}}$ and the two-photon state~$\ket{\Psi_{4}}$. The density matrix elements evolve~as 
\begin{eqnarray}
      &&\dot{\rho}_{as} = -\left(\gamma +2i\Omega_{12}\right)\rho_{as} ,\nonumber \\
      &&\dot{\rho}_{14} = -\left(\gamma -2i\omega_{0}\right)\rho_{14} .\label{5.15}
\end{eqnarray}
The other set contains four coupled equations for the one-photon coherence of the cascade transitions from the upper two-photon state $\ket{\Psi_{4}}$ through the intermediate states~$\ket{\Psi_{s}}$ and $\ket{\Psi_{a}}$ to the ground state $\ket{\Psi_{1}}$. The density matrix elements evolve as
\begin{eqnarray}
      &&\dot{\rho}_{s4} = -\Gamma_{1}\rho_{s4}  ,\quad  \dot{\rho}_{a4} = -\Gamma_{2}\rho_{a4} ,\nonumber \\
      &&\dot{\rho}_{1s} = -\Gamma_{3}\rho_{1s}  + \left(\gamma +\gamma_{12}\right)\rho_{s4} ,
      \nonumber \\
      &&\dot{\rho}_{1a} = -\Gamma_{4}\rho_{1a} - \left(\gamma -\gamma_{12}\right)\rho_{a4} ,\label{5.16}
\end{eqnarray}
where, for simplicity, we have introduced the notation  
\begin{eqnarray}
&&\Gamma_{1,2} = \frac{1}{2}\left(3\gamma\pm\gamma_{12}\right) -i\left(\omega_{0}\mp\Omega_{12}\right) ,\nonumber\\
&&\Gamma_{3,4} = \frac{1}{2}\left(\gamma\pm\gamma_{12}\right) -i\left(\omega_{0}\pm\Omega_{12}\right) .\label{5.17}
\end{eqnarray}
The one-photon coherences (\ref{5.16}) oscillate with frequencies shifted from the unperturbed frequency $\omega_{0}$ 
by the dipole-dipole interaction energy $\Omega_{12}$. Similar to the decay of the populations, the coherence involving the symmetric state decay with an enhanced rate whereas the coherence involving the antisymmetric state decays with a reduced rate. 

The collective states of the system are shown in Fig.~\ref{figc1}, where it is seen that the intermediate states are shifted 
from their unperturbed energies by the dipole-dipole interaction energy, and there are two transition channels $\ket{\Psi_{4}} \rightarrow \ket{\Psi_{s}}
\rightarrow \ket{\Psi_{1}}$ and $\ket{\Psi_{4}} \rightarrow \ket{\Psi_{a}} \rightarrow \ket{\Psi_{1}}$, each with two cascade nondegenerate transitions. The transition channels are uncorrelated, but the transitions inside these channels are damped with different rates.

Our objective is to calculate the time evolution of the concurrence under the spontaneous decay of an initial excitation. To this end, we need the time evolution of the density matrix elements, which is obtained by solving the equations of motion (\ref{5.14})-(\ref{5.16}). The general solution valid for arbitrary initial conditions is given by
\begin{eqnarray}
  \rho_{44}(t) &=& \rho_{44}(0){\rm e}^{-2\gamma t}  ,\nonumber \\
  \rho_{ss}(t) &=& \rho_{ss}(0){\rm e}^{-(\gamma+\gamma_{12}) t} \nonumber \\
   &+& \rho_{44}(0)\frac{\gamma +\gamma_{12}}
   {\gamma -\gamma_{12}}\!\left[{\rm e}^{(\gamma-\gamma_{12}) t} -1\right]\!{\rm e}^{-2\gamma t} ,\nonumber \\
  \rho_{aa}(t) &=& \rho_{aa}(0){\rm e}^{-(\gamma-\gamma_{12}) t} \nonumber \\ 
   &+& \rho_{44}(0)\frac{\gamma -\gamma_{12}}
   {\gamma +\gamma_{12}}\!\left[{\rm e}^{(\gamma +\gamma_{12})t} -1\right]\!{\rm e}^{-2\gamma t} ,\label{5.18}
\end{eqnarray}
and $\rho_{11}(t)= 1- \rho_{44}(t)-\rho_{ss}(t)-\rho_{aa}(t)$. 

We see from Eq.~(\ref{5.18}) that the decay of the populations depends strongly on the initial state of the system. When the system is initially prepared in the state $\ket{\Psi_{s}}$, the population of the initial state decays exponentially with an enhanced rate $\gamma +\gamma_{12}$, while the initial population of the antisymmetric state decays with a reduced rate $\gamma -\gamma_{12}$. This occurs because the photons emitted from the excited atom can be absorbed by the atom in the ground state, so that the photons do not escape immediately from the system. For a general initial state that includes in the state~$\ket{\Psi_{4}}$, the populations of the symmetric and the antisymmetric states do not decay with a single exponential. 

Similarly, Eqs.~(\ref{5.15}) and (\ref{5.16}) are straightforward to solve, giving
\begin{eqnarray}
 && \rho_{sa}(t) = \rho_{sa}(0){\rm e}^{-(\gamma +2i\Omega_{12})t} ,\nonumber\\
 && \rho_{14}(t) = \rho_{14}(0){\rm e}^{-(\gamma -2i\omega_{0})t}  ,\nonumber \\
 && \rho_{s4}(t) = \rho_{s4}(0){\rm e}^{-\Gamma_{1}t} ,\ 
      \rho_{a4}(t) = \rho_{a4}(0){\rm e}^{-\Gamma_{2}t} ,\nonumber \\
 && \rho_{1s}(t) = \rho_{1s}(0){\rm e}^{-\Gamma_{3}t} \nonumber \\
 &&\qquad \ \ +\, \rho_{s4}(0)\frac{\gamma+\gamma_{12}}{\gamma+2i\Omega_{12}}\left({\rm e}^{-\Gamma_{3}t} -{\rm e}^{-\Gamma_{1}t}\right) , \nonumber\\
 && \rho_{1a}(t) = \rho_{1a}(0){\rm e}^{-\Gamma_{4}t} \nonumber \\
 &&\qquad \ \ +\, \rho_{a4}(0)\frac{\gamma-\gamma_{12}}{\gamma-2i\Omega_{12}}\left({\rm e}^{-\Gamma_{2}t} -{\rm e}^{-\Gamma_{4}t}\right) .\label{5.19}
\end{eqnarray}

The density matrix elements depend on the initial state $\ket I$ of the system. Since we are interested in creation of entanglement from an initial separable state, we take for the initial state of our system the single-excitation state $\ket{I}=\ket{\Psi_{3}} =\ket{e_{1}}\otimes\ket{g_{2}}$, which corresponds to atom $1$ in the excited state and atom $2$ in the ground state. The initial state is, of course, a separable state, in other words there is no entanglement in the system at~$t=0$. For $\ket I =\ket{\Psi_{3}}$, it is easily verified that the only non-vanishing matrix elements are
\begin{eqnarray}
\rho_{ss}(0)=\rho_{aa}(0)=\rho_{sa}(0)=\rho_{as}(0)=\frac{1}{2} ,\label{5.20}
\end{eqnarray}
and then the initial density matrix has the following form
\begin{eqnarray}
\rho (0) = \left(
\begin{array}{cccc}
0 & 0 & 0 & 0 \\
0 & \rho_{ss}(0) & \rho_{sa}(0) & 0 \\
0 & \rho_{as}(0) & \rho_{aa}(0) & 0 \\
0& 0 & 0 & 0
\end{array}
\right) \ . \label{5.21}
\end{eqnarray}
According to the solutions for the density matrix elements, Eqs.~(\ref{5.18}) and (\ref{5.19}), the matrix elements which are zero at the initial time $t=0$ will remain zero for all time, except the population of the ground state $\ket{\Psi_{1}}$ which will buildup during the evolution. Hence, the diagonal form of the density matrix will be preserved during the evolution. Moreover, the dynamics of the systems can be confined to the subspace spanned by three state vectors only,~$\ket{\Psi_{1}}, \ket{\Psi_{s}}$ and~$\ket{\Psi_{a}}$. 

Our next and, in fact, the major problem is to determine if an entanglement can be generated in the system~\cite{ft03,tf04a,tf04,yz08,das08}. To examine the occurrence of entanglement, we must consider the concurrence, Eq.~(\ref{4.11}) which, on the other side, is described by the density matrix elements. Looking at the density matrix (\ref{5.21}), two conclusions can be made. First of all, since the coherence $\rho_{14}(t)$ is equal to zero for all times $t$, we see from Eq.~(\ref{4.12}) that the criterion ${\cal C}_{1}(t)$ is always negative. Consequently,~${\cal C}_{1}(t)$ will not contribute to the concurrence. Secondly, the two-photon state $\ket{\Psi_{4}}$ is not involved in the dynamics of the system which, according to Eq.~(\ref{4.13}) rules out a possibility for the threshold behavior of the criterion~${\cal C}_{2}(t)$. Under this situation, the concurrence is determined only by the criterion ${\cal C}_{2}(t)$ which would always be positive. From this, it follows that
\begin{eqnarray}
 {\cal C}(t)&\equiv& {\cal C}_{2}(t) = \sqrt{\left[\rho_{ss}(t)-\rho_{aa}(t)\right]^{2} 
 +4\left[{\rm Im}\rho_{sa}\right]^{2}} \nonumber\\
 &=& {\rm e}^{-\gamma t}\sqrt{\sinh^{2}(\gamma_{12}t) +\sin^{2}(2\Omega_{12}t)}  .\label{5.22}
\end{eqnarray}
There are two contributions to the time evolution of the concurrence. First, there is an imaginary part of the coherence between the states $\ket{\Psi_{s}}$ and $\ket{\Psi_{a}}$. Since this is off-resonance coupling, it leads to oscillation in the concurrence with frequency~$2\Omega_{12}$, the frequency difference between the two states. The second contribution is the difference between the populations of the states~$\ket{\Psi_{s}}$ and~$\ket{\Psi_{a}}$. This contribution is non-oscillatory. We see that the effect of the non-oscillatory term is apparently to lengthen the lifetime of the concurrence.
\begin{figure}[hbp]
\includegraphics[width=8cm,keepaspectratio,clip]{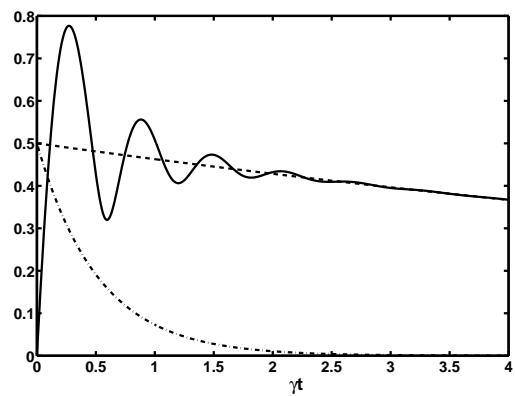}
\caption{Transient behavior of the concurrence (solid line) and the populations of the antisymmetric (dashed line) and the symmetric (dashed-dotted line) states for the atoms prepared initially in the
separable state $\ket{\Psi_{3}} =\ket{e_{1}}\otimes\ket{g_{2}}$, with
$\hat{\mu}\perp \hat{r}_{12}$, and $kr_{12}=\pi/5$.}
\label{figc2}
\end{figure}

The features described above are easily seen in Fig.~\ref{figc2}, where we plot the time evolution of the concurrence~${\cal C}(t)$, calculated from Eq.~(\ref{5.22}), together with the 
populations~$\rho_{ss}(t)$ and~$\rho_{aa}(t)$ of the symmetric and the antisymmetric states, respectively. The time evolution of the concurrence reflects the time evolution of the entanglement between the atoms. It is seen that the concurrence builds up immediately after $t=0$ and remains positive for all time, which indicates that spontaneous emission can indeed create entanglement between the initially unentangled atoms. The buildup of the concurrence in time generally consists of oscillatory and non-oscillatory components, so that two time scales of completely different behavior of the concurrence can be distinguished. At early times, the concurrence builds up in an oscillatory manner, and the oscillatory structure is smoothed out on a time scale $(\gamma +\gamma_{12})^{-1}$, the lifetime of the symmetric state. The oscillations vanish at time close to the point where the symmetric state becomes depopulated. At later times, the concurrence evolves in a non-oscillatory manner and overlaps with the population of the antisymmetric state. As a result, the concurrence decays slowly in time with the reduced rate $\gamma -\gamma_{12}$. Although the concurrence (\ref{5.22}) involves both symmetric and antisymmetric states, it is clear that crucial for the entanglement is the presence of the antisymmetric state.

The decay time of the population of the antisymmetric state, so that the transient entanglement seen in Fig.~\ref{figc2},  varies with the distance between the atoms. The time goes to infinity when $kr_{12}\rightarrow 0$. In this limit the transition rates to and from the antisymmetric state vanish and the state decouples from the remaining states. Hence, any initial population encoded into the state will remain there for all times. For example, with the initial state $\ket{\Psi_{3}}$ and $t\rightarrow \infty$, half of the population, that initially encoded into the antisymmetric state, still remains in the atomic system, and half of the population, that initially encoded into the symmetric state, is emitted into the field. As a result, the concurrence evolves to its stationary value of ${\cal C}(t)=0.5$ indicating that in the limit of $t\rightarrow\infty$, the spontaneous emission could produce steady state entanglement to the degree of $50$\% with the corresponding pure state of the system. 

The creation of the transient entanglement can be understood by considering the properties of the density 
matrix of the system. For the initial state of only one atom excited the density matrix is not diagonal 
due to the presence of coherences $\rho_{sa}(0)$ and $\rho_{as}(0)$. Since the form of the matrix is 
preserved during the evolution, it remains non-diagonal for all times. Consequently, the collective 
states are no longer the eigenstates of the system. The density matrix can be diagonalized to give new 
diagonal states. It is easy to verify that the states $\ket{\Psi_{1}}$ and $\ket{\Psi_{4}}$ remain 
unchanged, whereas the states $\ket{\Psi_{s}}$ and $\ket{\Psi_{a}}$ recombine into new diagonal symmetric
\begin{eqnarray}
\ket {\Psi_{+}} = \frac{\left[\rho_{++}(t)\!-\!\rho_{ss}(t)\right]\!\ket{\Psi_{a}}\!+\!
\rho_{as}(t)\ket{\Psi_{s}}}{\left\{\left[\rho_{++}(t)\!-\!\rho_{ss}(t)\right]^{2}
\!+\!\left|\rho_{as}(t)\right|^{2}\right\}^{\frac{1}{2}}}  ,
\end{eqnarray}
and antisymmetric
\begin{eqnarray}
\ket {\Psi_{-}} = \frac{\rho_{as}(t)\ket{\Psi_{a}}\!+\!
\left[\rho_{--}(t)\!-\!\rho_{aa}(t)\right]\!\ket{\Psi_{s}}}{\left\{\left[\rho_{--}(t)\!-\!\rho_{aa}(t)\right]^{2}
\!+\!\left|\rho_{as}(t)\right|^{2}\right\}^{\frac{1}{2}}}  ,\label{5.23}
\end{eqnarray}
states, with the eigenvalues $\rho_{++}(t)$ and $\rho_{--}(t)$, the populations of the diagonal states,
given by
\begin{eqnarray}
&&\rho_{++}(t) = \frac{1}{2}\left[\rho_{aa}(t)+\rho_{ss}(t)\right]\nonumber\\
&&\qquad \quad \ +\, \frac{1}{2}\left\{\!\left[\rho_{aa}(t)\!-\!\rho_{ss}(t)\right]^{2}
\!+\!4\!\left|\rho_{as}(t)\right|^{2}\right\}^{\!\frac{1}{2}}  ,\nonumber \\
&&\rho_{--}(t) =  \frac{1}{2}\left[\rho_{aa}(t)+\rho_{ss}(t)\right]\nonumber \\
&&\qquad \quad \ - \frac{1}{2}\left\{\!\left[\rho_{aa}(t)\!-\!\rho_{ss}(t)\right]^{2}
\!+\!4\!\left|\rho_{as}(t)\right|^{2}\right\}^{\!\frac{1}{2}}\! .\label{5.24}
\end{eqnarray}
It follows from Eq.~(\ref{5.23}) that the coherences $\rho_{sa}(t)$ and~$\rho_{as}(t)$ cause the
system to evolve between the ground state $ \ket{\Psi_{1}}$ and two "new" one-photon states $\ket {\Psi_{+}}$ and~$\ket{\Psi_{-}}$, which are linear combinations of the collective states $\ket{\Psi_{s}}$ and $\ket{\Psi_{a}}$. It is easy verified that for the initial condition (\ref{5.20}), the population $\rho_{--}(t)=0$ for all times, whereas the population $\rho_{++}(t)$ is different from zero and equals to the sum of the populations $\rho_{ss}(t)$ and~$\rho_{aa}(t)$. 
The lack of population in the states $\ket {\Psi_{-}}$ together with no population in the state $\ket
{\Psi_{4}}$ reduces the four-level system to an effective two-level system with the excited nonmaximally entangled state $\ket {\Psi_{+}}$ and the separable ground state $\ket{\Psi_{1}}$. In this case, the density matrix has a simple diagonal form
\begin{eqnarray}
     \rho (t) = \rho_{++}(t)\ket{\Psi_{+}}\bra{\Psi_{+}}
     +\rho_{11}(t)\ket{\Psi_{1}}\bra{\Psi_{1}} .\label{5.25}
\end{eqnarray}
Since $\rho_{--}(t)=0$ for all times $t$, we can easily find from~Eq.~(\ref{5.24}) that in this case
$|\rho_{as}(t)|^{2}=\rho_{aa}(t)\rho_{ss}(t)$, and then the state $\ket {\Psi_{+}}$ can be written as
\begin{eqnarray}
\ket {\Psi_{+}} = \frac{\sqrt{\rho_{ss}(t)}\ket{\Psi_{s}} +
\sqrt{\rho_{aa}(t)}\ket{\Psi_{a}}}{\sqrt{\rho_{aa}(t)+ \rho_{ss}(t)}} .\label{5.26}
\end{eqnarray}
When the collective states $\ket{\Psi_{s}}$ and $\ket{\Psi_{a}}$ are equally populated, the state $\ket {\Psi_{+}}$ reduces to a separable state 
\begin{eqnarray}
\ket {\Psi_{+}} = \ket {e_{1}}\otimes\ket {g_{2}} .\label{5.26a}
\end{eqnarray}
On the other hand, the state $\ket {\Psi_{+}}$ reduces to a maximally entangled state, $\ket {\Psi_{s}}$ or $\ket {\Psi_{a}}$, when either $\rho_{ss}(t)$ or $\rho_{aa}(t)$ is equal to zero. Since the population of the symmetric state decays faster than the antisymmetric state, see Eq.~(\ref{5.14}), at time when the state $\ket{\Psi_{s}}$ becomes depopulated, the 
state~$\ket {\Psi_{+}}$ reduces to the maximally entangled antisymmetric state~$\ket{\Psi_{a}}$. This explains why at later times the evolution of the concurrence follows the evolution of the population of the state $\ket{\Psi_{a}}$.

The above analysis give clear evidence that the creation of transient entanglement from the separable state by spontaneous emission depends crucially on the presence of the antisymmetric state.

\subsection{The Dicke model}

\noindent The most familiar model to study collective effects in spontaneous emission by a system of two or more identical atoms is the Dicke model~\cite{dic}. It was originally introduced by Dicke in his famous article published in 1954, and currently a very large literature exists on a wide variety of problems involving the model, in particular on the concepts of super-radiance and the directional propagation of light in atomic ensembles~\cite{ch03,hc07,se06}. Recently, the model has been employed in the studies of entanglement, in particular for creation of entanglement in atomic ensembles composed of a large number of trapped and cooled atoms~\cite{gr06,ps06,de07,lk09}.

The two-atom Dicke model is a simplified form of the two-atom system considered in the preceding section. It is often called, a small sample model and corresponds to a system of two atoms confined to a region much smaller than the radiation wavelength of the atomic transitions. In other words, the model assumes that the atoms are close enough that we can ignore any effects resulting from different spatial positions of the atoms. Mathematically, it is equivalent to set the phase factors $\exp(i\vec{k}\cdot{\vec r}_{i})$ associated with atomic positions, equal to one. In terms of the collective states of a two-atom system, the Dicke model involves only the triplet states $\ket{\Psi_{4}}$, $\ket{\Psi_{s}}$, and~$\ket{\Psi_{1}}$, with the antisymmetric state $\ket{\Psi_{a}}$ totally decoupled from the triplet states and thus not participating in the evolution of the two-atom system. The antisymmetric state appears as a metastable or trapping state. In the terminology of modern quantum optics, the state is an example of a decoherence-free state that any initial population will remain in the state for all~times. 

The trapping property of the antisymmetric state could have a significant effect on entanglement in the small sample model of two atoms. According to Eq.~(\ref{4.21}), the concurrence depends on the population of the antisymmetric state. Thus, an initially entangled or separable system would evolve to a steady state whose the entangled properties are determined by the initial population of the antisymmetric state. In this way, one could create a stable entanglement between the atoms simply by a suitable preparation of the atoms in the non-radiating antisymmetric state. 
For example, if the system is initially prepared in the antisymmetric state, ${\cal C}_{2}(0) = \rho_{aa}(0)=1$, and then due to the trapping property of the antisymmetric state that the population in the state does not change in time, the system could stay maximally entangled for all $t>0$ and would never disentangle.   

If the system is initially prepared in a state with no population in the antisymmetric state, $\rho_{aa}(0)=\rho_{aa}(t)=0$, and then the concurrence is determined only by the populations of the triplet states
\begin{eqnarray}
  {\cal C}(t) = \max\left\{0,\,\rho_{ss}(t) - 2\sqrt{\rho_{11}(t)\rho_{44}(t)}\right\} ,\label{5.27} 
\end{eqnarray}
Hence, one would expect that the absence of the population in the antisymmetric state could result in a large transient entanglement buildup through the population of the symmetric state. 

We demonstrate a somehow surprising result that in the Dicke model, entanglement cannot be created by spontaneous emission if one disregards the antisymmetric state. In the Dicke model, the time evolution of the diagonal density matrix elements under the spontaneous emission is determined by the following equations~\cite{ft03}
\begin{eqnarray}
 && \rho_{44}(t) = \rho_{44}(0)\,{\rm e}^{-2\gamma t}  ,\nonumber \\
 && \rho_{ss}(t) = \rho_{ss}(0)\,{\rm e}^{ -2\gamma t} 
  + 2\gamma t \rho_{44}(0)\,{\rm e}^{-2\gamma t} ,\nonumber \\
 && \rho_{aa}(t) = \rho_{aa}(0) . \label{5.28}
\end{eqnarray}
The population of the symmetric state decays with a rate double that of the single atom, and for an initial state with $\rho_{44}(0)\neq 0$, the population does not decay exponentially. The time evolution of the population is a convolution of a linear function of time and an exponential decay. 

Let us assume that the atoms are initially prepare in the separable state $\ket{\Psi_{4}}$, which implies that $\rho_{44}(0)=1$ and all other density matrix elements equal to zero. Since in the Dicke model, the population of the upper state can only decay through the symmetric state, the population will accumulate in this state during the evolution. As a result, an entanglement may emerge in the system. However, for the $\rho_{44}(0)=1$ initial condition, it follows from Eqs.~(\ref{5.27}) and (\ref{5.28}) that
\begin{eqnarray}
  {\cal C}(t) = \max\left\{0,\,\left(\tau  -2\sqrt{{\rm e}^{\tau} -1 -\tau}\right){\rm e}^{-\tau}\right\} ,\label{5.29} 
\end{eqnarray}
where $\tau =2\gamma t$. 

The exponent under the square root in Eq.~(\ref{5.29}) may be developed into series and one then obtains the following expression for the concurrence
\begin{eqnarray}
  {\cal C}(t) = \max\left\{0,\,\tau {\rm e}^{-\tau}\!\left[ 1\!-\!\sqrt{2}\left(1\!+\!\frac{1}{3}\tau
 \!+\!\cdots\right)^{\frac{1}{2}}\right]\right\} .\label{5.30} 
\end{eqnarray}
Since the term in the square brackets is always negative, this shows that no entanglement can be created during the spontaneous decay of the initial excitation. This is a surprising result, because the symmetric state can be significantly populated during the spontaneous decay and still no entanglement can be created in the system. It is easy to find from Eq.~(\ref{5.28}) that at early times the population $\rho_{ss}(t)$ increases linearly from its initial value of~$\rho_{ss}(0)=0$, attains the maximum of~$0.42$ at a time $t=1/(2\gamma)$, and then for $t>1/(2\gamma)$ decreases exponentially with the rate~$2\gamma$. 
 
The failure of the spontaneous creation of entanglement in the Dicke model is linked to the threshold behavior of the concurrence. Simply, the threshold term 
$2\sqrt{\rho_{11}(t)\rho_{44}(t)}$, which depends on the redistribution of the population between the two separable states is dominant in the concurrence for all time despite the fact that the symmetric state can be significantly populated during the evolution. In the coming sections, we shall explore in more details the role of the threshold in the evolution of initial entangled states and a delayed creation of entanglement from an initial separable state. In the terminology of sudden death of entanglement that will be introduced in the next section, we may conclude that the Dicke model is "dead" for creation of entanglement by spontaneous emission.

\section{Strange behaviors of entanglement}

\noindent In the preceding section we have developed a simple model for creation of entanglement in a system composed of two two-level atoms interacting with a common environment. We have showed how the spontaneous decay of an initial excitation encoded into a separable state of the system can create a transient entanglement between the atoms. We will now examine the opposite situation, a spontaneous disentanglement of initially entangled atoms. Typically, an initial entanglement is expected to decays exponentially in time. However, we point out some "strange" behavior of the disentanglement process of a two-atom system  undergoing spontaneous evolution that an initial entanglement encoded into the system can be lost in a very different way compared to an exponential decay. We provide a discussion of this unusual phenomenon in terms of the density matrix elements and show the connection of the phenomenon with the threshold behavior of the concurrence. We also discuss other unusual features of entanglement and disentanglement. Namely, we explore an interesting phenomenon of entanglement revival that spontaneous emission can lead to a revival of the entanglement that has already been destroyed.
In the connection to the threshold behavior of the concurrence, we will also return to the problem of the dynamical creation of entanglement from an initial separable state. We discuss a phenomenon of delayed sudden birth of entanglement that the spontaneous creation of entanglement may be postponed to later times even if the correlation between the atoms exists for all time.

\subsection{Sudden death of entanglement}

\noindent The most familiar of the "sudden" features of entanglement and disentanglement is the phenomenon of entanglement {\it sudden death}, i.e., abrupt disappearance of the entanglement at a finite time even if the correlation between the atoms exists for all time. The subject received its initial stimulus in an article by Yu and Eberly~\cite{eb04,eb07}, in which they for the first time introduced the concept of sudden death of entanglement. Many authors since have dealt with the entanglement sudden death in systems composed of two atoms or two harmonic oscillators. 
Most of the subsequent literature can be divided into two categories: (i) studies which deal with independent atoms interacting with local environments~\cite{loe}, and (ii) studies which deal with interacting atoms coupled to a common environment~\cite{coe}. We postpone the studies of the latter group to the following section, where we will mostly focus on the phenomenon of entanglement revival. Here, we focus on the original concept of Yu and Eberly, and discuss in details the phenomenon of sudden death of entanglement in a system of two independent atoms interacting with local environments. This model is also applicable to a situation of two distant atoms interacting with a common environment. At large distances, the collective parameters $\gamma_{12}$ and $\Omega_{12}$ are very small, so that the interaction between the atoms can be ignored and the atoms can be treated as independent sub-systems. 

Suppose that at $t=0$ the system of two independent atoms is prepared in a non-maximally entangled state of the form
\begin{eqnarray}
  \ket{\Upsilon_{0}} = \sqrt{q}\, \ket {e_{1}}\otimes\ket {e_{2}} 
  + \sqrt{1-q}\, \ket {g_{1}}\otimes\ket {g_{2}} ,\label{6.1}
\end{eqnarray}
where $q$ is a positive real number such that $0\leq q\leq 1$. The state corresponds to an excitation of the system into a coherent superposition of its product states in which both or neither of the atoms is excited. In the special case of $q=1/2$, the state~(\ref{6.1}) reduces to the maximally entangled Bell state~$\ket{\Psi_{s}}$. 

Let us next consider the time evolution of the concurrence when the system is initially prepared in the state~(\ref{6.1}). It is not difficult to verify that the initial values for the density matrix elements are
\begin{eqnarray}
 \rho_{44}(0) = q ,\  \rho_{14}(0) = \sqrt{q(1-q)} ,\ \rho_{11}(0) = 1-q ,\label{6.2}
\end{eqnarray}
and the other matrix elements, the populations of the symmetric and antisymmetric states, and all one-photon coherences are zero, i.e. $\rho_{ss}(0)=\rho_{aa}(0)=0$ and $\rho_{es}(0)=\rho_{ea}(0)=\rho_{sg}(0)=\rho_{ag}(0)=\rho_{as}(0)=0$. According to Eq.~(\ref{5.19}), the coherences will remain zero for all time, that they cannot be produced by spontaneous decay. However, the populations $\rho_{ss}(t)$ and $\rho_{aa}(t)$ can buildup during the evolution. This implies that for all times, the density matrix of the system spanned in the basis of the collective states (\ref{3.9}), is in the $X$-state form
\begin{eqnarray}
  \rho(t) = \left(
    \begin{array}{cccc}
      \rho_{11}(t) & 0 & 0 & \rho_{14}(t) \\
      0 & \rho_{ss}(t) & 0 & 0\\
      0 & 0 & \rho_{aa}(t) & 0\\
      \rho_{41}(t) & 0 & 0 &\rho_{44}(t)
    \end{array}\right) ,\label{6.3}
\end{eqnarray}
with the density matrix elements evolving as
\begin{eqnarray}
 && \rho_{44}(t) = q\,{\rm e}^{-2\gamma t}  ,\nonumber \\
 && \rho_{14}(t) = \sqrt{q(1-q)}\,{\rm e}^{-(\gamma -2i\omega_{0})t}  ,\nonumber \\
 && \rho_{ss}(t) = \rho_{aa}(t) = q \left(1 -{\rm e}^{-\gamma t}\right){\rm e}^{-\gamma t} ,\label{6.4}
\end{eqnarray}
subject to conservation of the trace of $\rho(t)$: $\rho_{11}(t)=1-\rho_{ss}(t)-\rho_{aa}(t)-\rho_{44}(t)$. It is to be noticed that the symmetric and antisymmetric states are equally populated for all time. This holds for any initial state and results from the fact that the atoms radiate independently from each other. In this case, the populations of the collective states decay with the same rate, equal to the single-atom damping rate. Therefore, the initial relation between the populations cannot be changed during the spontaneous emission.

The density matrix~(\ref{6.3}) leads to a particularly simple expression for the concurence. In general, the concurrence is given in terms of two entanglement criteria~${\cal C}_{1}(t)$ and ${\cal C}_{2}(t)$, as seen from Eq.~(\ref{4.11}). However, with the initial state (\ref{6.1}), the two-photon coherence $\rho_{14}(t)$ is different from zero and the symmetric and antisymmetric states are equally populated for all time. As a consequence, the criterion ${\cal C}_{2}(t)$ is always negative, irrespective of $q$ and times $t$. Therefore, entangled properties of the system are solely determined by the criterion ${\cal C}_{1}(t)$. On substituting from Eq.~(\ref{6.4}) into Eq.~(\ref{4.12}), we obtain the following expression for the concurrence
\begin{eqnarray}
  {\cal C}(t) = \max\left\{0,\,\Lambda(t)\, {\rm e}^{-\gamma t}\right\} ,\label{6.5} 
\end{eqnarray}
where
\begin{eqnarray}
  \Lambda(t)= 2\sqrt{q(1-q)}\left[1 -\sqrt{\frac{q}{1-q}}\left(1-{\rm e}^{-\gamma t}\right)\right] .\label{6.5a}
\end{eqnarray}
This shows that the major features of the entanglement are determined by the properties of $\Lambda(t)$ which, on the other hand, 
is dependent on the parameter~$q$. We see that there is a threshold for values of $q$; $q=1/2$, below which $\Lambda(t)$ is always 
positive. 

However, above the threshold,  $\Lambda(t)$ can take negative values indicating that the initial entanglement can vanish at a finite time. Consequently, the sudden death of the entanglement is possible for initial states with $q > 1/2$. Since $\rho_{44}(0)=q$, we can conclude that the entanglement sudden death is ruled out for the initially not inverted system.

\begin{figure}[hbp]
\includegraphics[width=8cm,keepaspectratio,clip]{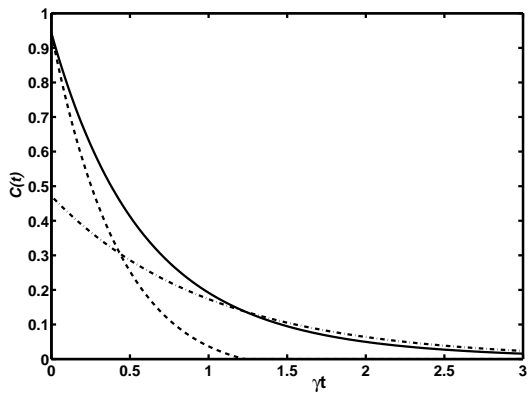}
\caption{Time evolution of the concurrence ${\cal C}(t)$ for two values of the parameter $q$: $q=1/3$ (solid line) and $q=2/3$ (dashed line). We also plot the time evolution of the two-photon 
correlation $|\rho_{14}(t)|$ (dashed-dotted line), with $q=2/3$. }
\label{figc3}
\end{figure}

Figure~\ref{figc3} shows the concurrence ${\cal C}(t)$, calculated from Eq.~(\ref{6.5}), as a function of time for two different values of the parameter $q$. It is evident from the figure that for $q<1/2$ the initial entanglement decays exponentially in time without any discontinuity. The entanglement sudden death appears for $q>1/2$ that the concurrence decays in a non-exponential way and vanishes at a finite time. In addition, we plot the two-photon coherence $|\rho_{14}(t)|$ for $q=2/3$. It is apparent that the coherence decays exponentially in time, which clearly illustrates that the entanglement disappear at finite time despite the fact that the two-photon coherence is different from zero for all time. 

As we have already stated, time at which the entanglement disappear is a sensitive function of the initial conditions determined by the parameter $q$. It is easily verified from Eq.~(\ref{6.5}) that the time $t_{d}$ at which the entanglement disappears is given by
\begin{eqnarray}
  t_{d} = \frac{1}{\gamma}\ln\left(\frac{q+\sqrt{q(1-q)}}{2q-1}\right)  . \label{6.6} 
\end{eqnarray}
The time $t_{d}$ gives the collapse time of the entanglement beyond which the entanglement disappears. The dead zone of the entanglement continues till infinity that the entanglement never revive. It may continue for a finite rather than infinite time that under some circumstances the already dead entanglement may revive after some finite time. A revival of the entanglement may occur when the atoms directly interact with each other. We leave the discussion of this problem to the following section. 

Here, we would like to point out that first demonstration of entanglement sudden death has recently been reported by Almeida~{\it et al.}~\cite{al07}.  The apparatus used in the experiment involved a tomographic reconstruction of the density matrix and from it the concurrence by measuring polarization entangled photon pairs produced in the process of spontaneous parametric down-conversion by a system composed of two adjacent nonlinear crystals. One of the crystals produced photon pairs with $V$-polarization and the other produced pairs with $H$-polarization. Parametric down conversion is a nonlinear process used to produce polarization entangled photon pairs, which are manifested by the simultaneous or nearly simultaneous production of pairs of photons in momentum-conserving, phase matched modes.
Since the pairs of polarized photons are spatially indistinguishable, they are described by a pure state
\begin{eqnarray}
  \ket\Phi = |\alpha|\ket{HH} +|\beta|{\rm e}^{i\delta}\ket{VV}  . \label{6.7} 
\end{eqnarray}
where the coefficients $|\alpha|$ and $|\beta|$, and the phase $\delta$ were adjusted by applying 
half- and quarter-wave plates in the laser beam pumping the crystals to control the creation of pairs of a desired polarization.
\begin{figure}[hbp]
\includegraphics[width=8cm,keepaspectratio,clip]{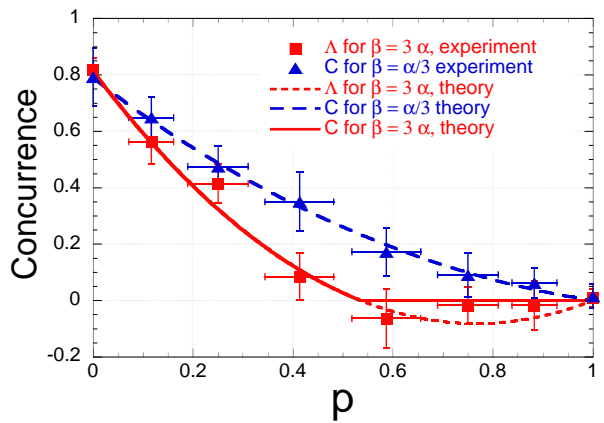}
\caption{Results of the Almeida {\it et al.} experiment demonstrating the sudden death of entanglement. The theoretical prediction for the concurrence $C$ is plotted as a function of $p=1-\exp(-\gamma t)$ for two values of the ratio $|\beta|/\alpha|$: $|\beta|/\alpha|=3$ (solid line) and $|\beta|/|\alpha|=0.333$ (dashed line). The squares and triangles are experimentally measured values for the concurrence. Reproduced from Ref.~\cite{al07} with permission.}
\label{figc4}
\end{figure}

In the experiment, they measured the decay of a single polarized beam, serving as a qubit, which was monitored by generating pairs of photons of the same polarization and registering coincidence counts with one photon propagating through the interferometer and the other serving as a trigger. 

Figure~\ref{figc4} shows the results of the measured concurrence for two different values of the ratio $|\beta|/|\alpha|$. The solid and dashed lines represent the theoretically 
predicted concurrence. The measured values for the concurrence are found to be in good agreement with the theoretical predictions. Thus, it was confirmed that entanglement may display the sudden death feature and that the spontaneous decay of the initial entanglement depends on the relation between the coefficients $|\alpha$ and~$|\beta|$. For~$|\beta|<|\alpha|$ the entanglement decays exponentially in time, while for~$|\beta|>|\alpha|$, the entanglement vanishes at finite time. 

In closing this section, we summarize the research on the entanglement sudden death. The phenomenon has been extensively studied in recent years and it has been demonstrated that the entanglement sudden death can be obtained in a variety of systems including atoms coupled to single-mode cavities, atoms coupled to local environments appearing as multi-mode reservoirs to the atoms, atoms interacting with a common environment. Other studies have been carried out for independent and also for interacting harmonic oscillators. Studies have also been carried out for non-Markovian and non-RWA situations where the sudden death of entanglement can be observed in completely different regime of the parameters. 

Finally, we briefly comment on the evolution of an entanglement initially encoded into spin-anticorrelated states. We have just shown that the phenomenon of entanglement sudden death is characteristic of two-photon or spin correlated entangled states. It is easy to conclude from Eq.~(\ref{4.21}) that an initial entanglement encoded in a spin anti-correlared state, the symmetric or antisymmetric states, would decay asymptotically in time without any discontinuity. Expression (\ref{4.21}) clearly shows that necessary condition for entanglement sudden death is a simultaneous population of the ground $\ket{\Psi_{1}}$ and the two-photon~$\ket{\Psi_{4}}$ states.

\subsection{Revival of entanglement}

\noindent We have already shown that under appropriate conditions, two initially entangled and afterwards not interacting qubits can become completely disentangled in a finite time. The term "entanglement sudden death" has been introduced to describe this unusual feature. 
In this section, we extend these analysis to include the direct interactions between the atoms  and study in details the time evolution of an initial entanglement. As we shall see, the inter-atomic interactions can lead to transient features of a spontaneously evolving entanglement which are quite distinct to those seen in their absence. The interactions impose strong qualitative changes to the time evolution of entanglement that the irreversible spontaneous decay can lead to a {\it revival} of the entanglement that has already been destroyed~\cite{ft06}. We pay particular attention to the role of the collective damping $\gamma_{12}$ and the dipole-dipole interaction $\Omega_{12}$ in entanglement revival and find that the feature depends critically on whether or not the collective damping is present. We study in detail the dependence of the revival time on the initial state of the system and on the separation between the atoms.

Consider again the initial state of the system given by Eq.~(\ref{6.1}), but now assume that the atoms can interact with each other. In this case, the initial values of the density matrix of the system as the same as for independent atoms, but the time evolution of the density matrix elements is different. It is now given by 
\begin{eqnarray}
 && \rho_{44}(t) = q\,{\rm e}^{-2\gamma t}  ,\nonumber \\
   && \rho_{14}(t) = \sqrt{q(1-q)}\,{\rm e}^{-(\gamma -2i\omega_{0})t}  ,\nonumber \\
  &&\rho_{ss}(t) = q\, \frac{\gamma +\gamma_{12}}
   {\gamma -\gamma_{12}}\!\left[{\rm e}^{(\gamma-\gamma_{12}) t} -1\right]{\rm e}^{-2\gamma t} ,\nonumber \\
  && \rho_{aa}(t) = q\,\frac{\gamma -\gamma_{12}}
   {\gamma +\gamma_{12}}\!\left[{\rm e}^{(\gamma +\gamma_{12})t} -1\right]{\rm e}^{-2\gamma t}  ,\label{6.8}
\end{eqnarray}
and $\rho_{11}(t)= 1- \rho_{44}(t)-\rho_{ss}(t)-\rho_{aa}(t)$. As before for independent atoms, the remaining density matrix elements are equal to zero.
By comparing Eq.~(\ref{6.8}) with Eq.~(\ref{6.4}) for independent atoms, we see that a major difference between the time evolution of the density matrix elements for independent and interacting atoms is that in the later the symmetric and antisymmetric states are populated with different rates. As a result, the transient buildup of the populations from $\rho_{ss}(0)=\rho_{aa}(0)=0$ at $t=0$ will lead to unequal populations of the states for all later times $t>0$. 

A direct consequence of unequal populations of the symmetric and antisymmetric states is seen in the transient evolution of an initial entanglement which may now depend on both criteria ${\cal C}_{1}(t)$ and ${\cal C}_{2}(t)$, that are needed to construct the concurrence ${\cal C}(t)$, rather than on one of them. According to Eqs.~(\ref{4.11}) and (\ref{6.2}), a two-atom system initially prepared in the state~(\ref{6.1}) can be entangled according to the criterion ${\cal C}_{1}(t)$, and the degree to which the system is initally entangled~is ${\cal C}_{1}(0)= 2\sqrt{q(1-q)}$. 
The other criterion, ${\cal C}_{2}(t)$, is negative at $t=0$. Nevertheless, during the spontaneous evolution, a population builds up in the symmetric and antisymmetric states and according to Eq.~(\ref{6.8}), $\rho_{ss}(t)\neq\rho_{aa}(t)$ for all $t>0$. As discussed in Sec.~4, an unequal  population of the states gives rise to positive values of the criterion~${\cal C}_{2}(t)$ which then may result in an entanglement of the atoms. Of course, in order to create the entanglement, the the population difference has to overweight the threshold term in the criterion~${\cal C}_{2}(t)$. If this is the case, we could see an entanglement having its origin in the correlation determined by the criterion~${\cal C}_{2}(t)$.

According to the above discussion, the concurrence, in general,  is determined by both, ${\cal C}_{1}(t)$ and ${\cal C}_{2}(t)$ criteria 
\begin{eqnarray}
  {\cal C}(t) = \max\left\{0,\,{\cal C}_{1}(t),\,{\cal C}_{2}(t)\right\} ,\label{6.9} 
\end{eqnarray}
where
\begin{eqnarray}
{\cal C}_{1}(t) = 2|\rho_{14}(t)| -\left[\rho_{ss}(t)+\rho_{aa}(t)\right] ,\label{6.10}
\end{eqnarray}
and
\begin{eqnarray}
{\cal C}_{2}(t) = |\rho_{ss}(t)-\rho_{aa}(t)| -2\sqrt{\rho_{11}(t)\rho_{44}(t)} .\label{6.11} 
\end{eqnarray}

We point out again, that initially at $t=0$, ${\cal C}_{2}(0)$ is negative and then for $t>0$ it may rise to positive values. Thus, an entanglement can be generated during the spontaneous evolution of the system. This kind of entanglement is an example of spontaneously generated entanglement. 

\begin{figure}[hbp]
\includegraphics[width=8cm,keepaspectratio,clip]{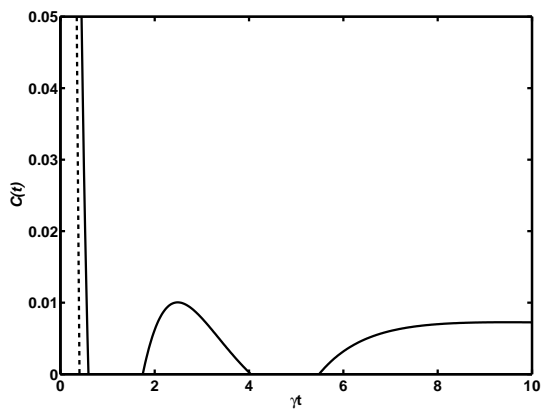}
\caption{Transient evolution of the concurrence ${\cal C}(t)$ for the initial state (\ref{6.1}) with $q=0.9$. The solid line represents ${\cal C}(t)$ for the collective system with the interatomic separation $r_{12}=\lambda/20$. The dashed line shows ${\cal C}(t)$ for independent atoms, $\gamma_{12}=0$.}
\label{figc5}
\end{figure}

Let us illustrate the time evolution of the concurrence for a collective system with $\gamma_{12}\neq 0$.  Figure~\ref{figc5} shows the deviation of the time evolution of the concurrence for two interacting atoms $(\gamma_{12}\neq 0)$ from that of independent atoms $(\gamma_{12}=0)$. In both cases, the initial entanglement falls sharply in time. For independent atoms we observe the collapse of the entanglement without any revivals. However, for interacting atoms, the system collapses over a short time and remains disentangled until a time $t_{r}\approx 1.5/\gamma$ at which, somewhat counterintuitively, the entanglement revives. This revival then decays asymptotically to zero, but after some period of time it revivals again. Thus, important features of the transient evolution of the entanglement is that there are two time intervals at which the entanglement vanishes and two time intervals at which the entanglement revives. 

\begin{figure}[hbp]
\includegraphics[width=8cm,keepaspectratio,clip]{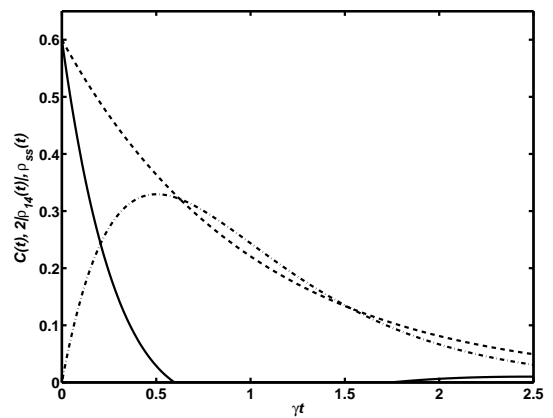}
\caption{The transient behavior of the entanglement criterion ${\cal C}_{1}(t)$ (solid line), the two-photon coherence $2|\rho_{14}(t)|$ (dashed line) and the population $\rho_{ss}(t)$ (dashed-dotted line) plotted for the same parameters as in Fig.~\ref{figc5}.}
\label{figc6}
\end{figure}

The origin of the sudden death and revivals of the entanglement can be explained in terms of the populations of the collective states and the rates with which the populations and the two-photon coherence decay. It is easily verified from Eq.~(\ref{6.8}) that at early times of~$\gamma t \ll 1$, the population $\rho_{aa}(t)\approx 0$, but $\rho_{ss}(t)$ is large. Moreover, the two-photon coherence $\rho_{14}(t)$ is also large at that short time. Thus, the entanglement behavior can be understood entirely in terms of the population $\rho_{ss}(t)$ and the coherence $\rho_{14}(t)$. 
In this connection, we compare in Fig.~\ref{figc6} the transient behavior of the concurrence ${\cal C}(t)$, with the transient properties of the population $\rho_{ss}(t)$, and the coherence $\rho_{14}(t)$ for the same choice of parameters as in Fig.~\ref{figc5}. As can be seen from the graphs, the entanglement vanishes at the time where $\rho_{ss}(t)$ is maximal, and remains zero until the time $t_{r}$ at which $\rho_{ss}(t)$ becomes smaller than $\rho_{14}(t)$. Clearly, the first dead zone of the entanglement arises due to the significant accumulation of the population in the symmetric state.

The reason for the first revival of the entanglement is in the unequal decay rates of the population $\rho_{ss}(t)$ and the coherence $\rho_{14}(t)$. According to the criterion~${\cal C}_{1}(t)$, entanglement would be generated if the sum of the populations of the single-photon collective states is small while at the same time the two-photon coherence is large. Since the coherence $\rho_{14}(t)$ decays more slowly than the population of the symmetric state, and keeping in mind that the population of the antisymmetric state is negligibly small at that time, the two-photon coherence can become significant and entanglement generated over some period of time during the decay. Note, that this is the same coherence that produced the initial entanglement.
Therefore, we may call the first revival as an "echo" of the initial entanglement that has been unmasked by destroying the population of the symmetric state. 

The revival of the entanglements depends on the initial state of the system, which is determined by the parameter $q$. There is a range of values of $q$ for which the concurrence is positive. This range defines the values of~$q$ for which the first revival occurs. The range depends on the values of $\gamma_{12}$ and the revival is most pronounced for~$q$ over the range $0.85\leq q\leq 0.95$. For values outside this range, no revival is observed. Note that the values of~$q$ at which a pronounced revival exists are close to one. This is not surprised because for $q>1/2$ the system is initially inverted that increases the rate of spontaneous~emission.

We may estimate approximate values for the first death and revival times of the entanglement.
Using Eqs.~(\ref{4.12}) and (\ref{6.8}) we find that in the case of small inter-atomic separations, at which $\gamma_{12}\approx \gamma$, the entanglement criterion ${\cal C}_{1}(t)$ attains  zeroth at times satisfying the relation
\begin{eqnarray}
     \gamma t\, \exp(-\gamma t) = \sqrt{\frac{1-q}{q}} , \label{6.12}
\end{eqnarray}
which for $q>0.85$ has two nondegenerate solutions,~$t_{d}$ and $t_{r}>t_{d}$. The solution $t_{d}$ gives the death time of the entanglement beyond which it disappears. At this time, the atoms becomes disentangled and remain separable until the time $t_{r}$, at which the entanglement revives. The analytic expression (\ref{6.12}) makes clear that for the parameters of Fig.~\ref{figc5}, the system disentangles at $t_{d}=0.6/\gamma$ and entangles again at the time~$t_{r}=1.7/\gamma$.

Figure~\ref{figc5} shows that entanglement revivals not once but twice, and the second revival occurs at long times. The long time entanglement has completely different origin than that at short times. At long times, all the density matrix elements are almost zero except $\rho_{aa}(t)$, which remains large even for long times due to the small decay rate of the antisymmetric state.
These considerations imply that the long time entanglement comes from the population $\rho_{aa}(t)$ which determines positive values of the criterion~${\cal C}_{2}(t)$. The entanglement decays asymptotically with the reduced rate $\gamma -\gamma_{12}$, and vanishes at 
\begin{eqnarray}
    t_{d2} \approx \frac{1}{\gamma}{\rm arcsinh}\left(\sqrt{\frac{1-q}{q}}\frac{2\gamma}{\gamma -\gamma_{12}}\right)  . \label{6.13}
\end{eqnarray}
This explicitly shows that lifetime of the entanglement depends on the collective damping parameter $\gamma_{12}$ and approaches infinity when $\gamma_{12}\rightarrow \gamma$.

We may summarize this section that the most interesting consequence of the collective damping is the possibility of entanglement revival. What is even more remarkable, the revival of entanglement occurs under the spontaneous evolution of the system without the presence of any external coherent fields. We may say that entanglement is reversible even if it evolves under the irreversible process. 

Finally, we would like to point out that entanglement revival is associated not only with the collective damping, but may occur 
in a number of other situations. For example, it has been found that entanglement revival may occur under non-Markovian dynamics 
of two independent atoms coupled to local reservoirs~\cite{nma}. Here, the memory effects of the non-Markovian dynamics may result 
in the "return" of the correlation from the reservoirs to the atoms. Entanglement revival has also been found in reversible 
systems whose dynamics are determined by the Jaynes-Cummings Hamiltonian~\cite{cr09,yy06,yy07}. In these models, an initial 
entanglement encoded into atoms undergoes a coherent transfer forth and back to the field modes, so it returns periodically 
to the atoms. Some examples of such coherent oscillations of entanglement will be discussed in details in~Sections~8 and~9.

\subsection{Sudden birth of entanglement}

\noindent We have already seen that entanglement can exhibit strange features, sudden death and revival, which are characteristic of the dynamics of a special class of initial two-photon entangled states. In this section, we discuss an another noteworthy strange feature of transient entanglement, a delayed creation ({\it sudden birth}) of entanglement~\cite{ft08,sbe}. The phenomenon arises for a different set of initial states of the system, and is characteristic of the transient dynamics of one-photon entangled states. As in the Sec.~5.1, we study the time evolution of the concurrence of a two-atom system initially prepared in a separable state. The model differs from that discussed in Sec.~5.1 in the initial state, where we considered only the case of the initial one-photon state $\ket{\Psi_{3}}$. The system, of course, can be prepared in other separable states. Here, we choose the separable two-photon state~$\ket{\Psi_{4}}$ as the initial state of the system and consider the transient spontaneous dynamics of the concurrence. The model is interesting in that it provides an example of the creation of entanglement on demand in the presence of a dissipative environment.  

Suppose that at $t=0$ the system is prepared in the separable two-photon state $\ket{\Psi_{4}}$. In this case, $\rho_{44}(0)  = 1$ and all the remaining density matrix elements are zero. According to Eq.~(\ref{5.19}), the off-diagonal terms (coherences) will remain zero for all time, but the populations $\rho_{ss}(t)$, $\rho_{aa}(t)$ and $\rho_{11}(t)$ will buildup during the evolution. This implies that for~$t>0$, the density matrix of the system spanned in the basis of the collective states~(\ref{3.9}), is of a diagonal form
\begin{eqnarray}
  \rho(t) = \left(
    \begin{array}{cccc}
      \rho_{11}(t) & 0 & 0 &0 \\
      0 & \rho_{ss}(t) & 0 & 0\\
      0 & 0 & \rho_{aa}(t) & 0\\
      0 & 0 & 0 &\rho_{44}(t)
    \end{array}\right) ,\label{6.15}
\end{eqnarray}
with the density matrix elements given by
\begin{eqnarray}
 && \rho_{44}(t) = {\rm e}^{-2\gamma t}  ,\nonumber \\
 &&\rho_{ss}(t) = \frac{\gamma +\gamma_{12}}
   {\gamma -\gamma_{12}}\!\left[{\rm e}^{(\gamma-\gamma_{12}) t} -1\right]{\rm e}^{-2\gamma t} ,\nonumber \\
 && \rho_{aa}(t) = \frac{\gamma -\gamma_{12}}
   {\gamma +\gamma_{12}}\!\left[{\rm e}^{(\gamma +\gamma_{12})t} -1\right]{\rm e}^{-2\gamma t} ,\label{6.16}
\end{eqnarray}
and $\rho_{11}(t)= 1- \rho_{44}(t)-\rho_{ss}(t)-\rho_{aa}(t)$.

After these preliminaries, we now turn to determine which of the two criteria, ${\cal C}_{1}(t)$ and/or ${\cal C}_{2}(t)$ could be positive. Looking at the density matrix (\ref{6.15}), one can easily find that in this case ${\cal C}_{1}(t)$ is always negative. Thus, positive values of the concurrence are  determined only by the criterion ${\cal C}_{2}(t)$:
\begin{eqnarray}
  {\cal C} = \max\left\{0,\,{\cal C}_{2}(t)\right\} ,\label{6.17} 
\end{eqnarray}
with
\begin{eqnarray}
 {\cal C}_{2}(t) = \left|\rho_{ss}(t) -\rho_{aa}(t)\right| - 2\sqrt{\rho_{11}(t)\rho_{44}(t)} .\label{6.18}
\end{eqnarray}

We now use Eq.~(\ref{6.17}) to discuss the ability of the system to generate entanglement in the process of spontaneous emission. It is clear from Eq.~(\ref{6.18}) that entanglement, if any, may only result from an unequal population of the symmetric and antisymmetric states. 
When the system is prepared in the state $\ket{\Psi_{4}}$, the resulting spontaneous transitions are cascades: The system decays first to the intermediate states $\ket{\Psi_{s}}$ and $\ket{\Psi_{a}}$, from which then decays to the ground state $\ket{\Psi_{1}}$. 
Since the transition rates to and from the states $\ket{\Psi_{s}}$ and $\ket{\Psi_{a}}$ are different when $\gamma_{12}\neq 0$, there appears unbalanced population distribution between these states. According to Eq.~(\ref{6.18}),  this may result in a transient entanglement between the atoms.
\begin{figure}[hbp]
\includegraphics[width=8cm,keepaspectratio,clip]{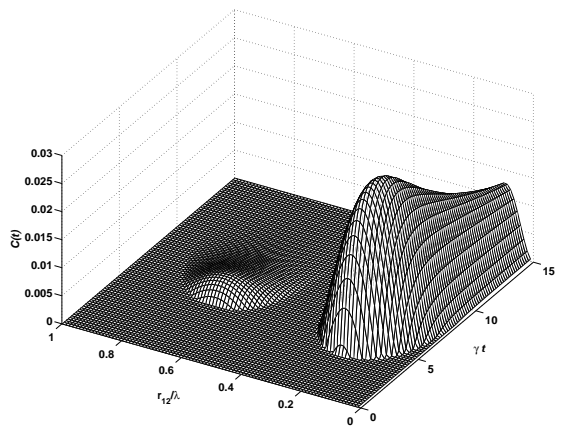}
\caption{The time evolution of the concurrence and its dependence on the distance between two, initially inverted qubits and $\hat{\mu}\perp \hat{r}_{12}$.}
\label{figc7}
\end{figure}

To visualize the behavior of the concurrence, we plot~${\cal C}(t)$ as a function of time and the distance between the atoms in Fig.~\ref{figc7}. 
We see that there is no entanglemnet at earlier times, but suddenly at some finite time an entanglement emerges. However, it happens only for a limited range of the distances $r_{12}$. It is easily verified that the "islands" of entanglement seen in~Fig.~\ref{figc7} appear at distances for which $\gamma_{12}$ is different from zero. Again, it reflects the role of the collective damping which is responsible for creation of entanglement by spontaneous emission from an initial separable state. This result also demonstrates how the collective damping leading to entanglement between the atoms becomes less important as the interatomic distance is increased. Atoms separated by more than just a few wave-lengths become separable.

We have seen that $\gamma_{12}$ modifies the damping rates of the transitions between the collective states and, on the other hand, introduces two time scales for the decay of the population of the system. 
One can easily find that the entanglement seen in Fig.~\ref{figc7} decays out on a time 
scale $(\gamma -\gamma_{12})^{-1}$ that is the time scale of the population decay from the antisymmetric state. Again, this shows that crucial for entanglement creation by spontaneous emission is the presence of the antisymmetric state. This is perhaps not surprising, as the population of the antisymmetric state builds up on a much longer time scale than the population of the symmetric state. At long times, the antisymmetric state will posse a large population with with no population left in the symmetric state. What is surprising and, in fact, is very similar to the situation found in the Dicke model that at early times $\gamma t\ll 1$ and $\gamma_{12}\neq 0$,  the symmetric state is significantly populated with almost no population in the antisymmetric state, and no entanglement is created. 
\begin{figure}[hbp]
\includegraphics[width=8cm,keepaspectratio,clip]{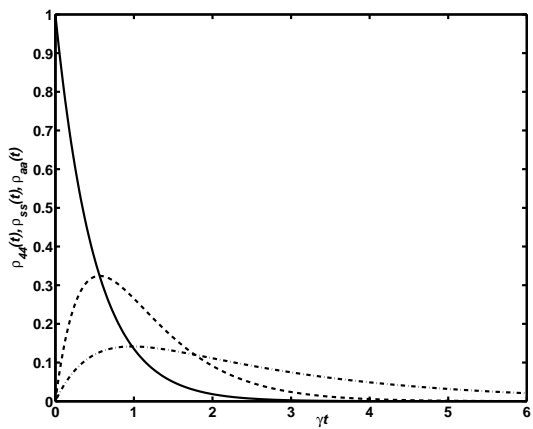}
\caption{The time evolution of the populations $\rho_{44}(t)$ (solid line), $\rho_{ss}(t)$ (dashed line) and $\rho_{aa}(t)$ (dash-dotted line) for $\theta =0$, $r_{12}/\lambda =0.25$ and $\vec{\mu}\perp \vec{r}_{12}$.}
\label{figc8}
\end{figure}

The above considerations are supported by the analysis of the time evolution of the population of the excited states of the system that is illustrated in Fig.~\ref{figc8}. It is quite evident from the figure that at the time $t\approx 4/\gamma$ when the entanglement starts to build up, the antisymmetric state is the only excited state of the system being populated. Clearly, the effect of the delayed creation of entanglement is attributed to the slow decay rate of the antisymmetric state. The state decays on the time scale of $(\gamma -\gamma_{12})^{-1}$ that is much shorter than the decay time of the symmetric and the upper states.

Concluding this section, we note that sudden birth of entanglement is characteristic of initial states which include the two-photon state with both qubits inverted. The sudden death of entanglement is not found in the small sample Dicke model that ignores the evolution of the antisymmetric state. It is also not found in a system with initially only one qubit excited. In experimental practice, the initial conditions of both atoms inverted that can be done by using
a standard technique of a short $\pi$ pulse excitation. One could also use a short $\pi/2$ pulse excitation which leaves atoms separable and simultaneously prepared in the superposition of their energy states that includes the state with both atoms inverted.

\section{Two atoms in a single-mode cavity}

\begin{figure}[hbp]
\includegraphics[width=6cm,keepaspectratio,clip]{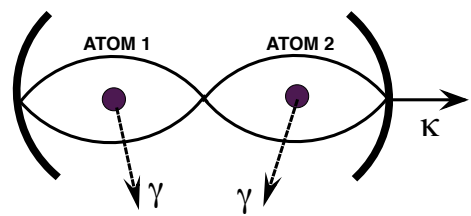}
\caption{Schematic diagram of the system composed of two atoms coupled to a single-mode standing-wave cavity. Each atom is located in a separate half-wavelength of the cavity mode.}
\label{figc9}
\end{figure}

\noindent The systems we have considered so far involved two atoms interacting in free space with a common environment that appears as multi-mode reservoirs to the atoms. In this section, we consider a different scheme where the atoms are confined to a single-mode cavity, as illustrated in Fig.~\ref{figc9}. It is reasonable to expect that inside the cavity the atomic dynamics will be strongly affected~\cite{os01,zg00,xl05}.

We discuss the entanglement of two identical atoms which, in general, could be in nonequivalent positions inside the cavity field~\cite{nf07}. We model this situation by assuming that the atoms experience different amplitudes of the cavity field which, on the other side, leads to different coupling strengths of the atoms to the cavity field. We derive the master equation for the density operator of the atoms by using the technique of adiabatic elimination of the cavity mode. We work in the limit of a large detuning of the cavity field frequency from the atomic resonance and/or a large cavity damping. With such an approximation, we can eliminate the dynamics of the cavity mode leaving the master equation for the atoms alone. The analogy of this system with that of a single two-level atom driven by a detuned coherent field is exploited and discussed. Next, we discuss in detail entangled properties of the system and the dependence of the entanglement on the evolution time and the position of the atoms inside the cavity mode.

We begin by a brief outline of the derivation of the master equation for the density operator of two atoms coupled to a cavity mode. As before, the atoms are modeled as two-level system (qubits) with upper levels $\ket{e_{i}}$,\ $(i=1,2)$, lower levels~$\ket{g_{i}}$, connected by an electric dipole transition and separated by energy $\hbar\omega_{0}$. The atoms are coupled to a standing-wave cavity mode with the coupling constants $g(\vec{r}_{i})$, having units of frequency and being proportional to the electric dipole matrix elements of the atoms. In addition, the atoms are damped at the rate $\gamma$ by spontaneous emission to modes other than the privileged cavity mode. Moreover, we assume that the atoms are separated by many optical wavelengths and therefore ignore the direct interaction between them, i.e. the collective parameters  $\gamma_{12}=0$ and $\Omega_{12}=0$. This also allows a selective preparation of the atoms such that a given set of initial conditions for the atomic states is achieved.
The cavity mode is damped with the rate $\kappa$ and its frequency $\omega_{c}$ is significantly detuned from the atomic transition frequency $\omega_{0}$, so there is no direct exchange of photons between the atoms and the cavity mode. The behavior of the total system, the atoms plus the cavity mode, is described by the density operator~$\rho_{T}$,  which in the interaction picture satisfies the master equation
\begin{eqnarray}
\frac{\partial \rho_{T}}{\partial t} =  -\frac{i}{\hbar}[H_{0},\rho_{T}] -\frac{1}{2}\kappa{\cal L}_{c}\rho_{T} -\frac{1}{2}\gamma{\cal L}_{a}\rho_{T} ,\label{7.1}
\end{eqnarray}
where
\begin{eqnarray}
     H_{0} = \sum_{j=1}^{2}\left[g(\vec{r}_{j})a S_{j}^{+}{\rm e}^{-i\Delta t} + {\rm H.c.}\right] \label{7.2}  
\end{eqnarray}
is the Hamiltonian describing the interaction, in the rotating-wave (RWA) approximation, between the cavity field and the atoms, and
\begin{eqnarray}
&&{\cal L}_{c}\rho_{T} =  a^{\dagger}a\rho_{T} + \rho_{T}a^{\dagger}a -2a\rho_{T} a^{\dagger} , \nonumber\\
&&{\cal L}_{a}\rho_{T} = \sum_{j=1}^{2}\!\left( \left[\rho_{T} S_{j}^{+},S_{j}^{-}\right]\!+\!\left[
      S_{j}^{+},S_{j}^{-}\rho_{T}\right]\right) ,\label{7.3} 
\end{eqnarray}
are the Liouvillian operators representing losses in the system by the cavity damping and by spontaneous emission, respectively. The operators $S_{j}^{+}\,(S_{j}^{-})$ and $S_{j}^{z}$ are the usual raising (lowering) and population difference operators of the $j$th atom, respectively, $a$ and $a^{\dagger}$ are the cavity-mode annihilation and creation operators,~$\Delta =\omega_{c}-\omega_{0}$ is the detuning of the cavity-mode frequency from the atomic transition frequency, and $\vec{r}_{j}$ is the position coordinate of the $j$th atom within the cavity mode.

To isolate the atomic dynamics, we introduce the photon number representation for the density operator with respect to the cavity mode
\begin{eqnarray}
 \rho_{T}  = \sum_{m,n=0}^{\infty}\rho_{mn}\ket m \bra n  , \label{7.4} 
\end{eqnarray}
where $\rho_{mn}$ are the density matrix elements in the basis of the photon number states of the cavity mode. To avoid population of higher levels we assume that the cavity mode is strongly detuned from the atomic transition frequency. In this case, we may assume that only the two lowest energy levels, the ground state $(n=0)$ and the one-photon state $(n=1)$ of the cavity mode can be populated. Thus, neglecting population in all but the two lowest levels, we find from the master equation~(\ref{7.1}) the following equations of motion for the diagonal field-matrix elements, the population of the photon-number states
\begin{eqnarray}
 && \dot{\rho}_{00} = -i\sum_{j=1}^{2}\left(g_{j}S_{j}^{+}\tilde{\rho}_{10} - g_{j}^{\ast}\tilde{\rho}_{01}S_{j}^{-}\right) 
  - \frac{1}{2}\gamma {\cal L}_{a}\rho_{00} ,\nonumber\\
&&\dot{\rho}_{11} = -i\sum_{j=1}^{2}\left(g_{j}^{\ast}S_{j}^{-}\tilde{\rho}_{01} - g_{j}\tilde{\rho}_{10} S_{j}^{+}\right) \nonumber\\
  && \quad \ \ -\, \kappa\rho_{11} - \frac{1}{2}\gamma {\cal L}_{a}\rho_{11} ,\label{7.5}
\end{eqnarray}
and the off-diagonal density matrix elements, the coherences  
\begin{eqnarray} 
&& \dot{\tilde{\rho}}_{01} = \left(\frac{1}{2}\kappa+i\Delta\right)\tilde{\rho}_{01} - i\sum_{j=1}^{2} g_{j}\left(S_{j}^{+}\rho_{11} 
  - \rho_{00}S_{j}^{+}\right) \nonumber\\
  &&\quad \ \  -\, \frac{1}{2}\gamma {\cal L}_{a}\rho_{01} ,\nonumber \\
 && \dot{\tilde{\rho}}_{10} = \left(\frac{1}{2}\kappa -i\Delta\right)\tilde{\rho}_{10} + i\sum_{j=1}^{2} g_{j}^{\ast}\left( \rho_{11}S_{j}^{-} 
  - S_{j}^{-}\rho_{00}\right) \nonumber\\
  &&\quad \ \ -\, \frac{1}{2}\gamma {\cal L}_{a}\rho_{10} .\label{7.6}
\end{eqnarray}
Here, $g_{j}\equiv g(\vec{r}_{j})$ and we have removed the fast oscillating terms by introducing a rotating frame through the relations
\begin{eqnarray}
  \tilde{\rho}_{01} = \rho_{01}\exp(i\Delta t) ,\quad  \tilde{\rho}_{10} = \rho_{10}\exp(-i\Delta t) . \label{7.7}
\end{eqnarray}
Equations (\ref{7.5}) and (\ref{7.6}) are cumbersome, although involving only two equations, because of the coupling which exists between the diagonal and off-diagonal matrix elements. Moreover, the field-matrix elements are still operators with respect to the atomic variables. 

In order to simplify the procedure and also to minimize spontaneous emission of photons to modes other than the privileged cavity mode, we assume that the cavity frequency is highly detuned from the atomic resonances and/or the cavity field is strongly damped,  i.e.
\begin{eqnarray}
\Delta ,\kappa \gg \gamma, g_{j} .\label{7.8}
\end{eqnarray}
Under such conditions, the most populated state of the cavity field is the ground state $\ket 0$, and then we can make a fundamental physical assumption: The coherence $\tilde{\rho}_{01}$ changes slowly in time, in the sense that $\dot{\tilde{\rho}}_{01}\ll \omega_{0}\tilde{\rho}_{01}$, so that we can apply the so-called adiabatic approximation, which states that for a large detuning and/or cavity damping
\begin{eqnarray} 
\dot{\tilde{\rho}}_{01}\approx 0 \quad {\rm and}\quad \dot{\tilde{\rho}}_{10}\approx 0 .\label{7.9}
\end{eqnarray}
With such an approximation, the coherences obey the following relations 
\begin{eqnarray}
 && \tilde{\rho}_{01} \approx \frac{-i}{{\cal K}}\sum_{j=1}^{2}g_{j}\left(S_{j}^{+}\rho_{11} - \rho_{00}S_{j}^{+}\right) 
  ,\nonumber \\
 && \tilde{\rho}_{10} \approx \frac{i}{{\cal K}^{\ast}}\sum_{j=1}^{2} g_{j}^{\ast}\left( \rho_{11}S_{j}^{-} - S_{j}^{-}\rho_{00}\right) 
  ,\label{7.10}
\end{eqnarray}
where ${\cal K} =\kappa/2 -i\Delta$.

Substitution of Eq.~(\ref{7.10}) into the equations of motion for the populations, $\dot{\rho}_{00}$ and $\dot{\rho}_{11}$, leads to
\begin{eqnarray}
 \dot{\rho}_{00}= \frac{1}{{\cal K}}\sum_{i,j=1}^{2}g_{i}g_{j}^{\ast}\left[S_{i}^{+}S_{j}^{-},\rho_{00}\right] - \frac{1}{2}\gamma {\cal L}_{a}\rho_{00} ,\label{7.11}
\end{eqnarray}
and
\begin{eqnarray}
 \dot{\rho}_{11} = \frac{1}{{\cal K}}\sum_{i,j=1}^{2} g_{i}^{\ast}g_{j}\left[S_{i}^{-}S_{j}^{+}, \rho_{11}\right] - \frac{1}{2}\gamma {\cal L}_{a}\rho_{11}.\label{7.12}
\end{eqnarray}
Since 
\begin{eqnarray}
\rho_{00} + \rho_{11} = {\rm Tr}_{F}(\rho_{T}) = \rho \label{7.13}
\end{eqnarray} 
is the reduced density operator of the atoms, we find by adding Eqs. (\ref{7.11}) and (\ref{7.12}) and neglecting the population~$\rho_{11}$, as the cavity mode will never be populated, that the adiabatic elimination procedure leads to the following master equation for the density operator of the atoms
\begin{eqnarray}
 && \frac{d\rho}{dt} = i\sum_{i=1}^{2}\delta_{i}\left[S_{i}^{+}S_{i}^{-},\rho\right] 
    + i\sum_{i\neq j=1}^{2}\Omega_{ij}\left[S_{i}^{+}S_{j}^{-},\rho\right] \nonumber \\
   &&\quad \ \ +\, \frac{1}{2}\!\sum_{i,j=1}^{2}\!\gamma_{ij}\!\left( \left[\rho S_{i}^{+},S_{j}^{-}\right]\!+\!\left[S_{i}^{+},S_{j}^{-}\rho\right]\right)\!-\!\frac{1}{2}\gamma{\cal L}_{a}\rho ,\nonumber\\
   \label{7.14}
\end{eqnarray}
where the coefficients of the equation are
\begin{eqnarray}
 && \delta_{i} = \frac{|g_{i}|^{2}\Delta}{\left(\frac{1}{4}\kappa^{2} +\Delta^{2}\right)} ,
  \quad \Omega_{ij} = \frac{g_{i}g_{j}^{\ast}\Delta}{\left(\frac{1}{4}\kappa^{2} +\Delta^{2}\right)} ,\nonumber \\
 && \gamma_{ij} = \frac{g_{i}g_{j}^{\ast}\kappa}{\left(\frac{1}{4}\kappa^{2} +\Delta^{2}\right)} .\label{7.15}
\end{eqnarray}
The master equation (\ref{7.14}) is the principal result of this section. The major importance is that it 
includes both, single-atom and two-atom collective terms. The coefficients $\gamma_{ii}$ and $\delta_{i}$ represent
respectively the decay rate induced by the cavity and the frequency shift 
of the energy levels of the $i$th atom. The shift is an analog of a dynamic Stark shift~\cite{nf07}. 
The multi-atom terms,~$\gamma_{ij}$ and  $\Omega_{ij}\, (i\neq j)$ represent, respectively, 
the collective damping and the collective shift in energy separation of the levels of atom $i$ due to its interaction 
with the atom~$j$ through the cavity mode. The term $\Omega_{ij}$ is an analog of the familiar dipole-dipole 
interaction between the atoms~\cite{dic,ft02,leh,ag74}. This shows that one can engineer the interaction between 
distant atoms by the adiabatic elimination of the cavity mode. It is interesting to note, that the engineering 
depends on what king of the cavity is used. The Stark shift $\delta_{i}$ and the collective shift~$\Omega_{ij}$ 
depend on the detuning $\Delta$ and vanish when $\Delta =0$. The damping terms $\gamma_{ij}$ depend on the cavity 
damping $\kappa$ and vanish in the good cavity limit of~$g_{i}\gg \kappa$, or in the limit of $\Delta \gg \kappa$. 
Consequently, no addition atomic damping and the collective damping result from the adiabatic elimination of the 
cavity mode. In other words, the interaction with a strongly detuned cavity mode does not create the collective 
incoherent interaction of the atoms with the environment.

The new feature of non-equivalent positions of the atoms inside the cavity mode is in the shift of the atomic levels $\delta_{i}$ and in the additional damping rates $\gamma_{ii}$ that can have different values for different atoms when~$g_{1}\neq g_{2}$. As a consequence of unequal coupling strengths, the system composed of two identical atoms located at different positions inside the cavity mode may behave as a system of two non-identical atoms of different transition frequencies and different damping rates. In what follows, we choose the reference frame such that 
\begin{eqnarray}
g(\vec{r}_{1}) = g_{0} \quad {\rm and} \quad g(\vec{r}_{2})= g_{0}\cos\left(kr_{12}\right) , \label{7.16} 
\end{eqnarray}
where $r_{12} =|\vec{r}_{2}-\vec{r}_{1}|$ is the distance between the atoms. This choice of the reference frame corresponds to a situation where atom $1$ is kept exactly at an antinode and the atom $2$ can be moved through successive nodes and antinodes of the standing wave. This choice, of course, involves no loss of generality as the dynamics of the system depends on the relative rather than the individual positions of the atoms.

The remaining term $\frac{1}{2}\gamma{\cal L}_{a}\rho$ represents spontaneous emission of the atoms into modes other than 
the cavity mode. For atoms located inside a cavity, the damping rate $\gamma$ depends on the cavity mode function, also known 
as the Airy function of the cavity~\cite{yariv,dem91,mk73,kle81}. The function depends on the solid angle subtended by the cavity 
mode, and is often modeled as a simple Lorentzian with a maximum value at the cavity resonance frequency. The Lorentzian has a half
width at half maximum of $\kappa$. In other words, the cavity ``tailors'' the vacuum modes surrounding the atoms to those orthogonal to the mirrors, which 
modifies the structure of modes available to the atoms for spontaneous emission, thus changing the spontaneous emission rates. 
For an open cavity in which the cavity mode subtends a small solid angle, $\gamma$ is approximately equal to the free-space damping 
rate. For a small size cavity, where the cavity mode subtends a large solid angle, the rate $\gamma$ is very small and can be 
neglected. In this case, the atoms interchange photons entirely with the cavity mode and any losses in the system occurs only 
through the cavity mirrors with the rate determined by the cavity damping $\kappa$.

\subsection{Case of a good cavity}

\noindent Consider first the good cavity limit of $\Delta\gg g_{i}\gg \kappa,\gamma$. Now because the cavity damping is negligible, the atoms are coupled only through the coherent dipole-dipole interaction term $\Omega_{ij}$ and there is no the collective damping, $\gamma_{ij}=0$. The dynamics between the atoms will simplify to the {\it coherent} exchange of an initial excitation. We shall explicitly demonstrate how the dynamics of the system can be expressed in terms of the Bloch vector components of a two-level system driven by a detuned coherent. Next, we will illustrate how the concurrence ${\cal C}(t)$ and its dependence on the relative position $r_{12}$ of the atoms inside the cavity mode can be readily interpreted in terms of the variables of the two-level system.  

As we have already learnt, the concurrence is determined in terms of matrix elements of the density operator of the system. Thus, we use the master equation~(\ref{7.13}) and find that in the product-state representation (\ref{3.6}), the equations of motion for the density matrix elements are in the form
\begin{eqnarray}
&&\dot{\rho}_{44} = -2\gamma\rho_{44} ,\  \dot{\rho}_{11} = \gamma\left(\rho_{22}+\rho_{33}\right) ,\nonumber \\
&&\dot{\rho}_{23} = -\left(\gamma -i\delta_{12}\right)\rho_{23} 
  + i\Omega_{12}\left(\rho_{22} - \rho_{33}\right) , \nonumber\\
 && \dot{\rho}_{32} = -\left(\gamma +i\delta_{12}\right)\rho_{32} 
  - i\Omega_{12}\left(\rho_{22} - \rho_{33}\right) , \nonumber\\
 && \dot{\rho}_{22} = -\gamma\rho_{22}  
  + i\Omega_{12}\left(\rho_{23} - \rho_{32}\right) , \nonumber\\
 && \dot{\rho}_{33} = -\gamma\rho_{33} 
  - i\Omega_{12}\left(\rho_{23} - \rho_{32}\right)  .\label{7.17} 
\end{eqnarray} 
The parameter $\delta_{12} =\delta_{1}-\delta_{2}$ is a difference between the single-atom Stark shifts. This parameter is of central importance here as it determines the relative variation of atomic transition frequencies with position of the atoms inside the cavity mode. Note that the parameter $\delta_{12}$ is present only when $g_{1}\neq g_{2}$, and vanish for $g_{1}=g_{2}$.

Equation (\ref{7.17}) can be converted into two separate sets of equations if we introduce the following linear combinations  
\begin{eqnarray}
&&u = \rho_{23}+\rho_{32}, \quad v= i(\rho_{23}-\rho_{32}), \nonumber \\
&&w = \rho_{22}-\rho_{33},\quad  s = \rho_{22}+\rho_{33} ,\label{7.18}
\end{eqnarray}
In terms of the new variables, Eq.~(\ref{7.17}) splits into a set of equations
\begin{eqnarray}
 && \dot{\rho}_{44} = -2\gamma \rho_{44} ,\ \dot{\rho}_{11} = \gamma\rho_{ss} , \nonumber \\
  &&\dot{\rho}_{ss} = -\gamma\rho_{ss} +2\gamma\rho_{44} ,\label{7.19}
\end{eqnarray}
involving only the populations, and
\begin{eqnarray}
  &&\dot{u} = -\gamma u + \delta_{12}v ,\nonumber \\
  &&\dot{v} = -\gamma v - \delta_{12} u  -2\Omega_{12} w ,\nonumber \\
  &&\dot{w} = -\gamma w +2\Omega_{12} v ,\label{7.20}
\end{eqnarray}
involving the coherences and the population difference between the states $\ket{\Psi_{2}}$ and $\ket{\Psi_{3}}$.
\begin{figure}[hbp]
\includegraphics[width=5cm,keepaspectratio,clip]{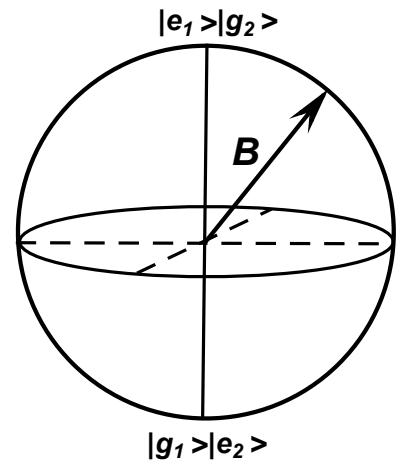}
\caption{The Bloch vector model of a system composed of two product states. The vector $\vec{B}$ rotates with frequency $\Omega_{12}$ on the sphere between the north pole corresponding to the excited state and the south pole corresponding to the ground state of the system. Any other point corresponds to a superposition of these states.}
\label{figc10}
\end{figure}

It is interesting to note that the equations of motion~(\ref{7.20}) are of exactly the same form as the optical Bloch equations of a two-level system driven by a detuned coherent field~\cite{al75}. Here, the dipole-dipole interaction~$\Omega_{12}$ plays the same role as the Rabi frequency of a driving  coherent field, and $\delta_{12}$ appears as a detuning of the field from the driven transition~\cite{al75}. However, the analogy is not absolute because, in contrast to the case of a "real" two-level atom driven by a detuned laser field, the parameters~$\Omega_{12}$ and $\delta_{12}$ in Eq.~(\ref{7.20}) are associated with the same quantities $g_{1}$ and $g_{2}$, i.e. both depend on the coupling constants of the atoms to the cavity field. Consequently, the parameters $\Omega_{12}$ and $\delta_{12}$ can not be independently adjusted. Moreover, the damping rate~$\gamma$ that appears in Eq.~(\ref{7.20}) represents spontaneous decay out of the two-level system, actually to the auxiliary state~$\ket{\Psi_{1}}$, not to the ground state of the two-level~system.

The dynamics of the effective two-level system can be easily visualized in the Bloch vector model, as shown in Fig.~\ref{figc10}. In this model, the system and the driving field are represented by vectors in a three-dimensional space, and the time evolution is simply visualized as a precession of the system-state vector about the driving field. In terms of the Bloch vector, Eq.~(\ref{7.20}) can be written as
\begin{eqnarray}
   \frac{d\vec{B}}{dt} = -\gamma \vec{B} +\vec{\Omega}_{B}\times \vec{B}  ,\label{7.21}
\end{eqnarray}
where $\vec{\Omega}_{B}= (-2\Omega_{12},0,\delta_{12})$ is the pseudofield vector and $\vec{B}= (u,v,w)$ is the Bloch vector. The quantities $u$ and~$v$ are, respectively, the real and imaginary parts of the coherence between the levels $\ket{\Psi_{2}}$ and $\ket{\Psi_{3}}$, and $w$ is the population inversion. The solutions of Eq.~(\ref{7.21}) describe the damped precession of the Bloch vector around $\vec{\Omega}_{B}$. The Bloch vector makes a constant angle $\theta =\tan^{-1}(-2\Omega_{12}/\delta_{12})$ with $\vec{\Omega}_{B}$, it rotates around it tracing out, in the absence of damping, a circle on the Bloch sphere. When the Bloch vector does {\it not} intersect the "north pole" $\vec{B}=(0,0,1)$ or the "south pole" $\vec{B}=(0,0,-1)$, the inversion $w(t)\neq \pm 1$ and then, as we shall see, an entanglement between the atoms occurs.

\begin{figure}[hbp]
\includegraphics[width=8cm,keepaspectratio,clip]{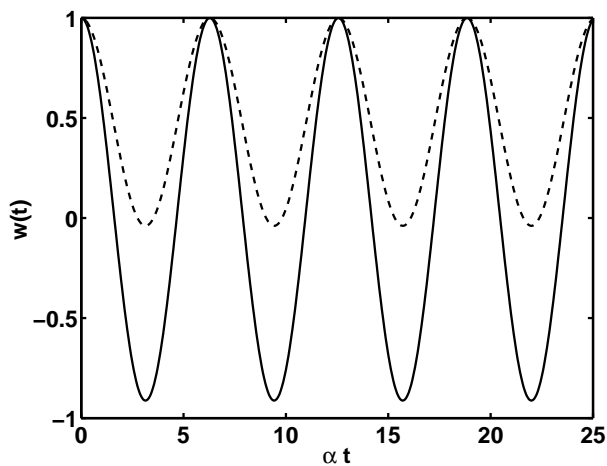}
\caption{Population inversion $w(t)$ as a function of time for $\gamma =0$ and different displacements $\delta r_{d}$ of the atom $2$ from an antinode of the standing wave: $\delta r_{d}=0.1\lambda$ (solid line), $\delta r_{d}= 0.18\lambda$ (dashed line). The system was initially in the state $\ket{\Psi_{2}}$.}
\label{figc11}
\end{figure}

Figure~\ref{figc11} shows an example of the Rabi nutation due to a detuned field. The system was initially in the state~$\ket{\Psi_{2}}$. We see that the Bloch vector periodically intersects the "north pole", $w(t) = 1$, but never intersects the "south pole", $w(t) = -1$, when the atoms are not in the equivalent positions inside the cavity mode. This means that complete population inversion does not occur during the evolution of the system. Viewing this result from the perspective of the population transfer between two atoms, no complete population transfer between the atoms can occur. 

Let us now consider how the concurrence varies in time and, in particular, with the relative position of the atoms inside the cavity mode. We assume that initially the system is prepared in an arbitrary one-photon state. In this case, it is enough to solve the set of equations (\ref{7.20}), and a simple calculation shows that he solutions, valid for arbitrary initial conditions, are of the form
\begin{eqnarray}
&&u(t) = \frac{2\Omega_{12}{\cal A}
       + \delta_{12}\left(v_{0}\alpha\sin\alpha t +{\cal B}\cos\alpha t\right)}{\alpha^{2}}{\rm e}^{-\gamma t} ,\nonumber \\
 && v(t) =  \frac{\left(v_{0}\alpha\cos\alpha t +{\cal B}\sin\alpha t\right)}{\alpha}{\rm e}^{-\gamma t} ,\label{7.22} \\
  && w(t) = \frac{ -\delta_{12}{\cal A} +  2\Omega_{12}\left(v_{0}\alpha\sin\alpha t +{\cal B}\cos\alpha t\right)}{\alpha^{2}}{\rm e}^{-\gamma t}  ,\nonumber
\end{eqnarray}
where  $\alpha =\sqrt{4\Omega_{12}^{2}+\delta_{12}^{2}}$ is the detuned Rabi frequency, and
\begin{equation}
  {\cal A} = 2\Omega_{12}u_{0}-\delta_{12}w_{0} ,\quad {\cal B}= \delta_{12}u_{0}+2\Omega_{12}w_{0} ,\label{7.23}
\end{equation}
are constants given in terms of the initial values $w_{0}\equiv w(0)$, $u_{0}\equiv u(0)$ and $v_{0}\equiv v(0)$ of the population inversion and coherence in the system. The density matrix elements oscillate in time with the detuned Rabi frequency~$\alpha$, and all are equally damped with the rate $\gamma$ due to the decay of the population to the ground state~$\ket{\Psi_{1}}$.

Since the $u(t)$ and $v(t)$ components of the coherence are related to the interaction between the atoms, their properties should be reflected in the entanglement between the atoms.
For an initial single-quantum excitation, the density matrix elements $\rho_{14}(t)$ and $\rho_{44}(t)$ are always zero and then, according to~Eq.~(\ref{4.11}), the concurrence is solely determined by the criterion ${\cal C}_{2}(t)$:
\begin{eqnarray}
  {\cal C}(t) \equiv {\cal C}_{2}(t) = 2|\rho_{23}(t)| .\label{7.24}
\end{eqnarray}
Non-zero values of ${\cal C}(t)$ are recognized as the signature of entanglement between the atoms. We see from Eq.~(\ref{7.24}) that the basic dynamical mechanism for entanglement creation in this system is the coherence~$\rho_{23}(t)$. That is, the origin of the entanglement in the system can be traced back to the time evolution of the coherence~$\rho_{23}(t)$. Utilizing the relation $\rho_{23}(t) = (u(t)-iv(t))/2$, and through the relation 
\begin{eqnarray}
  u^{2}(t) +v^{2}(t) +w^{2}(t) = {\rm e}^{-2\gamma t} ,\label{7.25} 
\end{eqnarray}
the concurrence can be written as
\begin{eqnarray}
  {\cal C}(t) = \sqrt{1-\bar{w}^{2}(t)}\, {\rm e}^{-\gamma t}  ,\label{7.26}
\end{eqnarray}
where the following notation has been introduced 
\begin{eqnarray}
\bar{w}(t) = \frac{ -\delta_{12}{\cal A} +  2\Omega_{12}\left(v_{0}\alpha\sin\alpha t +{\cal B}\cos\alpha t\right)}{\alpha^{2}} .\label{7.27}
\end{eqnarray}
Equation (\ref{7.26}) gives us a transparent way to interpret the entangled dynamics of the system. It shows that the concurrence can be completely determined {\it only} by evaluating the contribution of the population inversion. When the population is in the lower or upper level, $\bar{w}(t)=\pm 1$, and then ${\cal C}(t)=0$, whereas ${\cal C}(t)$ achieves its optimum value ${\cal C}(t)=1$ when $\bar{w}(t)=0$. Therefore, we can interpret the entanglement as a consequence of a distribution of the population among the energy levels.

It is easy to see from Eq.~(\ref{7.26}) that the evolution of the concurrence depends crucially on whether $g_{1}=g_{2}$ or~$g_{1}\neq g_{2}$. If $g_{1}=g_{2}$, then $\delta_{12}=0$ and one finds from Eq.~(\ref{7.26}) that the concurrence vanishes 
periodically at the particular times
\begin{eqnarray}
    t_{0} = n\pi/\alpha ,\qquad n=0,1,2,\ldots  .\label{7.28} 
\end{eqnarray}
On the other hand, if $g_{1}\neq g_{2}$ then $\delta_{12}\neq 0$, and we find that the concurrence behaves quite differently such that it vanishes only at times
\begin{eqnarray}
    t_{\delta} = 2t_{0} = 2n\pi/\alpha ,\qquad n=0,1,2,\ldots   ,\label{7.29} 
\end{eqnarray}
which shows that the entanglement exsists on the the time scale twice as long as for the case of equal coupling constants.

This rather surprising result can be understood in terms of spatial localization of the energy emitted by an initially excited atom. For equal coupling strengths $\delta_{12}=0$ because  the energy levels of the atoms are equally shifted by the interaction with the cavity mode. In this case the energy emitted by the first atom oscillates with frequency $2\Omega_{12}$ such that at  times $t_{n}=n\pi/\alpha \ (n=0,1,2,\ldots )$ it is fully absorbed by the second atom. Since at these times the energy is well localized in space as being completely absorbed by the localized atoms, the entanglement, which results from an unlocalized energy, is zero. The situation changes when $g_{1}\neq g_{2}$. In this case the energy levels of the atoms are unequally shifted by the interaction with the cavity mode. Consequently, due to the frequency mismatch, the energy emitted by the atom~$1$ is not fully absorbed by the atom $2$, leading to a partial spatial delocalization of the excitation at discrete times $t=n\pi/\alpha$, where $n=1,3,5,\ldots$. Consequently, at these times a partial entanglement is observed. The entanglement vanishes every time the excitation returns to its initial state, i.e. when it returns to atom~$1$.
\begin{figure}[hbp]
\includegraphics[width=8cm,keepaspectratio,clip]{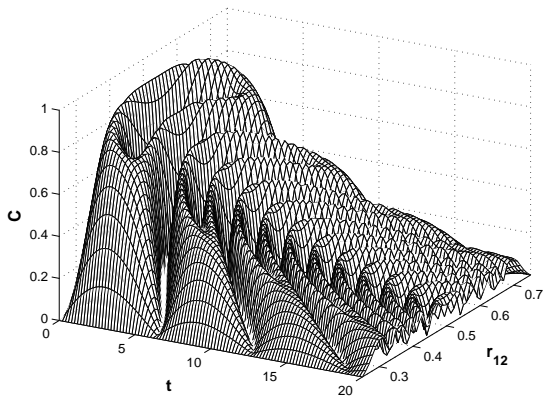}
\caption{Concurrence as a function of time $t$ and the position $r_{12}$ of the atom $2$ between two successive nodes of the standing wave cavity field for $\gamma =0.1$. The system was initially in the separable  state $\ket{\Psi_{2}}$.}
\label{figc12}
\end{figure}

Above considerations are illustrated in Fig.~\ref{figc12}, which shows the concurrence as a function of time and the position $r_{12}$ of the atom $2$ between two successive nodes of the standing wave cavity field. We see oscillations in the transient evolution of the entanglement that follow the Rabi flopping of population back and forth between the atoms. In other words, this oscillation reflects nutation of the atomic populations which, in turn, can be associated with the precession of the Bloch vector $\vec{B}$ about the driving field vector $\vec{\Omega}_{B}$ with frequency~$\alpha$. The most interesting feature of the transient entanglement seen in~Fig.~\ref{figc12} is that in the case of unequal coupling constants, the initially unentangled system evolves into an entangled state, and remains in this state longer than for the case of equal coupling constants.

\subsection{Entanglement diffraction pattern}

\noindent Let us now briefly consider the sensitivity of the entanglement on the relative position of the atoms inside the cavity mode.  We use the solution~(\ref{7.20}) to evaluate the concurrence  as a function of $r_{12}$ for a fixed time~$t$. We will discuss the variation of the concurrence with $r_{12}$ for two initial conditions. Our examples of the initial conditions are firstly the separable state $\ket{\Psi_{3}}$, and secondly, the maximally entangled symmetric state $\ket{\Psi_{s}}$. 

Our first example considers the initial separable state~$\ket{\Psi_{3}}$. In this case, we readily find from Eq.~(\ref{7.26}) that the variation of the concurrence with position of the atoms inside the standing wave is given by
\begin{eqnarray}
  {\cal C}(r_{12},t) &=& \left\{\!\left(\frac{\sin d}{d}\right)^{\!2}\!\tau^{2}\!
  +\!\left(\frac{\sin \frac{1}{2}d}{\frac{1}{2}d}\right)^{\!4}\!\tau^{4}\sin^{4}(kr_{12})
  \right\}^{\frac{1}{2}} \nonumber \\
  &\times&  |\cos kr_{12}|{\rm e}^{-\Gamma \tau} .\label{7.30} 
\end{eqnarray}
Here, $d = (1+\cos^{2}kr_{12})\tau/2$, we have make use of a scaled time variable $\tau =\left(2g_{0}^{2}/\Delta\right)t$ and
dimensionless quantity~$\Gamma =\left(\Delta/2g_{0}^{2}\right)\gamma$.

The concurrence depends on time, the relative distance between the atoms and decays exponentially in time with the rate $\Gamma$. It exhibits an interesting modulation of the amplitude of the harmonic oscillation. Somewhat surprisingly our results show that, in general,  the concurrence does not have a cosine shape, i.e. the shape of the cavity mode function. It is given by the product of two terms, one the absolute value of the cavity mode function $|\cos kr_{12}|$ and the other the time- and position-dependent diffraction structure. In other words, the concurrence is in the form of position and time dependent~{\it diffraction pattern}.
\begin{figure}[hbp]
\includegraphics[width=8cm,keepaspectratio,clip]{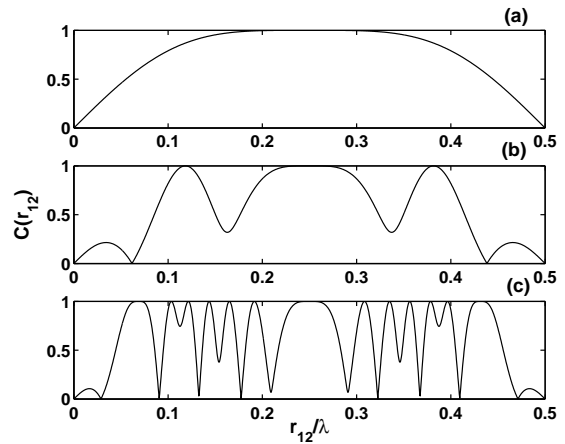}
\caption{Variation of the concurrence ${\cal C}(r_{12})$ with the position of the atom 2 inside a half-wavelength of the cavity field for $\gamma =0$ and different times $\tau$: (a) $\tau =\pi/2$, (b) $\tau =7\pi/2$, (c) $\tau =31\pi/2$. The system was initially in the state $\ket{\Psi_{3}}\, (w_{0}=1, u_{0}=v_{0}=0)$.}
\label{figc13}
\end{figure}

Figure~\ref{figc13} shows the concurrence as a function of $r_{12}$ for different times $\tau$ corresponding to the evolution intervals at which the entanglement is maximal for the idealized case of $g_{1}=g_{2}$.  At early times, the entanglement is seen to occur over the entire range of $r_{12}$ inside the half-wavelength of the cavity field.  In addition, the concurrence is a bell-shaped function of position without any oscillation. As time progresses, oscillations emerge and consequently the region of $r_{12}$ where the optimum entanglement occurs, becomes narrower. The evolution of the concurrence tends to become increasingly oscillatory with $r_{12}$ as time increases, and the optimum entanglement occurs in a still more restricted range of~$r_{12}$. As a result, the entanglement oscillates with position faster than the cosine function, and the oscillations are more dramatic for larger times. Only at very early times $(\tau\ll 1)$, the oscillations are not modulated by the diffraction pattern and the concurrence reduces to~$|\cos kr_{12}|$, but for longer times, ${\cal C}(r_{12})$ may vary slower or faster than the cosine functions. Another interesting observation is that within the diffraction structure itself, the magnitude of the concurrence exhibits a succession of nodes and of antinodes. As a consequence, the entanglement may be completely quenched even for locations of the atom close to the antinode of the cavity mode, and alternatively may achieve its optimum value even for locations of the atom close to a node of the cavity~mode. 

\begin{figure}[hbp]
\includegraphics[width=8cm,keepaspectratio,clip]{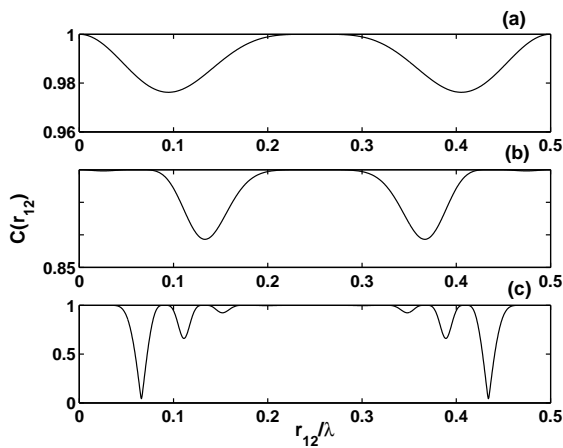}
\caption{Variation of the concurrence ${\cal C}(r_{12})$ with the position of the atom 2 inside a half-wavelength of the cavity field for $\gamma =0$ and different times $\tau$: (a) $\tau =\pi/2$, (b) $\tau =7\pi/2$, (c) $\tau =31\pi/2$. The system was initially in the state $\ket{\Psi_{s}}\, (u_{0}=1, v_{0}=w_{0}=0)$.}
\label{figc14}
\end{figure}

In the second case, we assume that initially the system was prepared in the maximally entangled state $\ket{\Psi_{s}}$. Now, the general solution (\ref{7.26}) leads to the concurrence that can be written~as
\begin{equation}
  {\cal C}(r_{12})\!=\!\left[1\!-\!\left(\frac{\sin \frac{1}{2}d}{\frac{1}{2}d}\right)^{\!4}\!\!\tau^{4}\sin^{4}(kr_{12})\cos^{2}(kr_{12})\right]^{\!\frac{1}{2}}\!\!{\rm e}^{-\Gamma \tau} .
  \label{7.32} 
\end{equation}
The concurrence exhibits only the second order diffraction pattern. This may result in a weaker dependence of the concurrence on the position of the atoms than that observed with the initial unentangled state. Figure~\ref{figc14} shows how the concurrence evolves with the position of the atom 2 inside the cavity mode and time $\tau$, in the absence of spontaneous emission $(\gamma=0)$. Notice that the initial entanglement is reduced for positions of the atom different from the antinode of the cavity mode. That is, at some points outside the antinode the concurrence may become zero indicating the complete reduction of the initial entanglement.

The strong modulation of the concurrence in time, seen in Figures~\ref{figc13} and \ref{figc14} can be understood as a consequence of the uncertainty relation between the evolution time and energy, $\Delta E\Delta t\geq \hbar$.  For the increasing time $\Delta t$, the uncertainty of the energy $\Delta E$ decreases which means that the energy (excitation) becomes more localized. The increase in the localization of the energy results in a degradation of the entanglement. In other words, with an increasing time, one can in principle obtain more information about the localization of the atoms inside the cavity mode.

In concluding this section, let us determine the explicit analytical form of single photon entangled states of the system. The calculations can be easy done using the density matrix of the system, which in the basis of the product states (\ref{3.6}) has the form
\begin{eqnarray}
 \rho(t) = \left(
  \begin{array}{cccc}
   \rho_{11}(t)  & 0 & 0 & 0\\
   0 & \rho_{22}(t)  & \rho_{23}(t) & 0\\
   0 & \rho_{32}(t) & \rho_{33}(t)  & 0\\
   0 & 0 & 0 & 0 
   \end{array}
 \right) ,\label{7.33}
\end{eqnarray} 
where we have assumed that only one photon is present in the system, i.e. $\rho_{44}(t)=0$ for all~times.

We see that the density matrix (\ref{7.33}) is not diagonal in the basis of the product states~(\ref{3.6}) due to the presence of the coherences $\rho_{23}(t)$ and $\rho_{32}(t)$. Thus, the product states (\ref{3.6}) are not the eigenstates of the system. To obtain the actual eigenstates of the system, we have to diagonalize the matrix (\ref{7.33}). We first find the eigenvalues of the matrix, from which we next determine eigenvectors, i.e. eigenstates of the system. Proceeding in the usual way, we calculate the determinant of the matrix~$\rho(t)$, and find a characteristic (polynomial) equation
\begin{eqnarray}
&& \lambda\left[\lambda -\rho_{11}(t)\right] \left\{\lambda^{2} - \left[\rho_{22}(t) + \rho_{33}(t)\right]\lambda\right. \nonumber\\
 &&\left. + \rho_{22}(t)\rho_{33}(t) - |\rho_{23}(t)|^{2}\right\} = 0 .\label{7.34}
\end{eqnarray}
This is a simple fourth order polynomial whose the roots, corresponding to the eigenvalues of the matrix (\ref{7.33}), are
\begin{eqnarray}
 && \lambda_{1}(t) = \rho_{11}(t), \quad  \lambda_{4}(t) = 0 ,\nonumber \\
 && \lambda_{2,3}(t) = \frac{1}{2}\left[\rho_{22}(t) + \rho_{33}(t)\right] \pm \frac{1}{2}U ,\label{7.35}
\end{eqnarray}
where $U= \left[(\rho_{22}(t)-\rho_{33}(t))^{2}  + 4|\rho_{23}(t)|^{2}\right]^{1/2}$.
Since the following relations are satisfied for all times
\begin{eqnarray}
 && \rho_{22}(t) + \rho_{33}(t) = 1 - \rho_{11}(t) ,\nonumber\\
 && \left[\rho_{22}(t)-\rho_{33}(t)\right]^{2}=w^{2}(t) ,\nonumber \\ 
 && 4|\rho_{23}(t)|^{2} =v^{2}(t) +u^{2}(t) ,\label{7.36}
\end{eqnarray}
and
\begin{eqnarray}
  u^{2}(t) +v^{2}(t) +w^{2}(t) = {\rm e}^{-2\gamma t} ,\label{7.37} 
\end{eqnarray}
we can write the roots (\ref{7.35}) in the form
\begin{eqnarray}
  && \lambda_{1}(t) = \rho_{11}(t) ,\quad \lambda_{4}(t) = 0 ,\nonumber \\
 && \lambda_{2,3}(t) = \frac{1}{2}\left[1-\rho_{11}(t) \pm {\rm e}^{-\gamma t} \,\right] .\label{7.38}
\end{eqnarray}
It is seen from Eq.~(\ref{7.38}) that in the presence of spontaneous emission $(\gamma\neq 0)$, three of the eigenvalues are non-zero, indicating that the system is in a mixed state. In the Bloch picture, this corresponds to the fact that the Bloch vector $\vec{B}$ is not constant. According to Eq.~(\ref{7.37}), the magnitude of the Bloch vector decays in time with the rate $2\gamma$.
However, in the absence of spontaneous emission, $\gamma=0$, and then the population of the ground state $\rho_{11}(t)=0$ for all times. Consequently, all the eigenvalues are zero except $\lambda_{2}(t)$, that is equal to one. In this case, the magnitude of the Bloch vector is conserved and the system remains in a pure state for all times.

Following the standard procedure of diagonalisation of matrices, we can easily find the eigenstates of the matrix~(\ref{7.33}) corresponding to the eigenvalues~(\ref{7.35}). The normalised eigenstates are of the form
\begin{eqnarray}
 && \ket{\vartheta_{1}(t)} = \rho_{11}(t)\ket{\Psi_{1}}  ,\nonumber \\
 && \ket{\vartheta_{2}(t)} = \frac{1}{{\cal N}_{2}}\left[\rho_{23}(t)\ket{\Psi_{2}} + \left(\lambda_{2}(t) - \rho_{22}(t)\right)\ket{\Psi_{3}}\right] ,\nonumber\\
&& \ket{\vartheta_{3}(t)} = \frac{1}{{\cal N}_{3}}\left[\rho_{23}(t)\ket{\Psi_{2}} + \left(\lambda_{3}(t) - \rho_{22}(t)\right)\ket{\Psi_{3}}\right] ,\nonumber\\
&&\ket{\vartheta_{4}(t)} = \lambda_{4}(t)\ket{\Psi_{4}} ,\label{7.39}
\end{eqnarray}
where ${\cal N}_{2,3}= \left\{|\rho_{23}(t)|^{2} +\left[\lambda_{2,3}(t) -\rho_{22}(t)\right]\right\}^{1/2}$. Thus, the system initially prepared in a one-photon state evolves into a mixed state determined by the states (\ref{7.39}).
\begin{figure}[hbp]
\includegraphics[width=8cm,keepaspectratio,clip]{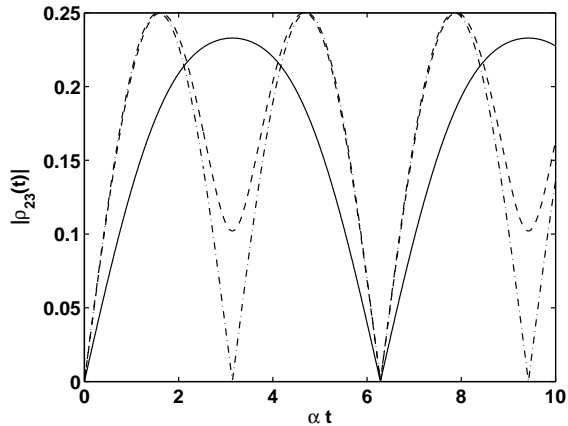}
\caption{Magnitude of the coherence $\rho_{23}(t)$ as a function of the normalised time $\alpha t$ for $\gamma =0$ and different~$r_{12}$: $r_{12} = 0.3\lambda$ (solid line), (b) $r_{12} = 0.4\lambda$ (dashed line), $r_{12} = 0.5\lambda$ (dashed-dotted line). The system was initially in the separable state $\ket{\Psi_{2}}$.}
\label{figc15}
\end{figure}

It is evident from Eq.~(\ref{7.39}) that the states $\ket{\vartheta_{2}(t)}$ and $\ket{\vartheta_{3}(t)}$ are in the form of entangled states that are produced by the coherence $\rho_{23}(t)$. Thus, a non-zero value of the coherence $\rho_{23}(t)$ leads to entangled states of the atoms. In other words, whenever $\rho_{23}(t)\neq 0$, the atoms are entangled.

In Fig.~\ref{figc15}, we plot $|\rho_{23}(t)|$ as a function of time for~$\gamma =0$ and different relative positions $r_{12}$ of the atoms inside the cavity mode. The figure demonstrates that the coherence develops in time and remains different from zero for all times except for some discrete 
times at which the population is completely in the states $\ket{\Psi_{2}}$ or~$\ket{\Psi_{3}}$. Interestingly, when the atom~2 is not located precisely at an antinode of the cavity field, the coherence remains non-zero for longer times indicating that in this case the atoms can be entangled on the time scale much longer than that predicted for the atoms precisely located at the antinodes of the cavity field. 

As above, we can explained this result in terms of the degree of localization or non-locality of the 
energy. One can see from Eq.~(\ref{7.17}) that for equal coupling strengths, $\delta_{12}=0$, and then the initial energy oscillates with frequency $2\Omega_{12}$ between the atoms. When absorbed by one of the atoms, the energy is then well localized in space as being at localised objects. Consequently, the entanglement, which results from a non-locality of the energy, is zero. The situation changes when $\delta_{12}\neq 0$. According to Eq.~(\ref{7.17}), in this case the energy levels of the atoms are unequally shifted. Due to the frequency mismatch, the energy radiated by the atom~$1$ is not fully absorbed by the atom~$2$, leading to a partial spatial delocalization of the excitation at discrete times corresponding to the time at which the atom $2$ is excited. Consequently, at these times an entanglement is observed.

\subsection{Case of a bad cavity}

\noindent In the preceding section, we have discussed the dynamics of two atoms coupled to a single-mode cavity field. 
We have considered the good cavity regime, where we have ignored the effects of the cavity damping. In the present section, 
we continue the analysis of the dynamics of the atoms coupled to a single-mode cavity, but now we consider the opposite limit, 
the bad cavity regime where the dynamics are primary determined by the cavity damping~$\kappa$. As we shall see, a fast damping of the cavity mode may result in a dark state geometry of the system that there might exist an entangled non-decaying (trapped) state. We demonstrate the occurrence of a stationary entanglement in the system.

If we include the cavity damping, $\kappa\neq 0$, we then readily find from the master equation (\ref{7.14}) that the density matrix elements satisfy a set of four coupled differential equations that can be written in a matrix form as 
\begin{eqnarray}
  \frac{d}{dt}\vec{Y} = {\bold A}\vec{Y} ,\label{7.40}
\end{eqnarray}
where $\vec{Y}$ is a column vector having components $(\rho_{22},\rho_{33},u, v)$, and $\bold A$ is a $4\times 4$ matrix
\begin{eqnarray}
  {\bold A} = \left(
    \begin{array}{cccc}
      -\gamma_{2} & 0 & -\frac{1}{2}\gamma_{12} & -\Omega_{12} \\
      0 & -\gamma_{1} & -\frac{1}{2}\gamma_{12} & \Omega_{12}  \\
      -\gamma_{12} & -\gamma_{12} & -\frac{1}{2}\gamma & -\delta_{12} \\
      -2\Omega_{12} & 2\Omega_{12} & -\delta_{12} & -\frac{1}{2}\gamma  
    \end{array}\right) ,\label{7.41}
\end{eqnarray}
with $\delta_{12}=\delta_{2}-\delta_{1}$, $\gamma =\gamma_{1} +\gamma_{2}$, and the remaining  coefficient given in Eq.~(\ref{7.15}).

The equations of motion (\ref{7.40}) are first-order differential equations with the coefficients dependent on the relative position of the atoms inside the standing-wave cavity mode. The set of equations (\ref{7.40}) is easily solved by matrix inversion. The method, however, depends critically on whether det$[{\bold A}]\neq 0$ or det$[{\bold A}]=0$. It is easily verified that the determinant of the matrix ${\bold A}$ is of the form
\begin{eqnarray}
 {\rm det}[{\bold A} ] = \Omega_{12}(\gamma_{2}-\gamma_{1}) +\gamma_{12}\delta_{12} .\label{7.42}
 \end{eqnarray}
Further, using the explicit forms of the coefficients, Eq.~(\ref{7.15}), we find the determinant is equal to zero. It is always zero independent of the values of the parameters involved. An immediate consequence of det$[{\bold A}]=0$ is that the density matrix elements can have non-zero steady-state values that depend on the initial values of the density matrix elements~\cite{ft86,fs02,be05}. 
This fact is connected with the existence of a linear combination of the density matrix elements 
\begin{eqnarray}
\rho_{cc} = \gamma_{1}\rho_{22} +\gamma_{2}\rho_{33} -\gamma_{12}u ,\label{7.43}
\end{eqnarray}
which is a constant of motion, i.e. $\dot{\rho}_{cc} =0$.

The existence of the constant of motion reflects the presence of a linear combination of the states of the system which does not decay. From Eq.~(\ref{7.40}), it is straightforward to show that the state corresponding to the constant of motion is of the form
\begin{eqnarray}
\ket{\Psi_{c}} =  \sqrt{\frac{\gamma_{2}}{\gamma_{1}+\gamma_{2}}}\ket{\Psi_{2}} 
-\sqrt{\frac{\gamma_{1}}{\gamma_{1}+\gamma_{2}}}\ket{\Psi_{3}} ,\label{7.44}
\end{eqnarray}
It is easily verified that the state (\ref{7.44}) satisfies the trapping condition of $\dot{\rho}_{cc}=0$. In other words, spontaneous emission from this state does not take place, so that any initial population in the state $\ket{\Psi_{c}}$ will stay there for all times. 

However, the channel $\ket{\Psi_{4}}\rightarrow \ket{\Psi_{c}}$ can be open for spontaneous emission,  i.e. transitions into the state~$\ket{\Psi_{c}}$ from the two-photon state $\ket{\Psi_{4}}$  could be allowed for spontaneous emission. Hence, the state $\ket{\Psi_{c}}$ could be populated by spontaneous emission from the upper state~$\ket{\Psi_{4}}$. It is easy to find from the master equation~(\ref{7.14}) that the inclusion of the state~$\ket{\Psi_{4}}$ into the dynamics of the system leads to the following equation of motion for the population of the state~$\ket{\Psi_{c}}$:
\begin{eqnarray}
\frac{d}{dt}\rho_{cc} = \frac{(\gamma_{2}-\gamma_{1})^{2}}{\gamma_{1}+\gamma_{2}}\rho_{44}
 .\label{7.45}
\end{eqnarray}
We see that the state $\ket{\Psi_{c}}$ is populated with the rate proportional to the difference between the damping rates induced by the cavity damping. According to Eq.~(\ref{7.15}), the damping rates are different from each other only when $g_{1}\neq g_{2}$, i.e. when the atoms are in nonequivalent positions inside the cavity mode. Otherwise, the transition rate from the state $\ket{\Psi_{4}}$ to $\ket{\Psi_{c}}$ is equal to zero indicating that the state becomes decoupled from all the states so that cannot be accessible by spontaneous emission.
In this case, the state $\ket{\Psi_{c}}$ can be regarded as a {\it dark} state in the sense that the state is decoupled from the environment. 
\begin{figure}[hbp]
\includegraphics[width=8cm,keepaspectratio,clip]{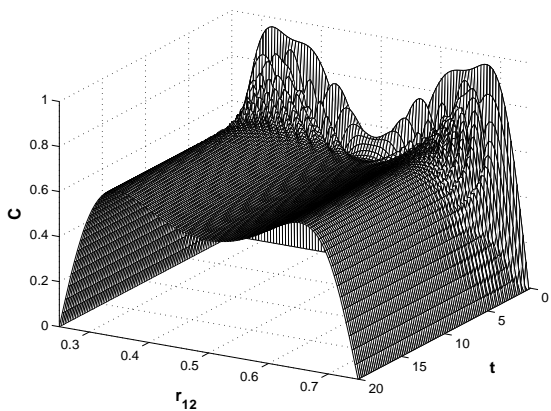}
\caption{Concurrence as a function of time $t$ and the position $r_{12}$ of the atom 2 
  between two successive nodes of the standing wave cavity field for $\gamma=0$ and $\Delta/\kappa =30$. The system was initially in the state $\ket{\Psi_{2}}$.}
\label{figc16}
\end{figure}

Let us now to compute the concurrence with the initial state $\ket I$ corresponding to an excitation of the system into the separable state $\ket{\Psi_{2}}$. On making use of Eqs~(\ref{4.11}) and~(\ref{7.40}), and remembering that with the initial state~$\ket{\Psi_{2}}$, the coherence $\rho_{14}(t)$ and the population~$\rho_{44}(t)$ are always zero, we obtain 
\begin{eqnarray}
  {\cal C}_{\ket{\Psi_{2}}}(t) &=& \frac{2g_{1}g_{2}}{\left(g_{1}^{2}+g_{2}^{2}\right)^{2}}\left| g_{1}^{2} -g_{2}^{2}
  {\rm e}^{-\left(\gamma_{1} + \gamma_{2}\right) t}\right. \nonumber\\
 &-&\left.  g_{1}^{2}{\rm e}^{-\frac{1}{2}\left(\gamma_{1}+\gamma_{2}\right)\left(1-i\eta\right)t}\right. \nonumber\\
 &+&\left. g_{2}^{2}{\rm e}^{-\frac{1}{2}\left(\gamma_{1}+\gamma_{2}\right)\left(1+i\eta\right) t}\right|
  ,\label{7.46} 
\end{eqnarray}
where $\eta=\Delta/\kappa$. 

If the system is initially prepared in the state $\ket{\Psi_{3}}$, the concurrence takes the following form 
\begin{eqnarray}
  {\cal C}_{\ket{\Psi_{3}}}(t) &=& \frac{2g_{1}g_{2}}{\left(g_{1}^{2}+g_{2}^{2}\right)^{2}}\left| g_{2}^{2} -g_{1}^{2}
  {\rm e}^{-\left(\gamma_{1} + \gamma_{2}\right) t}\right. \nonumber\\
 &-&\left.  g_{2}^{2}{\rm e}^{-\frac{1}{2}\left(\gamma_{1}+\gamma_{2}\right)\left(1-i\eta\right)t}\right. \nonumber\\
 &+&\left. g_{1}^{2}{\rm e}^{-\frac{1}{2}\left(\gamma_{1}+\gamma_{2}\right)\left(1+i\eta\right) t}\right|
  .\label{7.47} 
\end{eqnarray}
The most obvious characteristic of these results is that the concurrence depends on the initial state even in the stationary limit of $t\rightarrow \infty$. For the steady-state, we have
\begin{eqnarray}
 &&{\cal C}_{\ket{\Psi_{2}}}(t\rightarrow\infty) \rightarrow \frac{2g_{1}^{3}g_{2}}{\left(g_{1}^{2}+g_{2}^{2}\right)^{2}} ,\nonumber \\
&&{\cal C}_{\ket{\Psi_{3}}}(t\rightarrow\infty) \rightarrow \frac{2g_{1}g_{2}^{3}}{\left(g_{1}^{2}+g_{2}^{2}\right)^{2}} .\label{7.48} 
\end{eqnarray}
Evidently, the system decays to a stationary state which depends on the initial conditions. Only in the limit of~$g_{1}=g_{2}$, which corresponds to the atoms located at the equivalent positions inside the cavity mode, the concurrence decays to the same stationary value, 
${\cal C}_{\ket{\Psi_{2}}}(t\rightarrow\infty)={\cal C}_{\ket{\Psi_{3}}}(t\rightarrow\infty) =1/2$. Otherwise, when $g_{1}\neq g_{2}$, the concurrence decays to different stationary values, which strongly depend on whether~$g_{1}>g_{2}$ or $g_{1}<g_{2}$. 

\begin{figure}[hbp]
\includegraphics[width=8cm,keepaspectratio,clip]{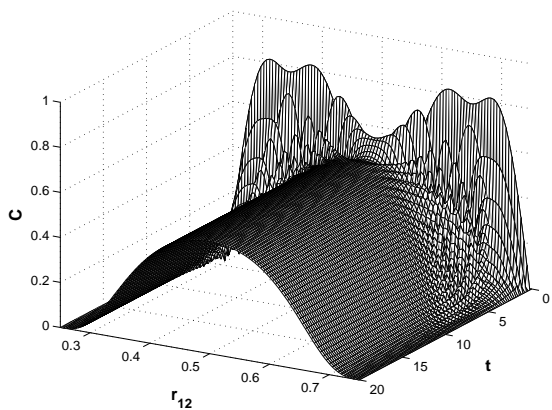}
\caption{Concurrence as a function of time $t$ and the position $r_{12}$ of the atom 2 
  between two successive nodes of the standing wave cavity field for $\gamma=0$ and $\Delta/\kappa =30$. The system was initially in the state~$\ket{\Psi_{3}}$.}
\label{figc17}
\end{figure}

Figures~\ref{figc16} and \ref{figc17}  show the time evolution of the entanglement and its dependence on the position of the atom~$2$ inside a half-wavelength of the standing-wave cavity mode. The most interesting feature is that the entanglement decays to a non-zero stationary state which depends on both, the relative position of the atoms inside the cavity mode and the initial state of the system. One can see from the figures that the stationary entanglement can be larger for positions of the atom different from the antinode of the cavity mode. Again, this effect can be explained in terms of the frequency shift of the atomic levels that leads to the frequency mismatch resulting to a spatial delocalization of the initial excitation.

So far we have considered only the evolution of the concurrence from initial one-photon separable states. We have demonstrated that a system of two atoms located inside a standing wave cavity mode and initially prepared in a separable one-photon state, can decay to an entangled state. The amount of entanglement depends on the relative position of the atoms inside the standing  wave mode. In fact, the amount of the stationary entanglement depends on the initial population of the trapping state $\ket{\Psi_{c}}$. Since the initial population of the state $\ket{\Psi_{c}}$ cannot day, it stays there for all times. 

A somehow different situation arises if the system is initially prepared in the separable state $\ket{\Psi_{4}}$. In this case, the initial population of the state $\ket{\Psi_{c}}$ is equal to zero. Nevertheless, the system could decay to a stationary entangled state determined by the population of the trapping state $\ket{\Psi_{c}}$. This possibility arises from the fact that under some circumstances the transition channel $\ket{\Psi_{4}}\rightarrow \ket{\Psi_{c}}$ can be allowed for spontaneous emission. Using Eqs.~(\ref{4.11}) and (\ref{7.14}), it is straightforward to show that for the initial state $\ket I =\ket{\Psi_{4}}$, the steady-state value for the concurrence is given by the expression
\begin{eqnarray}
 {\cal C}(\ket{\Psi_{4}},t\rightarrow\infty) \rightarrow 2\sqrt{\gamma_{1}\gamma_{2}}\ \frac{(\gamma_{2}-\gamma_{1})^{2}}
 {(\gamma_{1}+\gamma_{2})^{3}} .\label{7.49}
\end{eqnarray}
It is readily seen that the atoms can be entangled in the steady-state. However, it happens only if $\gamma_{1}\neq \gamma_{2}$, i.e., when the atoms are located in nonequivalent positions inside the cavity mode $(g_{1}\neq g_{2})$. Otherwise, when $\gamma_{1}=\gamma_{2}$, which corresponds to $g_{1}=g_{2}$, the system decays to a separable state.

\section{Two atoms in separate cavities}

\noindent Our third model for creation of entanglement between two atoms consists of two separated subsystems, each composed of a single mode cavity containing a single two-level atom~\cite{cr09,yy06,yy07,yy09}. In general, the two subsystems may not be identical in that the cavity frequencies and the coupling constants of the atoms to the cavity modes could be different. We study the entanglement evolution of two remote atoms interacting independently with a cavity field, as in the double Jaynes-Cummings model. We consider two cases: the first where there is only one excitation present in the entire system; the second where there is a superposition of two excitation levels, zero excitation and two excitations, one in each of the two remote systems. These two cases exhibit completely different properties of the entanglement evolution. As we shall see, a complete and abrupt loss of atom-atom entanglement, the entanglement sudden death, is found to occur for the two-photon Bell states provided the initial entanglement is not maximum, though the full entanglement is periodically regained. We present complete entanglement solutions, for the case of asymmetric atom-cavity couplings and off-resonant interactions. We will show that asymmetric cavities and cavity-atom detunings can prove advantageous in enabling a control of entanglement. For example, entanglement sudden death can be eliminated with the use of suitable detunings of the atomic transition frequencies from the cavity mode frequencies. Thus the general effect of detuning is to stabilize entanglement. The situations of asymmetric atom-cavity couplings and off-resonant interactions correspond to realistic experimental situations, where it may be difficult to produce identical cavities. The unequal coupling constants for example may arise when atoms are not in equivalent positions inside the cavities.

\begin{figure}[hbp]
\includegraphics[width=6cm,keepaspectratio,clip]{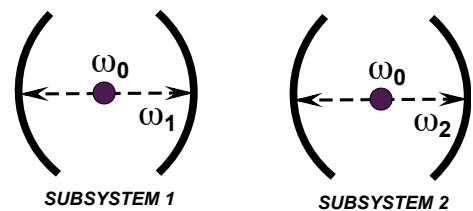}
\caption{Schematic diagram of a system composed of two distant and non-interacting
subsystems $1$ and $2$. Each subsystem is composed of a single-mode cavity containing a two-level atom.}
\label{figc18}
\end{figure}

We begin by introducing the Hamiltonian of the system, which now is composed of two separated single mode cavities each containing a single two-level atom, as shown in Fig.~\ref{figc18}. The total Hamiltonian $H$ for the atoms and the cavity fields, in the electric-dipole and rotating-wave approximations, can be written in the Jaynes-Cummings (JC) form as
\begin{equation}
H= H_{F}+H_{A}+H_{int} , \label{8.1}
\end{equation}
where 
\begin{equation}
H_{F}=\hbar\omega_{1}\left(a^{\dag}_{1}a_{1} +\frac{1}{2}\right)
+\hbar\omega_{2}\left(a^{\dag}_{2}a_{2}+\frac{1}{2}\right) \label{8.2}
\end{equation}
is the Hamiltonian of the cavity fields, 
\begin{equation}
H_{A}=\hbar\omega_{0}S_{1}^{z}+\hbar\omega_{0}S_{2}^{z} \label{8.3}
\end{equation}
 is the Hamiltonian of the atoms, and 
 \begin{equation}
H_{int}=\hbar g_{1}\!\left(a^{\dagger}_{1}S_{1}^{-}\!+\!a_{1}S_{1}^{+}\right)\!
+\!\hbar g_{2}\!\left(a^{\dagger}_{2}S_{2}^{-}\!+\!a_{2}S_{2}^{+}\right) \label{8.4}
\end{equation}
is the interaction Hamiltonian between the atoms and the cavity modes. Here, $g_{i}\, (i=1,2)$ is the coupling constant between the $i$th atom and the field mode of the $i$the cavity. As usual, we have denoted by $S_{i}^{+}$ and $S_{i}^{-}$ the raising and lowering operators of the~$i$th atom, and by $a^{\dag}_{j}\ (a_{j})$ the creation (annihilation) operators of the mode of the $j$th cavity. 

In contrast to what we have considered in the above sections, we now proceed to analyze the dynamics of the system that is completely isolated from the environment, so that the dissipation effects are not present. On the one side, this restricts the applicability of the results to the case of a good cavity. In fact, this is not an overly restrictive limitation regarding the recent advances in the cavity QED. On the other hand, the absence of the dissipation means that the evolution of the system is purely coherent and is governed by the Schr\"odinger equation. It also admits analytical treatment of the concurrence. 

We consider separately the cases of single and two-photon excitations of the system. In the first, we assume that only a single excitation is present and find the state vector of the system by solving the Schr\"odinger equation
\begin{equation}
i\hbar\frac{d}{dt}|\xi(t)\rangle= H|\xi(t)\rangle ,\label{8.5}
\end{equation}
where
\begin{eqnarray}
|\xi(t)\rangle =  d_{1}(t)\ket{\xi_{1}}\!+\!d_{2}(t)\ket{\xi_{2}}\!+\!d_{3}(t)\ket{\xi_{3}}\!+\!d_{4}(t)\ket{\xi_{4}} \label{8.6}
 \end{eqnarray}
is the state vector of the system given in terms of the product states (\ref{5.1}) of the atoms and the cavity modes at time~$t$. The coefficient $d_{i}(t)$ determines the probability amplitude of the $i$th state at time~$t$. The problem is to find the time evolution of the probability amplitudes provided that the initial state of the system is known. Of particular interest are the values of $d_{i}(t)$ for an imperfect matching of the atoms to the cavity fields. It is easily verified that with the Hamiltonian (\ref{8.1}), the coefficients satisfy the following differential equations 
\begin{eqnarray}
&&\dot{d_{1}} = -ig_{1}d_{3} ,\quad \dot{d_{3}} = 2i\Delta_{1}d_{3}-ig_{1}d_{1} ,\nonumber \\
&&\dot{d_{2}}=-ig_{2}d_{4} ,\quad \dot{d_{4}}=2i\Delta_{2}d_{4}-ig_{2}d_{2} ,\label{8.7}
\end{eqnarray}
where $2\Delta_{j}=(\omega_{0}-\omega_{j})$ represents the detuning between the atomic transition frequency and the $j$th cavity frequency.

Equations (\ref{8.7}) form decoupled pairs of simple differential equations that can be solved by using e.g. the Laplace transform technique. A solution of the equations, valid for an arbitrary initial state is readily found to be
\begin{eqnarray}
&&d_{1}(t) = {\rm e}^{i\Delta_{1}t}\left\{ d_{1}(0)\cos(\Omega_{1}t)\right. \nonumber\\
&&\left. \quad \quad \ -\,i\left[\delta_{1}d_{1}(0)+\beta_{1}d_{3}(0)\right]\sin(\Omega_{1}t)\right\} ,\nonumber \\
&&d_{2}(t) = {\rm e}^{i\Delta_{2}t}\left\{d_{2}(0)\cos(\Omega_{2}t)\right. \nonumber \\
&&\left. \quad \quad \ -\,i\left[\delta_{2}d_{2}(0)+\beta_{2}d_{4}(0)\right]\sin(\Omega_{2}t)\right\} ,\nonumber\\
&&d_{3}(t) = {\rm e}^{i\Delta_{1}t}\left\{d_{3}(0)\cos(\Omega_{1}t)\right. \nonumber \\ 
&&\left.\quad \quad \  +\,i\left[\delta_{1}d_{3}(0)-\beta_{1}d_{1}(0)\right]\sin(\Omega_{1}t)\right\} ,\nonumber \\
&&d_{4}(t) = {\rm e}^{i\Delta_{2}t}\left\{d_{4}(0)\cos(\Omega_{2}t)\right. \nonumber \\
&&\left. \quad \quad \ +\,i\left[\delta_{2}d_{4}(0)-\beta_{2}d_{2}(0)\right]\sin(\Omega_{2}t)\right\} ,\label{8.8}
\end{eqnarray}
where $\Omega_{i}=\sqrt{g_{i}^{2}+\Delta_{i}^{2}}\ (i=1,2)$ is a detuned Rabi frequency, $d_{j}(0)\, (j=1,2,3,4)$ are the initial 
values of the probability amplitudes, $\delta_{i}=\Delta_{i}/\Omega_{i}$, and $\beta_{i}=g_{i}/\Omega_{i}$ represent, respectively, 
the scaled (dimensionless) frequency detuning and coupling constant between the $i$th atom and the $i$th cavity. 

The probability amplitudes oscillate sinusoidally with the Rabi frequency $\Omega_{i}$, and their dynamics is strongly affected 
by the modulation term, proportional to $\sin(\Omega_{i}t)$, whose amplitude depends on the detuning $\delta_{i}$ and the coupling constant $\beta_{j}$ between the atom and the 
corresponding cavity mode. The modulation terms vanish periodically at times $t_{n}=n\pi/\Omega_{i},\ n=0,1,\ldots $. 

In general, the time evolution of the probability amplitudes is quite complicated and not easy to interpret. However, one can see 
from Eqs.~(\ref{8.8}) that the detuning enters the solutions in an antisymmetric way, whereas the coupling strength enters the 
solutions in a symmetric way. As we shall see, this difference will be evident in the features of the time evolution of 
entanglement in the system. 

Our interest is to evaluate the concurrence to quantify entanglement between the different parts of the two two-qubit subsystems $1$ and $2$. By denoting the two atoms as $A$ and $B$, and the corresponding cavity modes as $a$ and $b$, we may distinguish six pairs of sub-systems
$AB,ab,Aa,Ab,Ba$ and $Bb$. We shall quantify entanglement in each pair by the concurrence.
To calculate the entanglement between any two qubit pair $ij$, for example the atoms $A$ and $B$, we take the trace over the other sub-sytems, to evaluate the reduced density matrix, from which the concurrence can be calculated. In general, the reduced states are two qubit mixtures, and have a density matrix of the general $X$-form
 \begin{equation}
\rho_{ij}(t) =\left(\begin{array}{cccc}
a(t) & 0 & 0 & 0\\
0 & b(t) & z(t) & 0\\
0 & z^{\ast}(t) & c(t) & 0\\
0 & 0 & 0 & d(t)\end{array}\right) ,\label{8.9}
\end{equation}
which, according to Eq.~(\ref{4.11}) gives for the concurrence 
\begin{equation}
{\cal C}_{ij}(t) = 2\max\bigl\{0,|z(t)|-\sqrt{a(t)d(t)}\bigr\}.\label{8.10}
\end{equation}

These concurrences are found to be evaluated as
\begin{eqnarray}
&&{\cal C}_{AB}(t) = 2|d_{1}(t)||d_{2}(t)|,\quad {\cal C}_{ab}(t)=2|d_{3}(t)||d_{4}(t)| ,\nonumber \\
&&{\cal C}_{Aa}(t) = 2|d_{1}(t)||d_{3}(t)|,\quad {\cal C}_{aB}(t)=2|d_{2}(t)||d_{3}(t)| ,\nonumber \\
&&{\cal C}_{Ab}(t) = 2|d_{1}(t)||d_{4}(t)|,\quad {\cal C}_{Bb}(t)=2|d_{2}(t)||d_{4}(t)| .\nonumber\\
\label{8.11} 
\end{eqnarray}
Note that there is no the threshold term when only one-photon states are involved in the dynamics of the system~\cite{jam}.
 
\noindent As an example of the sensitivity of the system to imperfect matching of the atoms to the cavity modes, consider the evolution of the concurrences for the following initial~state 
\begin{eqnarray}
\ket{\xi_{0}} &=& \frac{1}{\sqrt{2}}\left(\ket{\xi_{1}}\pm\ket{\xi_{2}}\right)\nonumber\\
 &=&  \frac{1}{\sqrt{2}}\left( \ket{e_{1}}\otimes\ket{g_{2}}\pm \ket{g_{1}}\otimes\ket{e_{2}}\right)\otimes{\ket 0}_{1}\otimes{\ket 0}_{2} .\nonumber\\
 \label{11a}
\end{eqnarray}
Note, that the initial entanglement is restricted to just the two atoms. Hence, for the case of equal coupling constants, $g_{1}=g_{2}\equiv g$ but non-zero detunings, the bipartite concurrence measures are 
\begin{eqnarray}
{\cal C}_{AB}(t) & = & \cos^{2}\left(\Omega gt\right)+\left(\frac{\delta}{\Omega}\right)^{2}\sin^{2}\left(\Omega gt\right),\nonumber \\
{\cal C}_{ab}(t) & = & \frac{1}{\Omega^{2}}\sin^{2}\left(\Omega gt\right),\nonumber \\
{\cal C}_{Aa}(t) & = & {\cal C}_{Ab}(t)={\cal C}_{Ba}(t)={\cal C}_{Bb}(t)=\frac{1}{\Omega}\sin\left(\Omega gt\right)\nonumber \\
 &\times& \left[\cos^{2}\left(\Omega gt\right)+\left(\frac{\delta}{\Omega}\right)^{2}\sin^{2}\left(\Omega gt\right)\right]^{\frac{1}{2}} ,\label{8.12}
\end{eqnarray}
where we have introduced a dimensionless Rabi frequency, the same for both subsystems, $\Omega=(1+\delta^{2})^{1/2}$, and the scaled detuning $\delta=\Delta/g$. 
\begin{figure}[hbp]
\includegraphics[width=8cm,keepaspectratio,clip]{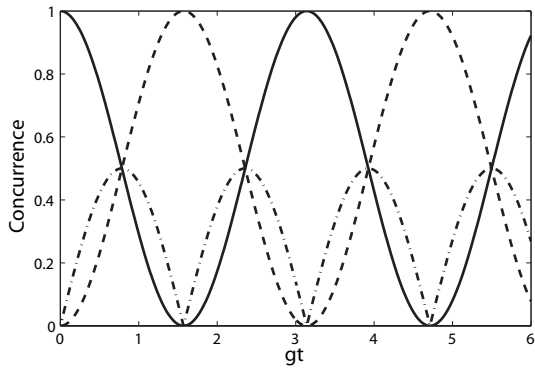}
\caption{Time evolution of the pair concurrences for equal coupling constants,
$g_{1}=g_{2}=g$ and exact resonance $\Delta_{1}=\Delta_{2}=0$: ${\cal C}_{AB}(t)$ (solid
line), ${\cal C}_{ab}(t)$ (dashed line), and ${\cal C}_{Aa}(t),{\cal C}_{aB}(t),{\cal C}_{Ab}(t),{\cal C}_{bB}(t)$ (dashed-dotted line). The atoms were initially in the entangled state~$\ket{\Psi_{s}}$.}
\label{figc19}
\end{figure}

The expressions (\ref{8.12}) for the concurrence measures are plotted in Fig.~\ref{figc19}. We see periodic oscillations of the concurrence, which reveal the periodic transfer of entanglement from atoms to cavity modes, and vice versa. It is interesting that the transfer process of entanglement from the atoms to the cavity modes does not involve just the pairs ${\cal C}_{AB}(t)$ and ${\cal C}_{ab}(t)$. By inspection of the time evolution of the concurrences in Fig.~\ref{figc19}, we find that the initial maximal entanglement between the atoms is not only totally transferred to the cavity modes, but at the same time an additional pairwise entanglement is created during the evolution. This can also be seen by summing the pair concurrence measures to find that at times $t_{n}=n\pi/4\Omega$, where $n=1,3,5...$, the total pair concurrence is larger than one. As time progresses, a part of the entanglement is transferred into the other pairs of the sub-systems. Then, after a further interval, the entanglement is completely transferred into the cavity modes. The additional entanglement vanishes at times $t_{n}=n\pi/2\Omega$, $n=1,2,3,...$, when the transfer process is completed. 

\begin{figure}[hbp]
\includegraphics[width=8cm,keepaspectratio,clip]{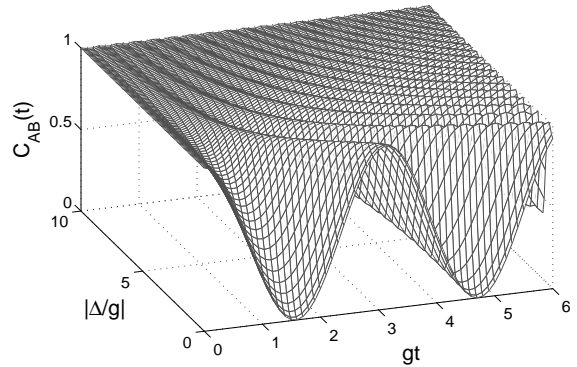}
\caption{The atom-atom concurrence ${\cal C}_{AB}(t)$ plotted as a function of the normalized time $gt$ and detuning $\Delta/g$ for equal coupling constants, $g_{1}=g_{2}\equiv g$.}
\label{figc20}
\end{figure}

Another interesting prediction of~(\ref{8.12}) is that complete entanglement transfer from the atoms to the cavity fields requires exact resonances. To examine the effect of detuning, we plot in Fig.~\ref{figc20} the atom-atom concurrence~${\cal C}_{AB}(t)$ as a function of time and the detuning $\Delta$. It is evident from the figure that in the case of non-zero detuning $\Delta$, the localized initial entanglement is not completely transferred to another pair of qubits. Detuning increases the oscillation frequency and decreases the minimum entanglement between the atoms. It is interesting, however, that the atomic entanglement returns to its initial maximum value periodically with the detuned Rabi frequency. This can be understood in terms of the localization of the initial energy. When $\Delta\neq 0$, only a part of the initial energy, proportional to $\omega_{0}$ is transferred to the cavity modes leaving the excess energy unlocalized. Thus, we may conclude that a non-resonant coupling of the atoms to the cavity modes can be used to stabilize entanglement between the atoms. 

We now turn to the problem of entanglement dynamics when two excitations (one in each subsystem) are present in the system. Using the Schr\"odinger equation, we set up equations of motion for the probability amplitudes and then solve them to study the dynamics of the entanglement. We define the state vector of the system in terms of the basis states (\ref{5.2}) as
\begin{eqnarray}
\ket\chi = d_{1}(t)\ket{\chi_{1}}\!+\!d_{2}(t)\ket{\chi_{2}}\!+\!d_{3}(t)\ket{\chi_{3}}\!+\!d_{4}(t)\ket{\chi_{4}} ,\label{8.13}
 \end{eqnarray}
and find from the Schr\"odinger equation that the probability amplitudes satisfy a closed set of four coupled equations of motion
\begin{eqnarray}
\dot{\tilde{d}}_{1} & = & -i(\Delta_{1}+\Delta_{2})\tilde{d}_{1}-i(g_{2}\tilde{d}_{2}+g_{1}\tilde{d}_{3}),\nonumber \\
\dot{\tilde{d}}_{2} & = & -i(\Delta_{1}-\Delta_{2})\tilde{d}_{2}-i(g_{2}\tilde{d}_{1}+g_{1}\tilde{d}_{4}),\nonumber \\
\dot{\tilde{d}}_{3} & = & i(\Delta_{1}-\Delta_{2})\tilde{d}_{3}-i(g_{1}\tilde{d}_{1}+g_{2}\tilde{d}_{4}),\nonumber \\
\dot{\tilde{d}}_{4} & = & i(\Delta_{1}+\Delta_{2})\tilde{d}_{4}-i(g_{1}\tilde{d}_{2}+g_{2}\tilde{d}_{3})
.\label{8.14}
\end{eqnarray}
To remove the fast oscillating terms in Eq.~(\ref{8.14}), we have introduced a rotating frame through the relations 
\begin{eqnarray}
\tilde{d}_{i} = d_{i}\,{\rm e}^{-i(\omega_{1}+\omega_{2})t} ,\quad i=1,2,3,4. \label{8.15}
\end{eqnarray}
As before for the single excitation, we allow for the possibility of imperfect couplings of the atoms to the cavity modes by introducing detunings $\Delta_{1}$ and $\Delta_{2}$, and unequal coupling  constants $g_{1}$ and $g_{2}$.

Although the set of equations of motion (\ref{8.14}) involves four coupled equations, it is analytically solvable. We can put Eqs.~(\ref{8.14}) into a matrix form and solve them by the matrix inversion. However, due to the complexity of the general solution, we study separately the time evolution of the probability amplitudes for two cases: In the first case, we consider non-zero detunings but equal coupling constants, i.e. $\Delta_{1}=\Delta_{2}\equiv\Delta$ and $g_{1}=g_{2}\equiv g$. In the second, we consider zero detunings, $\Delta_{1}=\Delta_{2}=0$ but unequal coupling constants, $g_{1}\neq g_{2}$.

We now investigate the evolution of the bipartite entanglement in the six pairs of the subsystems. We stress that the general state of the system, Eq.~(\ref{8.13}), cannot give any entanglement in the atom-atom subsystem, regardless of the values of the coefficients involved. This is readily seen by evaluating the reduced density operator for the atom systems. By tracing the density operator of the total system over the cavity modes, we obtain the reduced density operator of the atoms alone 
\begin{eqnarray}
\rho_{AB}={\rm Tr}_{cavity}\rho=\sum_{m,n=0,1}\langle n|\langle m|\rho|m\rangle\ket n ,\label{8.19}
\end{eqnarray}
where $\ket n$ and $\ket m$ refer to cavity modes $a$ and $b$ respectively, and find that
\begin{eqnarray}
\rho_{AB} &=& |d_{1}(t)|^{2}\ket{\Psi_{4}}\bra{\Psi_{4}}\!+\!|d_{2}(t)|^{2}\ket{\Psi_{2}}\bra{\Psi_{2}}\nonumber\\
&+& |d_{3}(t)|^{2}\ket{\Psi_{3}}\bra{\Psi_{3}}\!+\!|d_{4}(t)|^{2}\ket{\Psi_{1}}\bra{\Psi_{1}} .\label{8.20}
\end{eqnarray}
Evidently, the density operator represents is a mixture of separable states and hence cannot be entangled. The atom-atom bipartite entanglement is always zero, ${\cal C}_{AB}(t)=0$. The same conclusion applies to the bipartite entanglements ${\cal C}_{ab}(t),{\cal C}_{Ab}(t)$ and ${\cal C}_{Ba}(t)$. The only non-zero concurrence possible is for the qubit pairs ${\cal C}_{Aa}(t)$ and ${\cal C}_{Bb}(t)$. In other words, an entanglement can be created during the evolution between the atoms and the cavity modes to which they are coupled. The amount of entanglement that can
be created in these pairs can be determined from the relations 
\begin{eqnarray}
{\cal C}_{Aa}(t)=2|\tilde{d}_{1}(t)\tilde{d}_{3}^{\ast}(t)+\tilde{d}_{2}(t)\tilde{d}_{4}^{\ast}(t)|,\label{8.21}
\end{eqnarray}
 and \begin{eqnarray}
{\cal C}_{Bb}(t)=2|\tilde{d}_{1}(t)\tilde{d}_{2}^{\ast}(t)+\tilde{d}_{3}(t)\tilde{d}_{4}^{\ast}(t)|.\label{8.22}
\end{eqnarray}
Equations show that the source of entanglement between the atoms and the corresponding cavity modes is in the interference between the probability amplitudes.
\begin{figure}[hbp]
\includegraphics[width=8cm,keepaspectratio,clip]{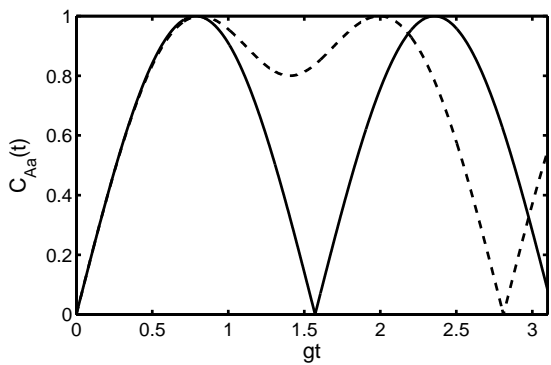}
\caption{Time evolution of the concurrence ${\cal C}_{Aa}(t)$ for the initial separable
state~$\ket{\chi_{1}}$ and different detunings: $\Delta=0$ (solid line) and $\Delta=g$ (dashed line).}
\label{figc21}
\end{figure}

Figure \ref{figc21} shows the entanglement creation and evolution between the atom $A$ and the cavity mode~$a$, with an initial separable state $\ket{\chi_{1}}$. Similar
entanglement properties appear between the atom $B$ and the cavity mode~$b$. We see that an entanglement between the atom and the cavity mode is created during the transfer process of the excitation between them. The maximum entanglement is observed at times when the excitation
is equally shared between the atom and the cavity mode, and vanishes when the excitation is completely located at either the atom or the cavity mode. In terms of the Rabi oscillations, the system becomes disentangled at every half of the Rabi cycle of the oscillation, i.e. at $gt=n\pi/\Omega\ (n=0,1,2,\ldots)$.

Similar to the case of single excitation states, the system becomes entangled for a longer time when the frequencies of the atom and the cavity field are detuned from each other. In other
words, an imperfect matching between the atom and the cavity field leads to a more stable entanglement than in the case of the perfect matching. In terms of the Rabi oscillations, the system becomes disentangled only at every Rabi cycle of the oscillation, i.e. at $gt=2n\pi/\Omega$. This is easy to understand if one considers entanglement as resulting from a superposition corresponding to a delocalization of energy. When the frequencies of the atom and the cavity mode are different, the initial amount of energy localized in the atom is not completely
transferred to the cavity mode. A part of the energy, not absorbed by the cavity mode remains delocalized, which results in a nonzero entanglement.

The above analysis has showed that no pairwise entanglement is possible between any parts of the two subsystems $1$ and $2$ when we have with certainty a single excitation in each of the subsystems. Nevertheless, we can generate entanglement in the qubit pairs by including in the wave function an auxiliary state $\ket{\chi_{0}}$, the ground state for which there is no excitation, and consider the time evolution of a state vector of the system containing the auxiliary state
\begin{eqnarray}
\ket{\chi (t)} &=& d_{1}(t)\ket{\chi_{1}} +d_{2}(t)\ket{\chi_{2}}
+d_{3}(t)\ket{\chi_{3}} \nonumber\\
&+& d_{4}(t)\ket{\chi_{4}} +d_{0}(t)\ket{\chi_{0}} .\label{8.23}
\end{eqnarray}
With the auxiliary state included, we find that the bipartite concurrence measures between the subsystems are of the forms 
\begin{eqnarray}
&&{\cal C}_{AB}(t) = 2\max\bigl\{0,|d_{1}(t)||d_{0}(t)|-|d_{2}(t)||d_{3}(t)|\bigr\} ,\nonumber \\
&&{\cal C}_{ab}(t) = 2\max\bigl\{0,|d_{4}(t)||d_{0}(t)|-|d_{2}(t)||d_{3}(t)|\bigr\} ,\nonumber \\
&&{\cal C}_{Ab}(t) = 2\max\bigl\{0,|d_{2}(t)||d_{0}(t)|-|d_{1}(t)||d_{4}(t)|\bigr\} ,\nonumber \\
&&{\cal C}_{aB}(t) = 2\max\bigl\{0,|d_{3}(t)||d_{0}(t)|-|d_{1}(t)||d_{4}(t)|\bigr\} . \nonumber\\
\label{8.24}
\end{eqnarray}
Evidently, an entanglement between the pairs of subsystems is possible \emph{only} if the ground state is included. Since the state $\ket{\chi_{0}}$ is not an eigenstate of the Hamiltonian of the system, the probability amplitude $d_{0}(t)$ is a constant of motion, i.e. the amplitude does not evolve in time, $d_{0}(t)=d_{0}(0)$.

Equation~(\ref{8.24}) shows several interesting properties of the concurrence and differences between the single and double-excitation cases. Firstly, the concurrences appear as differences of products of the absolute values of the probability amplitudes, i.e. they involve the threshold term for entanglement. This gives a possibility for sudden death and revival in the behavior of entanglement that were absent in the case of the one-excitation. Secondly, there is evident a competition in the creation of atom-atom entanglement between two pairs of states $(\ket{\chi_{1}}, \ket{\chi_{4}})$ and $(\ket{\chi_{2}},\ket{\chi_{3}})$. For example, entanglement creation
in concurrence pairs involving the states $\ket{\chi_{1}}$ and $\ket{\chi_{4}}$ is diminished by the presence of population in the states $\ket{\chi_{2}}$ and $\ket{\chi_{3}}$, and vice versa creation of entanglement involving the states $\ket{\chi_{2}}$ and $\ket{\chi_{3}}$ is diminished by the presence of population in the states $\ket{\chi_{1}}$ and $\ket{\chi_{4}}$. In terms of the population transfer, entanglement creation by a simultaneous exchange of two photons is diminished by one-photon exchange processes, and vice versa. The competition between these one and two-photon processes is the source of the phenomenon of entanglement sudden death.
\begin{figure}[hbp]
\includegraphics[width=8cm,keepaspectratio,clip]{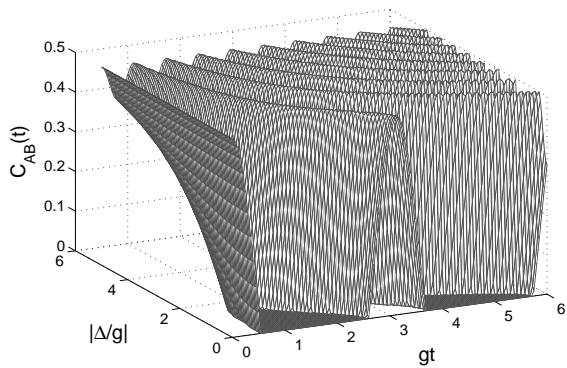}
\caption{The atom-atom concurrence ${\cal C}_{AB}$ as a function of the normalized time $gt$ and detuning $\Delta/g$ for $g_{1}=g_{2}\equiv g$ and the initial state $\ket{\chi_{sd}}$, given by Eq.~(\ref{5.4}) with $\alpha=\pi/12$ and $\beta =0$. The effect of the detuning is to remove the sudden death of entanglement.}
\label{figc22}
\end{figure}

Figure~\ref{figc22} illustrates the atom-atom concurrence as a function of time and the detuning $\Delta$, calculated from an initial condition of the atoms prepared in the non-maximally entangled state (\ref{5.4}). We can see that in the case of $\Delta=0$, the initial entanglement is lost in a discontinuous fashion and can revive periodically during the evolution. On the other hand, when $\Delta\neq 0$, the amplitude of the oscillation decreases and the entanglement sudden death dissapears. Thus, an imperfect matching of the cavities to the atomic transition can prevent the onset of the entanglement sudden death, to provide a stabilization of entanglement. The principle of the effect, which is similar to that showed in Fig.~\ref{figc20}, may again be readily be understood by realizing that a nonzero detuning inhibits the full entanglement transfer from the atoms to the cavity modes.

\section{Triggered entanglement evolution}

\noindent We now proceed to analyze the problem of how one could switch on (trigger) an evolution of stable or "frozen" entanglement.  This is an important problem for controlled transmission on demand of an initial entanglement stored, for example, in a trapped state. As we have seen, trapped states have the property that an initial population stored in these states cannot decay due to zero transition dipole moments to and from the remaining states of a given system. Then an interesting practical question arises, how one could access the trapping state to trigger an evolution of the encoded information. In the following, we address this question and consider two schemes introduced in the previous sections. In the first, we consider two atoms located inside a single mode cavity and initially prepared in a maximally entangled state. In the second, the atoms are located in separate cavities and the system is initially prepared in an entangled state such that the initial entanglement is equally shared between all the pairs of the sub-systems. The concurrence is then monitored as a function of time and atom-cavity-field detunings. 

Let us begin with the scheme, introduced in Section~7, that involves two atoms located inside a single mode cavity. Suppose that the atoms are initially prepared in the collective single excitation symmetric state 
\begin{equation}
\ket{\Psi_{s}} = \frac{1}{\sqrt{2}}\left( \ket{e_{1}}\otimes\ket{g_{2}} +\ket{g_{1}}\otimes\ket{e_{2}}\right) ,\label{9.1} 
\end{equation}
which is a maximally entangled state. Alternatively, we may say that an information has been initially encoded into the system and is stored in the maximally entangled symmetric state.

For the initial state (\ref{9.1}), the initial values of the Bloch vector components are $u_{0}=1,\, v_{0}=w_{0}=0$. It follows from Eqs.~(\ref{7.22}) and (\ref{7.26}) that initially at $t=0$, the concurrence ${\cal C}(0)=1$, and then for $t>0$ the time evolution of the concurrence is of the form
\begin{eqnarray}
  {\cal C}(t) = \left| 1 -\frac{2\delta_{12}^{2}}{\alpha^{2}}\sin^{2}
  \left(\frac{1}{2}\alpha t\right) -i\frac{\delta_{12}}{\alpha}\sin\alpha t\right|{\rm e}^{-\gamma t}  ,\label{9.2}
\end{eqnarray}
where $\gamma$ is the spontaneous emission rate and $\delta_{12}$ results from a non-equivalent position of the atoms inside the cavity mode. We see from Eq.~(\ref{9.2}) that in the absence of the spontaneous emission and with $\delta_{12}=0$, the concurrence does not evolve in time. In other words, the concurrence is a constant of motion that an initial entanglement encoded into the system at $t=0$ will remain there for all times. The evolution of the concurrence can be triggered by switching on a non-zero detuning $\delta_{12}$. In this case, the  concurrence oscillates in time and the amplitude of the oscillation increases with 
$\delta_{12}$. 
\begin{figure}[hbp]
\includegraphics[width=8cm,keepaspectratio,clip]{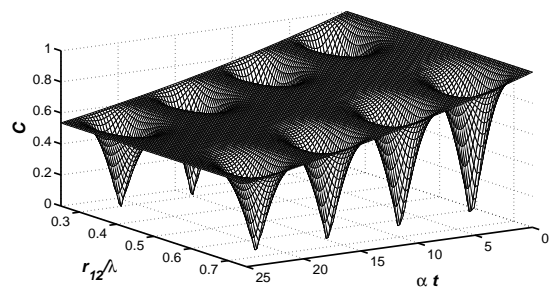}
\caption{Concurrence as a function of the normalized time $\alpha t$ and the position $r_{12}$ of the second atoms inside a half-wavelength of the standing-wave cavity mode. The overall slow decay of the concurrence is due to spontaneous emission with the rate $\gamma/\alpha =0.1$.}
\label{figc23}
\end{figure}

Figure~\ref{figc23} displays the concurrence as a function of time and the position $r_{12}$ of the atom~$"2"$ inside a half-wavelength of the cavity mode. We see that the initial entanglement does not evolve in time when the atom is located at $r_{12}/\lambda =0.5$, the antinode of the cavity field. This situation corresponds to the case of both atoms located in equivalent positions,  antinodes of the standing wave of the cavity field. Consequently, $g_{1}=g_{2}$ and then $\delta_{12}=0$. The absence of the oscillations in the concurrence versus time is linked to the fact that the Bloch vector $\vec{B}$ and the $\Omega_{B}$ vector are initially aligned at $t=0$ and remain aligned for all times when $r_{12}/\lambda =0.5$. By breaking the symmetry between the coupling constants, i.e. by a dislocation of the atom $"2"$ from the antinode of the standing wave, $g_{1}\neq g_{2}$, and then $\delta_{12}\neq 0$. As a result, the concurrence oscillates in time indicating a continuous oscillation of the excitation between the atoms. Here, the oscillation of the concurrence comes from the precession of the Bloch vector about the vector $\Omega_{B}$ with frequency $\alpha$. Thus, by an imperfect coupling of the atoms to the cavity mode, one can trigger an evolution of the "frozen" entanglement. 

Consider now the second scheme, introduced in Section~8 that involves two atoms each located in separate single-mode cavity. We concentrate on the $g_{1}=g_{2}$ case and will demonstrate the effect of non-zero detunings 
$\Delta_{1}=\Delta_{2}=\Delta$ on the evolution of an initial entanglement. We use the same strategy as in the above example and assume that initially the system was prepared in a superposition state such that the concurrence remains a constant of motion under the condition that the modes of the two cavities are resonant with the atomic transition frequencies.
\begin{figure}[hbp]
\includegraphics[width=8cm,keepaspectratio,clip]{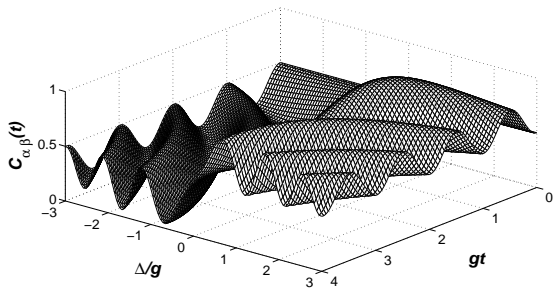}
\caption{Concurrence as a function of the normalized time $gt$ and the detuning $\Delta$ of 
the cavity modes from the atomic resonances for $g_{1}=g_{2}$.}
\label{figc24}
\end{figure}

Assume that the system is initially prepared in a superposition state with a uniform population distribution over the available energy states
\begin{eqnarray}
|\xi_{0}\rangle=\frac{1}{2}\left(\ket{\xi_{1}}+{\rm e}^{i\theta}\ket{\xi_{2}} \pm(\ket{\xi_{3}}
-{\rm e}^{i\phi}\ket{\xi_{4}})\right) ,\label{9.3}
\end{eqnarray}
where $\theta$ and $\phi$ are arbitrary phase factors. In this case, $|d_{i}(0)|=1/2\, (i=1,\ldots,4)$ and then, according to Eq.~(\ref{8.11}), the initial entanglement is equally shared between the six qubit pairs, so ${\cal C}_{\alpha\beta}(0)={\cal C}_{ab}(0)={\cal C}_{\alpha a}(0)={\cal C}_{\alpha b}(0)={\cal C}_{\beta a}={\cal C}_{\beta b}(0)=1/2$.

It is easy to show from Eqs.~(\ref{8.8}) and (\ref{8.11}) that with the initial state (\ref{9.3}) and zero detunings, the two-qubit entanglement between all of the pairs of the sub-systems remains constant in time. In other words, the initial entanglement is frozen in the initial state. It is interesting to note that the entanglement remains frozen in the initial state independent of whether $g_{1}=g_{2}$ or $g_{1}\neq g_{2}$.

The dependence of the entanglement of an arbitrary qubit pair on time $t$ and the detuning $\Delta$ is illustrated 
in Fig.~\ref{figc24}. The evolution of the concurrence exhibits several interesting properties. First of all, the concurrences 
vary in time only for nonzero detuning. Thus, a nonzero detuning $\Delta$ triggers an evolution of the entanglement. Secondly, 
the time evolution of the concurrences is not symmetric with respect to the sign of the detunings. A large and even \emph{maximal} 
entanglement can be created when the detuning is \emph{positive}, whereas the initial atomic entanglement is reduced and can even 
be suppressed when the detuning is \emph{negative}. When, for example, the atom-atom entanglement is maximum, ${\cal C}_{AB}(t)=1$, 
then the entanglement between the other qubit pairs is zero and vice versa, when the atom-atom entanglement is zero then entanglement 
between one of the other qubit pairs is maximal. This result also implies a possibility of a controlled evolution of the system 
towards the maximum entanglement between one of the qubit pairs. Finally, the results show that the maximum entanglement of an 
arbitrary qubit pair can be created when the detuning is matched to the coupling constant,~$\Delta =g$. Otherwise, the entanglement is reduced.

\section{Steered entanglement transfer}

\noindent The possibility of triggering an asymmetric evolution of an initial entanglement by changing the sign of the detuning $\Delta$ implies that one can also engineer the direction of evolution of entanglement by controlling the detuning. By this we mean that an initial entanglement can be transferred to a desired ``localized'' pair of the qubits by a suitable choice of the detunings. The localized atom-atom entanglement in this context means that the entanglement exists solely between the two atoms. As it is clear from Fig.~\ref{figc24}, a suitably chosen positive value of the detuning can channel entanglement entirely into the atoms. Therefore, the entanglement transfer can be controlled by varying the frequency of the cavity mode.

Another method of transferring an entanglement between qubit pairs in a controlled way is to create an asymmetry between the coupling constants $g_{1}$ and $g_{2}$. We will describe how to achieve a steered evolution of an initial entanglement to a desired pair of qubits using the scheme involving two cavities each containing a single two-level atom. We assume that the system is prepared initially in a superposition state 
\begin{eqnarray}
|\xi_{0}\rangle = \frac{1}{\sqrt{2}}\left(\ket{\xi_{1}} +\ket{\xi_{2}}\right) ,\label{10.1}
 \end{eqnarray}
where $\ket{\xi_{1}}$ and $\ket{\xi_{2}}$ are the single excitation states of the basis (\ref{5.1}). The initial state involves the superposition of the atoms only, so that initially at $t=0$ the atoms are maximally entangled, ${\cal C}_{AB}(0)=1$, and the other pairs of qubits are disentangled. During the evolution, the atom-atom entanglement can be completely transferred only to one of the pairs of qubits or can be redistributed between different pairs. A couple questions then arises; Can we steer the transfer process such that the entanglement could be transferred to a desired pair of qubits? Can we engineer the transfer process by changing only the system's parameters, such as the coupling constants of the atoms to the cavity modes?

Let us proceed to answer these questions by considering the time evolution of the concurrence. Provided the system is prepared initially in the state $\ket{\xi_{0}}$, we find from Eqs.~(\ref{8.8}) and (\ref{8.11}) that for the case of exact resonances $\Delta_{1}=\Delta_{2}=0$ and unequal coupling constants $g_{1}\neq g_{2}$, the entanglement measures take the forms
\begin{eqnarray}
{\cal C}_{AB}(t) & = & \bigl|\cos\left(\Omega_{1}gt\right)\bigl|\bigl|\cos\left(\Omega_{2}gt\right)\bigl| ,\nonumber\\
{\cal C}_{ab}(t) &=& \bigl|\sin\left(\Omega_{1}gt\right)\bigl|\bigl|\sin\left(\Omega_{2}gt\right)\bigl| ,\nonumber\\
{\cal C}_{Aa}(t) &=& \bigl|\cos\left(\Omega_{1}gt\right)\bigl|\bigl|\sin\left(\Omega_{1}gt\right)\bigl| ,\nonumber\\
{\cal C}_{Ab}(t) &=& \bigl|\cos\left(\Omega_{1}gt\right)\bigl|\bigl|\sin\left(\Omega_{2}gt\right)\bigl|,\nonumber \\
{\cal C}_{Ba}(t) & = & \bigl|\cos\left(\Omega_{2}gt\right)\bigl|\bigl|\sin\left(\Omega_{1}gt\right)\bigl| ,\nonumber\\
{\cal C}_{Bb}(t) &=&\bigl|\sin\left(\Omega_{2}gt\right)\bigl|\bigl|\cos\left(\Omega_{2}gt\right)\bigl| ,\label{10.2}
\end{eqnarray}
Here we have introduced dimensionless Rabi frequencies $\Omega_{1}=(1+u/g)$ and $\Omega_{2}=(1-u/g)$, with $g=(g_{1}+g_{2})/2$ and $u=(g_{1}-g_{2})/2$. 

We see that the entanglement measures are in the form of the product of two harmonic functions oscillating with different frequencies. At $t=0$, ${\cal C}_{AB}(0) =1$ and it is our purpose to determine which of the qubit pairs can be maximally entangled at $t>0$. Since $\sin\left(\Omega_{1(2)}gt\right)$ and $\cos\left(\Omega_{1(2)}gt\right)$ cannot be simultaneously equal to one, it rules out the possibility of the maximum concurrence equal to unity for the pairs $Aa$ and $Bb$. To further distinguish which of the remaining four concurrence measures can be equal to one at $t>0$, we note that if the ratio $g_{2}/g_{1}$ is not an integer number or a fraction of an integer number, none of the concurrence measures can be equal to one at $t>0$. It implies that no complete transfer of the initial maximal entanglement is possible to any of the qubit pairs. The complete transfer is possible only if the ratio is an integer number or a fraction of an integer number. 

However, the destination to where the initial entanglement can be completely transfered depends on whether the ratio is an even or an odd integer number. If the ratio $g_{2}/g_{1}$  is an even number, the initial maximal entanglement between the atoms  can be completely transferred {\it only} to the atom-field qubit pair $Ba$. On the other hand, if the ratio is an odd integer number, the initial entanglement between the atoms can be completely transferred only to the field-field qubit pair $ab$. From this it follows~that
\begin{eqnarray}
{\cal C}_{AB}(0)=1 \Longrightarrow \left\{
\begin{array}{cc}
     {\cal C}_{Ba}(t) =1 \quad {\rm for}\ g_{2}/g_{1}\ {\rm even} ,  \label{10.3} \\
{\cal C}_{ab}(t)=1 \quad {\rm for}\  g_{2}/g_{1}\ {\rm odd} .
\end{array}
\right.
\end{eqnarray}

\begin{figure}[hbp]
\includegraphics[width=9cm,keepaspectratio,clip]{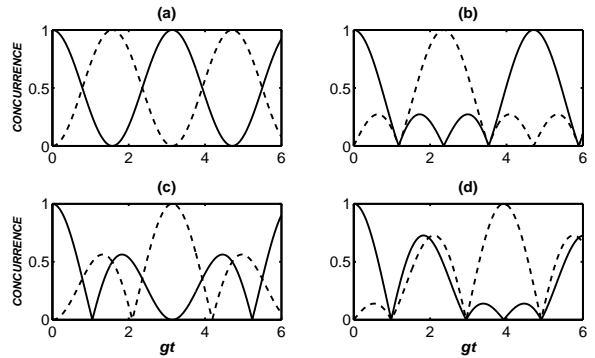}
\caption{Concurrence for qubit pairs plotted as a function of the normalized time 
$gt=\frac{1}{2}(g_{1}+g_{2})t$ for $\Delta =0$ and different ratios of the coupling constants 
$g_{2}/g_{1}$: (a) $g_{2}/g_{1}=1$, (b) $g_{2}/g_{1}=2$, (c)~$g_{2}/g_{1}=3$, (d) $g_{2}/g_{1}=4$. In all figures the solid line is for the atom-atom concurrence ${\cal C}_{AB}(t)$. The dashed line is for the concurrence measure (a) ${\cal C}_{ab}(t)$,  (b) ${\cal C}_{Ba}(t)$, (c) ${\cal C}_{ab}(t)$ and (d) ${\cal C}_{Ba}(t)$.}
\label{figc25}
\end{figure}

We illustrate this situation in Fig.~\ref{figc25}, where we plot the concurrence of the different qubit pairs as a function of time for exact resonances but unequal coupling constants. When $g_{1}=g_{2}$, we obtain the complete transfer of the entanglement only between the $AB$ and $ab$ qubit pairs. For $g_{1}\neq g_{2}$, the initial entanglement between the atoms can be completely transferred between $AB$ and the qubit pairs $ab$ and $Ba$. The transfer of the entanglement between $AB$ and the other pairs also occurs, but is not complete. 

The reason for this feature of the entanglement transfer can be understood intuitively by noting that, for example, for $g_{2}/g_{1}=2$ the Rabi frequency $g_{2}$ of the population oscillation in
the cavity system $2$ is twice that of the Rabi frequency $g_{1}$ for the population oscillation in the system $1$. This means that over a complete Rabi cycle $g_{2}t=\pi$, the initial population
in the system $2$ returns to the atom, but at the same time the population makes a half Rabi cycle in the system $1$, i.e. the excitation in system $1$ will be in the cavity mode. Thus, ${\cal C}_{Ba}(t)=1$ at that time, with the concurrence in the other qubit pairs equal to zero.

In summary of this section, we have seen that in the scheme involving two separate cavities, each containing a single two-level atom, we can achieve a controlled (steered) transfer of an initial entanglement to a desired pair of qubits. The results show that the initial entanglement can be completely transferred not to all but only to some of the qubit pairs. The complete steered transfer of entanglement can be done by properly adjusting the cavity frequencies to the atomic transition frequencies or by a proper engineering of the coupling constants between the atomic dipole moments and the cavity modes.

\section{Beyond the rotating-wave approximation}

\noindent The most interesting feature of transient entanglement is the phenomenon of entanglement sudden death. We have already seen that an entanglement encoded in spin correlated states can undergo an abrupt discontinuous evolution, while an initial entanglement encoded in a spin anti-correlared (single excitation) state decays asymptotically in time without any discontinuity. The lack of discontinuity has been referred to zero population of the two-photon state involved in the threshold term for the concurrence. If initially only a single excitation 
is present in the system, the population of the two-photon state is always zero. As a result, no threshold term is present in the concurrence measure and consequently an initial entanglement evolves continuously without any discontinuity~\cite{jam}. However, the analysis were restricted to the evolution of the system under the RWA Hamiltonian~\cite{ct92}.  In this section, we return to the problem of the transient evolution of entanglement in the system of two atoms located inside a single mode cavity. In contrast to what we have discussed in Section~7, we now 
examine the transient evolution of entanglement beyond the~RWA. We shall demonstrate that under the non-RWA interaction, an entanglement initially encoded into a single excitation state can undergo a discontinuity during the evolution of the system~\cite{jl09,yyc09,jz08}. We interpret this result as a consequence of the principle of complementarity between the evolution time and energy.

Let us reconsider the system composed of two identical two-level atoms located inside a single-mode standing-wave cavity. Our purpose is to compare the dynamics of an entanglement encoded initially in a spin anti-correlated state of the system under the RWA and non-RWA Hamiltonians. We assume that the atoms are strongly coupled to the cavity mode and consider the situation where the strength of the coupling is controlled by varying the position of the atoms  inside the standing~wave. Moreover, we will work in the short time regime of $\omega_{0}t\approx 1$, which is much shorter than a typical spontaneous emission time. Hence, we will neglect the spontaneous emission from the atoms.

The dynamics of the system are determined by the master equation of the density operator $\rho_{s}$ of the total, qubits plus the cavity field system
\begin{eqnarray}
\frac{\partial\rho_{s}}{\partial t} &=& -\frac{i}{\hbar}[H_{{\rm nRWA}},\rho_{s}] \nonumber \\
&-& \frac{1}{2}\kappa\left(a^\dagger a\rho_{s} + \rho_{s} a^\dagger a -2a\rho_{s} a^\dagger\right) ,\label{11.1}
\end{eqnarray}
where $\kappa$ is the damping rate of the cavity mode, and
\begin{eqnarray}
H_{{\rm nRWA}} &=& \frac{1}{2}\hbar\omega_{0}\sum_{j=1}^{2}S^z_j+\hbar\omega_{c} a^\dagger a \nonumber \\
&+& \hbar\sum_{j=1}^2\left[g(r_{j})S^{x}_{j}a^\dagger +g^{\ast}(r_{j})aS^{x}_{j}\right] \label{11.2}
\end{eqnarray}
is the non-RWA Hamiltonian of the system. The first term on the right-hand side of Eq.~(\ref{11.2}) is the Hamiltonian of the atoms, the second term is the Hamiltonian of the cavity field, and the third term the interaction Hamiltonian between the atoms and the cavity field in the electric-dipole approximation. The operators $a$ and $a^{\dagger}$ are the usual annihilation and creation operators of the cavity field, $S^{x}_{j}$ and $S^{z}_{j}$ are the atomic spin operators for the $j$th atom, and $\omega_{c}$ is the cavity mode frequency which, in general, can be different from the atomic transition frequency $\omega_{0}$. 

\begin{figure}[hbp]
\includegraphics[width=5cm,keepaspectratio,clip]{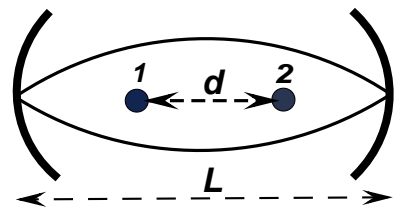}
\caption{Schematic diagram of the system composed of two distant atoms symmetrically located about the antinode of the standing-wave cavity mode.}
\label{figc26}
\end{figure}

The parameter $g(r_{j})$, which appears in Eq.~(\ref{11.2}) is the coupling constant between the cavity mode and the~$j$th atom located at a position $r_{j}$ along the cavity axis. The coupling constant varies with the position of the atoms inside the cavity mode~as
\begin{eqnarray}
g(r_{1,2}) = g_{0}\sin\left[\pi\left(n \mp\frac{d}{\lambda}\right)\right]  ,\label{11.3} 
\end{eqnarray}
where $n=L/\lambda$ determines the length of the cavity in units of the cavity wavelength~$\lambda$, and we have assumed that the atoms are placed symmetrically about the center of the cavity such that $r_{1}+r_{2}=L$ and $r_{2}-r_{1}=d$ is the distance between the atoms, as illustrated in  Fig.~\ref{figc26}. When the atoms are close to antinodes of the cavity mode, $d\approx (n-\frac{1}{2})\lambda$, and then both atoms experience the peak coupling strength $g_{0}$ to the cavity mode.  We shall call this situation, a strong coupling regime. The coupling constants decrease with the displacement of the atoms from the antinodes and become very small, $g(r_{j})\approx 0$ for distances $d\approx (n-1)\lambda$. In this case, the atoms are weakly coupled to the cavity mode, and we shall call this situation, a weak coupling regime. Thus, the strength of the coupling of the atoms to the cavity mode can be controlled by varying the positions of the atoms inside the standing~wave.

The interaction part of the Hamiltonian (\ref{11.2}) is in the non-RWA form. It contains both, the energy conserving terms~$S^{+}_{j}a$ and~$S^{-}_{j}a^{\dagger}$ as well as energy
non-conserving terms $S^{-}_{j}a$ and $S^{+}_{j}a^{\dagger}$, also known as the counter-rotating terms. The counter-rotating terms describe processes in which a photon is annihilated as the atom makes a downward transition, determined by the $S^{-}_{j}a$ term, or a photon is created as the atom makes an upward transition, determined by the $S^{+}_{j}a^{\dagger}$ term.

The counter-rotating terms, although at a first look seem to be unphysical since could lead to a violation of energy conservation, continue to create a debate on their significance and importance in the interaction of atoms with the electromagnetic field~\cite{lw09,guo,zr09}, since they have proven to produce important dynamical effects. Examples include the Bloch-Siegert shift of atomic frequencies~\cite{bs40}, quantum chaos~\cite{ma83}, bifurcations in the phase space~\cite{p91}, a fine structure in the optical Stern-Gerlach effect~\cite{l08}, generation of photons from the vacuum field~\cite{wdd08} and entanglement between atomic ensembles with no any initial excitation present~\cite{nb08}. We should point out that the effects are notable for strong couplings, $g_{0}\approx \omega_{0}$, and appear over short times,~$t\leq 1/\omega_{0}$.

Consider the time evolution of the concurrence when the dynamics of the system is determined by the non-RWA Hamiltonian. The required atomic density matrix elements are found by solving numerically the master equation (\ref{11.1}) in the combined $n$ photon basis of the cavity field and the product state basis of the atoms, Eq.~(\ref{3.6}). We compute the atomic density matrix elements by tracing the density operator~$\rho_{s}$ over the cavity field assumed to be in the vacuum state~$\ket 0$. Provided that the atoms are prepared initially in the single excitation collective symmetric state~$\ket I =\ket{\Psi_{s}}$, the concurrence is then determined by the ${\cal C}_{1}(t)$ criterion
\begin{eqnarray}
  {\cal C}(t) &=& \max\left\{0,\, {\cal C}_{1}(t)\right\}\nonumber \\
  &=&  2\max\!\left\{0,\!|\rho_{23}(t)|\!-\!\sqrt{\rho_{11}(t)\rho_{44}(t)}\right\} ,\label{11.4} 
\end{eqnarray}
It is apparent that the discontinuity or threshold behavior of the concurrence requires a nonzero population of both, the ground and the two-photon states of the system. For the initial state $\ket{\Psi_{s}}$, $\rho_{11}(0)=\rho_{44}(0)=0$, and it is predicted that $\rho_{44}(t)=0$ for all times as there are no external sources present. In this case, the initial entanglement is expected to evolve asymptotically without any discontinuity. 
\begin{figure}[hbp]
\includegraphics[width=8cm,keepaspectratio,clip]{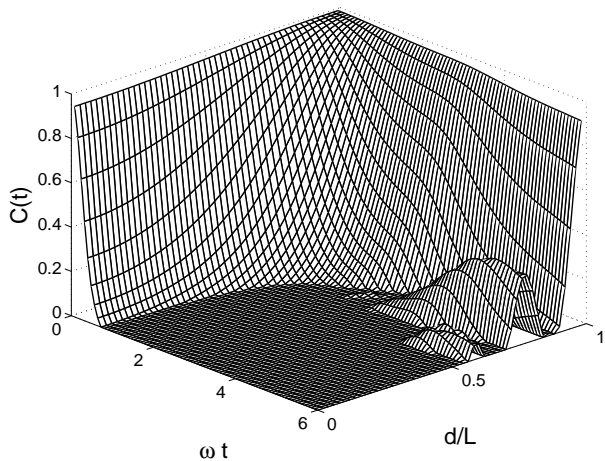}
\caption{Concurrence as a function of the dimensionless time $\omega t$ and the distance $d/\lambda$ between the atoms for  $g_{0}=\omega$, $(\omega -\omega_{0}) =0.01\omega$, $\kappa =0.1\omega$ and $n= 1/2$. The atoms were prepared initially in the single
excitation collective symmetric state $\ket{\Psi_{s}}$.}
\label{figc27}
\end{figure}

Let us examine if the expectations of the asymptotic evolution of the initial entanglement are manifested when the dynamics of the system undergoes the non-RWA Hamiltonian. 
Figure~\ref{figc27} displays the time evolution of the concurrence for the initial single excitation collective symmetric state~$\ket{\Psi_{s}}$ and for the peak coupling strength of the atoms to the cavity mode $g_{0}=\omega$. We see that for small distances between the atoms, corresponding to a strong coupling of the atoms to the cavity mode, $g(r_{1,2})\approx g_{0}$, the initial entanglement undergoes the discontinuity, the sudden death behavior. The entanglement disappears quite rapidly over a very short time, $\omega t\approx 1$. For large distances, where the coupling constants $g(r_{j})$ are very small, the discontinuity disappears and the entanglement decays asymptotically in time, the behavior predicted before under the~RWA. 
Clearly, the counter-rotating terms which become important in the strong coupling regime, are responsible for the entanglement sudden death.

\begin{figure}[hbp]
\includegraphics[width=8cm,keepaspectratio,clip]{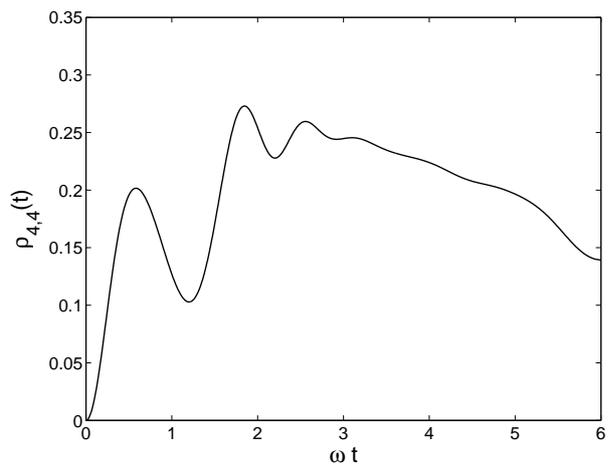}
\caption{Time evolution of the population $\rho_{44}(t)$ calculated for both atoms located at the antinode of the standing wave $(d=0)$, with $g_{0}=\omega$, $(\omega -\omega_{0}) =0.01\omega$, $\kappa =0.1\omega$ and $n= 1/2$. The atoms were prepared initially in the single
excitation collective symmetric state~$\ket{\Psi_{s}}$.}
\label{figc28}
\end{figure}

The discontinuity in the time evolution of entanglement, the sudden death effect seen in Fig.~\ref{figc27}, seems puzzling at first, because it appears to contradict the predictions based upon a simple argument that with a single excitation present in the system, it is impossible to achieve the discontinuity as the population $\rho_{44}(t)=0$ for all times. To resolve this problem, we plot in Fig.~\ref{figc28} the time evolution of the population of the two-photon state~$\ket{\Psi_{4}}$ for the same parameters as in Fig.~\ref{figc27} with $d=0$. It is seen that initially unpopulated two-photon state $\ket{\Psi_{4}}$ becomes populated in a comparatively short time. Thus, it is clear now that the entanglement sudden death is associated with the non-zero value of the population $\rho_{44}(t)$ which is developed during the evolution governed by the non-RWA Hamiltonian. 

To explain more clearly the origin of the population of the two-photon state, we offer two complementary views of the underlying physics. The first is obtained by the derivation of the reduced density operator for the atoms using state vectors of the combined atoms plus the cavity field system. The second is provided by the principle of complementarity between the evolution time~$\Delta t$ and uncertainty $\Delta E$ in the energy of the system.

Assume that there is only a single excitation in the system that evolves under the non-RWA Hamiltonian, that includes the counter-rotating terms $aS^-_j$ and $S^\dagger_j a^\dagger$. In this case, we must include the processes that do not strictly conserve excitation number, when one of the atoms goes to the excited (ground) state by emitting (absorbing) a photon. If these processes are included, then the Hilbert space of the system, the atoms plus the cavity field, should be spanned by six-state vector~$\ket{\Psi_{2}}\otimes{\ket 0},  \ket{\Psi_{3}}\otimes{\ket 0},  \ket{\Psi_{1}}\otimes{\ket 1},  \ket{\Psi_{2}}\otimes{\ket 2},  \ket{\Psi_{3}}\otimes{\ket 2}$ and $\ket{\Psi_{4}}\otimes{\ket 1}$. If we evaluate the reduced density operator $\rho(t)$ for the atoms by tracing over the cavity mode, we~find
\begin{eqnarray}
\rho (t) &=& {\rm Tr}_{cavity}[\rho_{s}(t)] =\sum_{i=0}^{2}\langle i|\rho_{s}(t)|i\rangle \nonumber\\
&=&  \rho_{11}(t)\ket{\Psi_{1}}\bra{\Psi_{1}}
+ \rho_{22}(t)\ket{\Psi_{2}}\bra{\Psi_{2}} \nonumber \\
&+& \rho_{33}(t)\ket{\Psi_{3}}\bra{\Psi_{3}}
+ \rho_{44}(t)\ket{\Psi_{4}}\bra{\Psi_{4}} \nonumber \\
&+&\rho_{32}(t)\ket{\Psi_{3}}\bra{\Psi_{2}}
+ \rho_{23}(t)\ket{\Psi_{2}}\bra{\Psi_{3}} \nonumber \\
&+& \rho_{14}(t)\ket{\Psi_{1}}\bra{\Psi_{4}}
+ \rho_{41}(t)\ket{\Psi_{4}}\bra{\Psi_{1}} ,\label{11.5}
\end{eqnarray}
where $|i\rangle$ refers to the cavity mode and $\rho_{ij}(t)$ are populations $(i=j)$ of the atomic states and coherences $(i\neq j)$ between them. It is evident from Eq.~(\ref{11.5}) that the evolution of the atoms under the non-RWA Hamiltonian involves the two-photon state, that population of the state~$\ket{\Psi_{4}}$ becomes possible.

An alternative explanation of the physical origin of the population of the two-photon state is provided by  the principle of complementarity between the evolution time $\Delta t$ and uncertainty $\Delta E$ in the energy
\begin{eqnarray}
\Delta t\Delta E \geq \hbar .\label{11.6}
\end{eqnarray}
Clearly, for the evolution time of the order $1/\omega$, that has been considered here, the energy of the initial excitation is spread over on energy interval of order $\hbar \omega$, the order required to achieve a non-zero population of the upper state $\ket{\Psi_{4}}$ of the two-atom system.

In summary of this section, we have demonstrated a somewhat surprising result that an entanglement initially encoded into a single excitation state can undergo the phenomenon of sudden death. This non-intuitive behavior of the entanglement occurs in the very short time of the evolution and arises solely from the counter-rotating terms in the interaction Hamiltonian of the system.  When the dynamics of the system are governed by the RWA Hamiltonian, the entanglement decays asymptotically without any discontinuity. A simple explanation of this unexpected feature is provided in terms of the quantum property of complementarity, which is manifested as a tradeoff between the knowledge of energy versus the evolution time of the system.

\section{Entangling atomic ensembles}

\noindent The phenomena described in the previous sections arise from entangled and disentangled dynamics of simple systems composed of only two atoms. Particular attention has been paid to the anomalous "strange" behavior of entanglement. Exact analytical results have been obtained which allowed for the complete understanding of the underlying physics. 

It has become apparent that a large scale quantum information processing and quantum computation are, in general, much more complex and require to create special entangled states involving a large number of strongly coupled atoms. Naturally, when more than two atoms are involved, the situation is much more complex, and requires completely different procedures than those already considered. In this connection, we now turn to multi-atom systems and address in this and the following sections the question of a simple procedure for creation of entangled states in multi-atom systems, such as macroscopic ensembles of cold atoms. The reason for the interest in atomic ensembles is twofold. On the one side, atomic ensembles are macroscopic systems that are easily created in the laboratory. On the other hand, the collective behavior of the atoms enables to achieve almost a perfect coupling of the ensembles to external fields without the need to achieve a strong field-single atom coupling.

Our main effort will be devoted to the study of the creation of entanglement between $M$ atomic ensembles. We shall make use a simple but powerful procedure, initiated by Gao-xiang Li~\cite{li06}, which constructs multi-mode entangled states from the vacuum by internal dynamical processes rather than being injected into the ensembles from external sources. The procedure is associated with the collective dynamics of the atomic ensembles subjected to driving lasers of a suitably adjusted amplitudes and phases. It employs the Holstein-Primakoff representation of angular momentum operators, which expresses atomic systems in terms of orthogonal collective bosonic modes, and then prepares separately each of the modes in a desired entangled state. 

\begin{figure}[hbp]
\includegraphics[width=5cm,keepaspectratio,clip]{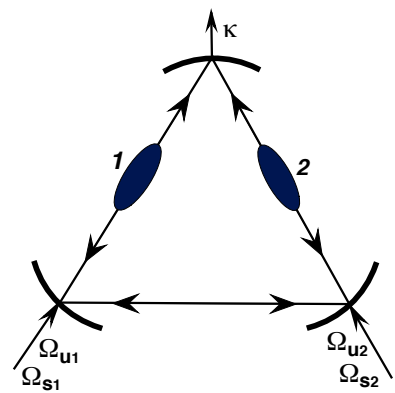}
\caption{Configuration of a ring cavity with two atomic ensembles, $1$ and $2$, for the preparation of multi-mode entangled states. The cavity modes are damped with the same rate $\kappa$. The driving laser fields are injected through the cavity mirrors and co-propagate with one of the cavity modes.}
\label{figc29}
\end{figure}

The physical system we first consider for creation of multi-mode entangled states consists of $M=2$ one-dimensional atomic ensembles trapped along the axis of a high-$Q$ ring cavity, as illustrated in Fig.~\ref{figc29}. The cavity is composed of three mirrors that create two mutually counter-propagating modes, called {\it clockwise} and {\it anti-clockwise} modes, to which the atoms are equally coupled. The cavity modes are degenerate in frequency, i.e.  $\omega_{+}=\omega_{-}=\omega_{c}$. The cavity is damped with the rate~$\kappa$ that is assumed small to achieve a high finesse at a relatively large size of the cavity. 

\begin{figure}[hbp]
\includegraphics[width=6cm,keepaspectratio,clip]{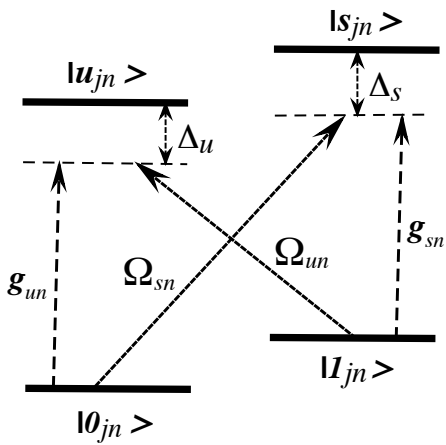}
\caption{Atomic level scheme. Two laser fields of the Rabi frequencies $\Omega_{u}$ and $\Omega_{s}$ drive the atomic transitions $|1_{jn}\rangle\rightarrow |u_{jn}\rangle$ and $|0_{jn}\rangle\rightarrow |s_{jn}\rangle$, respectively. The atomic transitions $|1_{jn}\rangle\rightarrow |s_{jn}\rangle$ and $|0_{jn}\rangle\rightarrow |u_{jn}\rangle$ are coupled to the single-mode cavity with the coupling strengths $g_{un}$ and $g_{sn}$, respectively.}
\label{figc30}
\end{figure}

The atomic ensembles contain a large number of identical four-level atoms, each composed of two stable ground states, $|0_{jn}\rangle$, $|1_{jn}\rangle$, and two excited states $|u_{jn}\rangle$, $|s_{jn}\rangle$, where the subscript $jn$ labels $j$th atom of the $n$th ensemble. The ground state $|0_{jn}\rangle$ of energy $E_{0}=0$  is coupled to the excited state $|s_{jn}\rangle$ by a laser field of the Rabi frequency $\Omega_{sn}$ and frequency $\omega_{Ls}$ that is detuned from the atomic transition frequency by $\Delta_{s} = (\omega_{s}-\omega_{Ls})$, where $\omega_{s}=E_{s}/\hbar$ and $E_{s}$ is the energy of the state $|s_{jn}\rangle$. Similarly, the ground state $|1_{jn}\rangle$ of energy $E_{1}=\hbar\omega_{1}$ is coupled to the excited state $|u_{jn}\rangle$ of energy $E_{u}=\hbar\omega_{u}$ by an another laser field with the Rabi frequency $\Omega_{u}$, and the angular frequency $\omega_{Lu}$ that is detuned from the atomic transition $|1_{jn}\rangle\rightarrow|u_{jn}\rangle$ by
$\Delta_{u}= (\omega_{u}-\omega_{1}-\omega_{Lu})$. The frequencies $\omega_{Lu}$ and $\omega_{Ls}$ of the laser fields are matched close to the cavity frequency, so that the wave numbers of the laser fields $k_{u}$ and $k_{s}$ are approximated by $k_u\approx k_s\approx k$. A schematic diagram of the atomic levels together with the configuration of the driving lasers and cavity couplings is shown in Fig.~\ref{figc30}. 

To secure some simplification in notation, we shall assume that both ensembles contain the same number~$N$ of the atoms. 
The atoms interact with the cavity modes of degenerate frequencies $\omega_{c}$ that simultaneously couple to the $|u_{jn}\rangle\leftrightarrow|0_{jn}\rangle$ and $|s_{jn}\rangle\leftrightarrow|1_{jn}\rangle$ transitions, with the coupling strengths  $g_{un}$ and $g_{sn}$, respectively. We assume that the atom-field coupling strengths are uniform through the atomic ensembles. This is consistent with current experiments involving ring cavities and and large samples of trapped atoms~\cite{kl06}. Practical sizes of the atomic samples are $\sim10^{3}$\ nm that is smaller than the cavity mode radius $w_{0}=130\mu$m, and also are much smaller than the practical length of a single arm of the~cavity.

In what follows, we will assume that the laser fields and the cavity frequencies are far detuned from the atomic transition frequencies. This will allow us to adiabatically eliminate the upper states of the atoms and obtain an effective dispersive type interaction of the driving lasers and the cavity modes with the atomic transitions.

\subsection{Effective Hamiltonian of the system}

\noindent The Hamiltonian of the system, in the rotating-wave approximation, and transformed into the interaction picture has the following form
\begin{eqnarray}
H=H_{0}+H_{AL}+H_{AC} ,\label{12.1}
\end{eqnarray}
where
\begin{eqnarray}
&& H_{0} =  \hbar\omega_c (a_+^\dag a_{+} +a_{-}^{\dag} a_{-}) 
+ \hbar\sum\limits_{n=1}^{2}\sum\limits_{j=1}^N  \left\{\omega_u|u_{jn}\rangle\langle
u_{jn}|\right. \nonumber\\
&&+\left. \omega_s|s_{jn}\rangle\langle s_{jn}|+\omega_1|1_{jn}\rangle\langle
1_{jn}|\right\} \label{12.2}
\end{eqnarray}
is the free Hamiltonian of the atomic ensembles and the cavity modes,
\begin{eqnarray}
&& H_{AL} =  \frac{1}{2}\hbar\sum\limits_{n=1}^{2}\sum\limits_{j=1}^N \left\{\Omega_{un}(x_{jn}) {\rm e}^{-i(\omega_{Lu}+\phi_{un})t}|u_{jn}\rangle\langle 1_{jn}|\right. \nonumber\\
&& +\left. \Omega_{sn}(x_{jn}) {\rm e}^{-i(\omega_{Ls}+\phi_{sn})t}
|s_{jn}\rangle\langle 0_{jn}|\!+\!{\rm H.c.}\right\} \label{12.3}
\end{eqnarray}
is the interaction Hamiltonian between the atoms and the driving fields, and
\begin{eqnarray}
&& H_{AC} = \hbar\sum\limits_{n=1}^{2}\sum\limits_{j=1}^N\left\{
g_{un}^{+}(x_{jn})|u_{jn}\rangle\langle0_{jn}|a_{+}\right. \nonumber\\
&&\left. + g_{un}^{-}(x_{jn})|u_{jn}\rangle\langle0_{jn}|a_{-}
+g_{sn}^{+}(x_{jn})|s_{jn}\rangle\langle1_{jn}|a_{+}\right. \nonumber \\
&&\left. + g_{sn}^{-}(x_{jn})|s_{jn}\rangle\langle1_{jn}|a_{-} +{\rm H.c.}\right\}  \label{12.4}
\end{eqnarray}
is the interaction Hamiltonian between the atoms and the two cavity modes. 
Here, $a_{\pm}$ and $a^{\dagger}_{\pm}$ are the annihilation and creation operators associated with the two counter-propagating 
modes of the cavity; clockwise $(+)$ and anti-clockwise $(-)$ propagating modes, and $k$ is the wave number of the cavity modes. 
The Hamiltonian retains only the energy conseving terms, which play the dominant role in the interaction. We have denoted the energies of the atomic levels by $\hbar\omega_{i} \, (i=1,u,s)$ and have set the energy of the ground state $|0_{jn}\rangle$ equal to zero. The parameters $\Omega_{un}(x_{jn})$ and $\Omega_{sn}(x_{jn})$ are spatially varying Rabi frequencies of the driving laser fields, and $\phi_{un},\phi_{sn}$ are their phases. The coupling constants of the atomic transitions to the cavity fields, $g_{un}^{\pm}(x_{jn})$ and $g_{sn}^{\pm}(x_{jn})$, are also dependent on the atomic position. In what follows, we will use the plane traveling wave representation for the laser fields and the cavity modes, in which
\begin{eqnarray}
&&\Omega_{un}(x_{jn}) = \Omega_{un}{\rm e}^{ik_{u}x_{jn}} ,\ \Omega_{sn}(x_{jn}) 
= \Omega_{sn}{\rm e}^{ik_{s}x_{jn}} ,\nonumber \\
&&g_{un}^{\pm}(x_{jn}) = g_{un}{\rm e}^{\pm ikx_{jn}} ,\ g_{sn}^{\pm}(x_{jn}) 
= g_{sn}{\rm e}^{\pm ikx_{jn}} ,\nonumber\\
\label{12.5}
\end{eqnarray}
where we have assumed that the coupling constants $g_{un}$ and~$g_{sn}$ are the same for the two cavity modes. This is acceptable if the modes have the same polarization and geometry, which is feasible with the practical situations~\cite{krb03,na03,kl06}.

We now make an unitary transformation and few standard approximations on the Hamiltonian (\ref{12.1}) to eliminate the explicit time dependence and the excited states to neglect atomic spontaneous emission and to obtain an effective two-level Raman-coupled Hamiltonian. Then, we will make the transformation of the atomic operators into the field (bosonic) representation. These operations will be done in the following steps. In the first step, we make the unitary transformation $U =\exp(iH_{0}^{\prime}t/\hbar)$, with
\begin{eqnarray}
H_{0}^{\prime} &=& \hbar(\omega_{Ls}-\omega_1)(a_+^\dag a_++a_-^\dag a_-) \nonumber\\
&+& \hbar\sum\limits_{n=1}^{2}\sum\limits_{j=1}^N\left\{ (\omega_u+\omega_1)|u_{jn}\rangle\langle u_{jn}|\right. \nonumber \\
&+&\left. \omega_{Ls}|s_{jn}\rangle\langle s_{jn}| + \omega_1|1_{jn}\rangle\langle 1_{jn}|\right\} ,\label{12.6}
\end{eqnarray}
and assume that the laser frequencies satisfy the resonance condition 
$\omega_{Ls}-\omega_{Lu}=2\omega_1$. This specifically chosen transformation allows us to eliminate the explicit time dependence of the Hamiltonian.

In the second step, we introduce detunings of the laser fields from the atomic transition frequencies
\begin{eqnarray}
\Delta_u=\omega_u-(\omega_{Lu}+\omega_1),\quad \Delta_s=\omega_s-\omega_{Ls} .\label{12.7}
\end{eqnarray}
In order to eliminate spontaneous scattering of photons to modes other than the privileged cavity mode, we assume assume that the detunings are much larger than the Rabi frequencies, cavity coupling constants and the atomic spontaneous emission rates, i.e.
\begin{eqnarray}
\Delta_u, \Delta_s\gg  g_{sn},g_{un}, \Omega_{sn}, \Omega_{un}, \gamma_{s},\gamma_{u} , \label{12.8}
\end{eqnarray}
where $\gamma_{s}$ and $\gamma_{u}$ are the total spontaneous emission rates from the states $|s_{jn}\rangle$ and $|u_{jn}\rangle$, respectively.

The assumption about large detunings allows us to perform the standard adiabatic elimination of the atomic excited states, $|s_{jn}\rangle$, $|u_{jn}\rangle$, and obtain an effective two-level Hamiltonian involving only the ground states of the atoms
\begin{eqnarray}
H_{e} &=& \left[\delta_{c}\!+\!\frac{N}{2}\sum\limits_{n=1}^{2}\!\left( 
\frac{g_{un}^{2}}{\Delta_{un}}\!+\!\frac{g_{sn}^{2}}{\Delta_{sn}}\right)\!\right]\!\left(a_+^\dag a_+\!+\!a_-^\dag a_{-}\right)\nonumber\\
&+&\sum\limits_{n=1}^{2}\left( \frac{g_{un}^{2}}{\Delta_{un}} -\frac{g_{sn}^{2}}{\Delta_{sn}}\right)
J^{(n)}_{z}\left(a_+^\dag a_+ +a_-^\dag a_{-}\right) \nonumber\\
&+& \frac{1}{\sqrt{N}}\sum\limits_{n=1}^{2}\left\{\, 
\beta_{un}{\rm e}^{-i\phi_{un}}\left(J_{0k}^{(n)}a_+^\dag +J_{2k}^{(n)}a_-^\dag \right)\right. \nonumber\\
&+&\left. \beta_{sn}{\rm e}^{-i\phi_{sn}}\left(J^{(n)\dagger}_{0k} a_+^\dag\!+\!J^{(n)\dagger}_{-2k}a_-^\dag\right)\!+\!{\rm H.c.}\right\} ,\label{12.9}
\end{eqnarray}
where $\delta_{c} =\omega_{c}-(\omega_{Ls}-\omega_{1})$ is the detuning of the cavity frequency from the Raman coupling resonance,
\begin{eqnarray}
&&J^{(n)}_{z} = \frac{1}{2}\sum\limits_{j=1}^N\left(|1_{jn}\rangle\langle 1_{jn}|-|0_{jn}\rangle\langle 0_{jn}|\right) = \sum\limits_{j=1}^N\sigma_{zj}^{(n)} ,\nonumber\\ 
&&J_{mk}^{(n)} = \sum\limits_{j=1}^{N}|0_{jn}\rangle\langle 1_{jn}|{\rm e}^{imkx_{jn}} 
=\sum\limits_{j=1}^{N}\sigma^{(n)}_{j}{\rm e}^{imkx_{jn}} \nonumber\\
\label{12.10}
\end{eqnarray}
are the spatially dependent macroscopic inversion and polarization operators, and
\begin{eqnarray}
\beta_{un}=\frac{\sqrt{N}\Omega_{un}g_{un}}{2\Delta_{un}} ,\quad 
\beta_{sn}=\frac{\sqrt{N}\Omega_{sn}g_{sn}}{2\Delta_{sn}} \label{12.11}
\end{eqnarray}
are the coupling strengths of the effective two-level system to the cavity modes. 

The first line of Eq.~(\ref{12.9}) represents the free energy of the atomic ensembles. The second line represents an intensity dependent (Stark) shift of the atomic energy levels, and the remaining two lines represent the interaction between the cavity field and the atomic ensembles. The three collective atomic operators, $J^{(n)}_{0k}, J^{(n)}_{2k}$ and $J^{(n)}_{-2k}$, which appear in Eq.~(\ref{12.9}), arise naturally for the position dependent atomic transition operators and appear in a cavity with two mutually counter-propagating modes. In the case of a single-mode cavity, the Hamiltonian involves only the $J^{(n)}_{0k}$ operator~\cite{ps06,de07,cb09}. The operators $\sigma^{(n)}_{j}, \sigma^{(n)\dagger}_{j}$ and $\sigma_{zj}^{(n)}$ are the standard Pauli spin$-1/2$ operators, which satisfy the well known commutation relations $[\sigma_{j}^{(n)},\sigma_{\ell}^{(m)\dag}]=2\sigma_{zj}^{(n)}\delta_{j\ell}\delta_{mn}$. However, the collective operators $J^{(n)}_{mk}, J^{(n)\dagger}_{mk}$ and $J^{(n)}_{z}$ do not in general satisfy the angular momentum commutation relations. The reason is in the presence of the phase factors $\exp(imkx_{jn})$, so that the commutation relations are satisfied only in the small sample limit of $kx_{jn}\ll 1$, at which $\exp(imkx_{jn})\approx 1$.

The essential feature of the effective Hamiltonian (\ref{12.9}) is that the atoms interact dispersively with the cavity mode. This means that the cavity mode will remain unpopulated during the evolution. Note the presence of nonlinear terms, $a^{\dagger}J_{j}^{\dagger},\ aJ_{j}^{-}$, which are analogous to the counter-rotating terms. These terms will play the crucial role in creation of a squeezed state between the atomic ensembles.

The Hamiltonian (\ref{12.9}) is general in the parameter values. To avoid unessential complexity due to the presence of the free energy and the Stark shift terms, we will work with a simplified version of the Hamiltonian by choosing the frequencies of the driving lasers and the cavity field such that
\begin{eqnarray}
\frac{g_{un}^{2}}{\Delta_{un}} = \frac{g_{sn}^{2}}{\Delta_{sn}} , \qquad 
\delta_{c} +\frac{Ng_{un}^{2}}{\Delta_{un}}= 0 .\label{12.12}
\end{eqnarray}
With this simplification, we find that the Hamiltonian~(\ref{12.9}) reduces to
\begin{eqnarray}
H_{e} &=& \frac{1}{\sqrt{N}}\sum\limits_{n=1}^{2}\left\{\, 
\beta_{un}{\rm e}^{-i\phi_{un}}\left(J_{0k}^{(n)}a_+^\dag +J_{2k}^{(n)}a_-^\dag \right)\right. \nonumber\\
&+&\left. \beta_{sn}{\rm e}^{-i\phi_{sn}}\!\left(\!J^{(n)\dagger}_{0k} a_+^\dag\!+\!J^{(n)\dagger}_{-2k}a_-^\dag\!\right)\!+\!{\rm H.c.}\right\} .\label{12.13}
\end{eqnarray}
This equation is in the form of a non-RWA Hamiltonian of two extended atomic ensembles independently coupled to two counter-propagating cavity modes. The parameters of the Hamiltonian are a function of the detunings and Rabi frequencies of the two highly detuned laser fields co-propagating with the clockwise cavity mode and thus could be controlled through the laser frequencies and intensities. In the small sample case of $kx_{jn}\ll 1$ and under the single-mode approximation, the Hamiltonian simplifies to the standard Dicke model~\cite{dic,ft02}.

In the final step we shall reformulate the Hamiltonian (\ref{12.13}) in terms of bosonic variables by adopting the Holstein-Primakoff representation of angular momentum operators~\cite{hp40}. In this representation, the collective atomic operators, $J^{(n)\dagger}_{mk}, J^{(n)}_{mk}$ and $J^{(n)}_{z}$ are expressed in terms of annihilation and creation operators $C^{(n)}_{mk}$ and $C^{(n)\dagger}_{mk}$ of a single bosonic mode. Provided the atoms in each ensemble are initially prepared in their ground states $\{|0_{jn}\rangle\}$, and taking into account that due to large detunings of the driving fields, the excitation probability of each atom is low during the laser-atom-cavity coupling, i.e., $\langle \sigma_{zj}^{(n)}\rangle \approx -1/2$, the collective atomic operators can be well approximated by
\begin{eqnarray}
J^{(n)}_{mk} = \sqrt{N} C^{(n)}_{mk} ,\quad J^{(n)}_{z} =  -\frac{N}{2} ,\label{12.14}
\end{eqnarray}
where
\begin{eqnarray}
C^{(n)}_{mk} = \frac{1}{\sqrt{N}}\sum\limits_{j=1}^N b^{(n)}_{j}{\rm e}^{ imkx_{jn}} ,\quad m=0,\pm 2 ,\label{12.15}
\end{eqnarray}
are collective bosonic operators with the operators $b^{(n)}_{j}$ and~$b^{(n)\dagger}_{j}$ obeying the standard bosonic commutation relation $[b^{(n)}_{j}, b^{(m)\dagger}_{\ell}]=\delta_{j\ell}\delta_{nm}$.

Note that the collective bosonic operators do not in general commute, i.e.
\begin{eqnarray}
\left[C^{(n)}_{mk},C^{(n^\prime)\dag}_{m^\prime k}\right] = \frac{1}{N}\sum\limits_{j=1}^N
\exp[i(m-m^\prime)kx_j] \delta_{nn^{\prime}} .\label{12.16}
\end{eqnarray}
Since, in the limit of $N\gg 1$:
\begin{eqnarray}
\sum_{j=1}^{N}{\rm e}^{i(m-m^\prime)kx_{j}} = N \delta_{m-m^{\prime},0} ,\label{12.17}
\end{eqnarray}
we obtain
\begin{eqnarray}
\left[C^{(n)}_{mk},C^{(n^\prime)\dag}_{m^\prime k}\right] \approx \delta_{m,m^\prime}\delta_{nn^{\prime}} ,\label{12.18}
\end{eqnarray}
which shows that in the limit of small separations between the atoms, the collective bosonic operators are orthogonal to each other.

In terms of the collective bosonic operators $C^{(n)}_{0k}$ and~$C^{(n)}_{\pm 2k}$, the effective Hamiltonian (\ref{12.13}) can be written~as
\begin{eqnarray}
H_{e} &=& \sum\limits_{n=1}^{2}\left\{ \left(\, \beta_{un} {\rm e}^{-i\phi_{un}}C^{(n)}_{0k}+\beta_{sn}{\rm e}^{-i\phi_{sn}}C_{0k}^{(n)\dag}\right)a_+^\dag\right. 
\nonumber\\
&+&\left. \left( \beta_{un}{\rm e}^{-i\phi_{un}}C^{(n)}_{2k}\!+\!\beta_{sn}{\rm e}^{-i\phi_{sn}}C^{(n)\dag}_{-2k}\right)\!a_-^\dag\!+\!{\rm H.c.}\right\} .\nonumber\\
\label{12.19}
\end{eqnarray}
The important property of the bosonic representation is the fact that effective Hamiltonian of ensembles of cold atoms trapped inside a ring cavity can be expressed now as the interaction between the cavity modes and three orthogonal field modes; a collective mode $C^{(n)}_{0k}$ solely coupled to the cavity mode~$a_+$, which co-propagates with the driving lasers, and two modes $C^{(n)}_{\pm 2k}$ that are solely coupled to the cavity counter-propagating mode $a_{-}$.

The Hamiltonian (\ref{12.19}) holds for the laser fields co-propagating with the clockwise mode only. Following the same procedure as above, we can easily show that in the case of the driving fields co-propagating with the anti-clockwise mode $a_-$,  the effective Hamiltonian takes the form
\begin{eqnarray}
H_{e} &=& \sum\limits_{n=1}^{2} \left\{ \left(\, \beta_{un}{\rm e}^{-i\phi_{un}}C^{(n)}_{0k}+\beta_{sn}{\rm e}^{-i\phi_{sn}}C_{0k}^{(n)\dag}\right)a_-^\dag\right. \nonumber\\
&+&\left. \left(\beta_{un}{\rm e}^{-i\phi_{un}}C^{(n)}_{-2k}\!+\!\beta_{sn}{\rm e}^{-i\phi_{sn}}C_{2k}^{(n)\dag}\right)\!a_+^\dag\!+\!{\rm H.c.}\right\} .\nonumber\\
\label{12.20}
\end{eqnarray}
We see the complete symmetry between the two cases that reversing the direction of the propagation of the laser fields from clockwise to anti-clockwise is equivalent to the exchange of $a_{+}\leftrightarrow a_{-}$ and $k\rightarrow -k$ in the Hamiltonian (\ref{12.19}).

Our objective is to prepare the atomic ensembles in a desired entangled state. To achieve it, we consider the evolution of the system under the effective Hamiltonian~(\ref{12.19}) including also a possible loss of photons due to the damping of the cavity mode. This is the only damping which we will consider as we have already eliminated spontaneous emission by choosing large detunings of the driving lasers.

With the cavity damping included, the properties of the system are determined by the density operator $\rho$ whose the time evolution is governed by the master equation
\begin{eqnarray}
\dot{\rho} = -i[H_{e} ,\rho]+{\cal L}_c\rho ,\label{12.21}
\end{eqnarray}
where
\begin{equation}
{\cal L}_{c}\rho = \frac{1}{2}\kappa \sum\limits_{i=\pm}\left(2a_{i}\rho
a^\dag_{i}-a^\dag_{i} a_{i}\rho-\rho a^\dag_{i} a_{i}\right) ,\label{12.22}
\end{equation}
is the Liuivilian operator representing the damping of the cavity field modes with the rate~$\kappa$.

In what follows, demonstrate how to generate on demand multi-mode entangled states in ensembles of cold atoms located inside a two-mode ring cavity.

\subsection{One and two-mode entangled states}

\noindent First we illustrate a procedure which constructs entangled states in a single $(M=1)$ ensemble of cold atoms located inside a two-mode ring cavity. We focus on the creation of single and two-mode entangled states and attempt to characterize the entanglement in terms of unitary operators called single and two-mode squeezed operators. The squeezed operators are defined as~\cite{cs85}
\begin{eqnarray}
S_0(\xi_0) &=& \exp\left[-\frac{1}{2}\left(\xi_0C_{0k}^{\dag 2}-\xi_{0}^{\ast}C_{0k}^2\right)\right] ,\nonumber\\
S_{\pm k}(\xi_1) &=& \exp\!\left(\!\xi_{1}^{\ast} C_{2k}C_{-2k}\!-\!\xi_{1}C_{2k}^\dag C_{-2k}^\dag\!\right) ,\label{12.23}
\end{eqnarray}
where $\xi_{0}$ and $\xi_{1}$ are complex one and two-mode squeezing parameters, respectively. 

We adopt the procedure of Li~\cite{li06},  which constructs squeezed states from the vacuum by a unitary transformation associated with the realistic dynamical process determined by the master equation (\ref{12.19}). In the procedure the atomic ensembles, initially in the ground state, are subjected to different sequences of laser pulses of specifically chosen Rabi frequencies $\Omega_{un}$ and $\Omega_{sn}$ and phases~$\phi_{un}$ and~$\phi_{sn}$. The laser pulses prepare the ensembles in a desired state that then decays to a steady-state with the rate $\kappa$. We will assume in all our considerations that the initial ground state of the atomic ensembles corresponds to all the atoms being in their ground states~$|0_{jn}\rangle$.

The squeezed operators can be easily associated with evolution operators for the effective 
Hamiltonians~(\ref{12.19}) and (\ref{12.20}), which are already expressed in terms of the collective bosonic operators $C_{0k}^{(n)}$ and $C_{\pm 2k}^{(n)}$. The squeezing parameters are then given in terms of the coupling strengths $\beta_{un}$ and $\beta_{sn}$ and the phases~$\phi_{un}, \phi_{sn}$, so they can be adjusted and controlled by the driving laser fields.

The construction procedure is done in two steps. In the first step, we adjust the driving lasers to propagate in the clockwise direction, along the cavity mode $a_{+}$. In this case, the dynamics of the system are determined by the Hamiltonian~(\ref{12.19}). We then send series of laser pulses of phases $\phi_{u1}=\phi_{s1}=0$ and arbitrary Rabi frequencies~$\Omega_{u1}$ and $\Omega_{s1}$, but such that $\beta_{u1} >\beta_{s1}$. With this choice of the parameters of the driving lasers and under the unitary squeezing transformation
\begin{eqnarray}
S_0(-\xi_0)S_{\pm k}(-\xi_{1})\rho S_0(\xi_0)S_{\pm k}(\xi_{1}) = \tilde{\rho} ,\label{12.24}
\end{eqnarray}
with
\begin{eqnarray}
\xi_0=\xi_{1}=\frac{1}{2}\ln\left(\frac{\beta_{u1}+\beta_{s1}}{\beta_{u1}-\beta_{s1}}\right) ,\label{12.25}
\end{eqnarray}
the master equation (\ref{12.21}) becomes
\begin{equation}
\frac{d}{dt}\tilde{\rho}=-i[\tilde{H_e},\tilde{\rho}]+ {\cal L}_{c}\tilde{\rho} ,\label{12.26}
\end{equation}
where
\begin{eqnarray}
\tilde{H_e} &=& S_0(-\xi_0)S_{\pm k}(-\xi_{1})H_eS_0(\xi_0)S_{\pm k}(\xi_{1})\nonumber\\
&=&\sqrt{\beta_{u1}^2\!-\!\beta_{s1}^2}\!\left(a_+^\dag C_{0k}\!+\!a_-^\dag C_{2k}\!+\!{\rm H.c.}\right) .\label{12.27}
\end{eqnarray}
It is seen that under the squeezing transformation, the Hamiltonian represents a simple system of two independent linear mixers, where the collective bosonic modes~$C_{0k}$ and $C_{2k}$ linearly couple to the cavity modes~$a_{+}$ and~$a_{-}$, respectively. The mode $C_{-2k}$ is decoupled from the cavity modes and therefore does not evolve. In other words, the state of the mode $C_{-2k}$ cannot be determined by the evolution operator for the Hamiltonian~(\ref{12.27}). The important property of the transformed system is that the master equation~(\ref{12.26}) is fully soluble, i.e. all eigenvectors and eigenvalues can be obtained exactly. Hence, we can monitor the evolution of the bosonic modes towards their steady-state values. Since we are interested in the steady-state of the system, we confine our attention only to the eigenvalues of Eq.~(\ref{12.26}), which are of the form
\begin{eqnarray}
\eta_\pm = -\frac{\kappa}{2}\pm\left[\left(\frac{\kappa}{2}\right)^2-\sqrt{\beta_{u1}^{2}-\beta_{s1}^{2}}\right]^\frac{1}{2} .\label{12.28}
\end{eqnarray}
Evidently, both eigenvalues have negative real parts which means that the system subjected to a series of laser pulses up to a short time $t$ will then evolve (decay) to a stationary state that is a vacuum state. Thus, as a result of the interaction given by the Hamiltonian (\ref{12.27}), and after a sufficiently long evolution time, the modes $a_{\pm}$, $C_{0k}$ and $C_{2k}$ will be found in the vacuum state, whereas the mode $C_{-2k}$ will remain in an undetermined state. The state of the mode $C_{-2k}$ will be determined in the next, second step of the preparation process.

In order to estimate the time scale for the system to reach the steady-state, we see from Eq.~(\ref{12.28}) that as long as $\sqrt{\beta_{u1}^{2}-\beta_{s1}^{2}}>\kappa/2$, the time scale for the system  to reach the steady state is of order of $\sim 2/\kappa$. Thus, as a result of the cavity damping the system, after a sufficient long time, will definitely be found in the stationary state.

In summary of the first step of the preparation, we find that in the steady-state, the density matrix representing the state of the transformed system is in the factorized form
\begin{eqnarray}
\tilde{\rho}(\tau\sim 2/\kappa)= \tilde{\rho}_{v}\otimes\tilde{\rho}_{C_{-2k}} ,\label{12.29}
\end{eqnarray}
where
\begin{eqnarray}
\tilde{\rho}_{v} =|0_{a_+},0_{a_-},0_{C_{0k}},0_{C_{2k}}\rangle
\langle 0_{a_+},0_{a_-},0_{C_{0k}},0_{C_{2k}}|  \label{12.30}
\end{eqnarray}
is the density matrix of the four modes prepared in their vacuum states, and $\tilde{\rho}_{C_{-2k}}$ is the density matrix of the mode $C_{-2k}$ whose the state has not been determined in the first step of the procedure. The ket $|0_{a_+},0_{a_-},0_{C_{0k}},0_{C_{2k}}\rangle$ represents the state with zero photons in each of the modes.

Thus, we are left with the problem of the preparation of the remaining collective mode $C_{-2k}$ in a desired squeezed vacuum state. This is done in what we call the second step of the preparation, in which we first adjust the driving lasers to propagate along the anti-clockwise mode $a_{-}$. We then send series of pulses of frequencies, phases and amplitudes the same as in the above first stage. As a result of the coupling to the cavity mode~$a_{-}$,  the interaction is now governed by the Hamiltonian (\ref{12.22}), and therefore after the unitary squeezing transformation the Hamiltonian of the system takes the form
\begin{eqnarray}
\tilde{H_e} &=& S_0(-\xi_0)S_{\pm k}(-\xi_{1})H_eS_0(\xi_0)S_{\pm k}(\xi_{1})\nonumber\\
&=&\sqrt{\beta_{u1}^2\!-\!\beta_{s1}^2}\!\left(\!a_-^\dag C_{0k}\!+\!a_+^\dag C_{-2k}\!+\!{\rm H.c.}\!\right) .\label{12.31}
\end{eqnarray}
As above in the case of the coupling to the cavity mode~$a_{+}$, the Hamiltonian (\ref{12.31}) describes a system of two independent linear mixers. Hence, the state of the system will evolve during the interaction   towards its stationary value, and after s suitably long time, $\sim 2/\kappa$, the transformed system will be found in the vacuum state.

Thus, after the second step of the preparation, the transformed system is found in the pure vacuum state determined by the density matrix of the form
\begin{equation}
\tilde{\rho}(\tau\sim 4/\kappa)=|\tilde{\Psi}\rangle\langle\tilde{\Psi} | ,\label{12.32}
\end{equation}
where
\begin{eqnarray}
|\tilde{\Psi}\rangle &=&   S_0(-\xi_0)S_{\pm k}(-\xi_{1})|\Psi\rangle\nonumber\\
&=& |0_{a_{+}},0_{a_{-}},0_{C_{0k}},0_{C_{2k}},0_{C_{-2k}}\rangle \label{12.33}
\end{eqnarray}
represents the vacuum state of the transformed system and the ket $|\Psi\rangle$ represents the final stationary state of the system.

If we now perform the inverse transformation from~$|\tilde{\Psi}\rangle$ to $|\Psi\rangle$, we find that the system is in the multi-mode pure squeezed state
\begin{eqnarray}
|\Psi\rangle = S_0(\xi_0)S_{\pm k}(\xi_{1})|0_{C_{0k}},\!0_{C_{2k}},\!0_{C_{-2k}}\rangle\!\otimes\!|0_{a_{+}},\!0_{a_{-}}\!\rangle .\label{12.34}
\end{eqnarray}
The density operator representing the multi-mode squeezed state (\ref{12.34}) is
\begin{eqnarray}
\rho = S_0(\xi_0)S_{\pm k}(\xi_{1})|\{0\}\rangle\langle \{0\}|S_0(-\xi_0)S_{\pm k}(-\xi_{1}) ,\label{12.34a}
\end{eqnarray}
where $|\{0\}\rangle =|0_{C_{0k}},\!0_{C_{2k}},\!0_{C_{-2k}}\rangle\!\otimes\!|0_{a_{+}},\!0_{a_{-}}\!\rangle $. 

It shows that in the steady-state the cavity modes are left in the vacuum state and the atomic ensemble is prepared in single and two-mode squeezed states. In other words, we have found that the collective mode $C_{0k}$ is prepared in the one-mode squeezed vacuum state $S_0(\xi_0)|0_{C_{0k}}\rangle$, whereas the collective modes~$C_{\pm2k}$ are in the two-mode squeezed vacuum state $S_{\pm k}(\xi_{1})|0_{C_{2k}},0_{C_{-2k}}\rangle$ associated with the superposition of two, position dependent counter-propagating modes. It is interesting to note that only a {\it single} set of the laser parameters, the Rabi frequencies and phases, was required to create the pure squeezed states.

The most interesting of the results of this section is that the state (\ref{12.34}) cannot be factorized into the product of states of the individual modes. Thus, we may conclude that the atomic ensemble, after the interaction with the sequences of the laser pulses is prepared in single and two-mode entangled states.

\subsection{Four-mode entangled state}

\noindent Thus far our discussion has been limited to creation of entangled states inside a single atomic ensemble. It is particularly interesting to generate entangled states between two $(M=2)$  atomic ensembles. Since the cavity is composed of two counter-propagating modes, the interaction between the collective modes and the pair of cavity modes can produce squeezing of the
multi-mode variety, with a possibility to create a four-mode squeezed state~\cite{ls87,msf07}. 

We now proceed to illustrate how it is possible to create a four-mode squeezed (entangled) state between two atomic ensembles by a separate addressing of the collective modes with sequences of laser pulses of suitably chosen Rabi frequencies and phases. We again adopt the procedure of Li and  demonstrate how to prepare the four collective bosonic modes $C_{\pm2k}^{(n)}$ in the following pure four-mode squeezed state
\begin{eqnarray}
|\Psi\rangle = S_{\pm k}(\xi)|0_{C_{2k}^{(1)}},0_{C_{-2k}^{(1)}},0_{C_{2k}^{(2)}},0_{C_{-2k}^{(2)}}\rangle ,\label{12.35}
\end{eqnarray}
where the unitary operator has the form
\begin{eqnarray}
S_{\pm k}(\xi) &=&  \exp\left\{-\xi\left(C_{2k}^{(1)\dag}C_{-2k}^{(1)\dag}
+C_{-2k}^{(1)\dag}C_{2k}^{(2)\dag}\right.\right.\nonumber\\
&&+ \left.\left. C_{2k}^{(2)\dag}C_{-2k}^{(2)\dag} - {\rm
H.c.}\right)\right\} ,\label{12.36}
\end{eqnarray}
and $\xi$ is the squeezing parameter which we take to be a real number. As in the above cases of one and two-mode squeezing, the squeezing parameter can be adjusted and controlled by the driving laser fields.

Before introducing the procedure, we would like to point out that a transformation of the density matrix  of the system with the unitary operator (\ref{12.36}) does not  lead to a density matrix describing a simple system of independent linear mixers. Therefore, we first transform the state $|\Psi\rangle$ to a new state $|\Phi\rangle = T|\Psi\rangle$, with a unitary operator $T$ chosen such that it transforms the field operators~$C_{m2k}^{(n)}$ to new operators 
\begin{eqnarray}
d_{m}^{(n)}=TC_{2mk}^{(n)}T^\dag ,\label{12.37}
\end{eqnarray}
which are linear combinations involving the operators of different ensembles only, i.e.
\begin{eqnarray}
&&d_+^{\,(1)} = \frac{C_{2k}^{(1)}\!+\!\lambda C_{2k}^{(2)}}{\sqrt{1+\lambda^2}} ,\quad
d_+^{\,(2)}=\frac{\lambda C_{2k}^{(1)}\!-\! C_{2k}^{(2)}}{\sqrt{1+\lambda^2}},\nonumber\\
&&d_-^{\,(1)} = \frac{\lambda C_{-2k}^{(2)}\!-\!C_{-2k}^{(1)}}{\sqrt{1+\lambda^2}} ,\quad
d_-^{\,(2)}=\frac{C_{-2k}^{(2)}\!+\!\lambda
C_{-2k}^{(1)}}{\sqrt{1+\lambda^2}} ,\nonumber \\
\label{12.38}
\end{eqnarray}
where  $\lambda = (1+\sqrt{5})/2$.

With this carefully chosen transformation, we find that the state $|\Phi\rangle$ takes a form
\begin{eqnarray}
|\Phi\rangle &=& \exp\left[-\xi\left(\lambda C_{2k}^{(1)\dag}
C_{-2k}^{(2)\dag}-\frac{1}{\lambda} C_{2k}^{(2)\dag} C_{-2k}^{(1)\dag}\right) +{\rm H.c.}\right]\nonumber\\
&&\times \,|0_{C_{2k}^{(1)}},0_{C_{-2k}^{(1)}},0_{C_{2k}^{(2)}},0_{C_{-2k}^{(2)}}\rangle \nonumber\\
&=& S_{\pm k}^{(1)}(\lambda \xi)S_{\pm k}^{(2)}\left(-\xi/\lambda\right)|0_{C_{2k}^{(1)}},0_{C_{-2k}^{(1)}},0_{C_{2k}^{(2)}},0_{C_{-2k}^{(2)}}\rangle ,\nonumber\\
\label{12.39}
\end{eqnarray}
where $S_{\pm k}^{(1)}(\lambda\xi)$ and $S_{\pm k}^{(2)}(-\xi/\lambda)$ are  squeezing operators involving field modes of the ensemble 1 and 2, respectively. 

As we shall see, the advantage of working with the transformed state~$|\Phi\rangle$ rather than the state $|\Psi\rangle$ is that the density matrix transformed with the operators $S_{\pm k}^{(1)}(\lambda \xi)$ and $S_{\pm k}^{(2)}(-\xi/\lambda)$ describes a simple system of a linear mixer of a cavity mode with one of the field modes.

We now proceed to perform the construction of the four-mode squeezed state (\ref{12.35}), which is done in four steps, each separately addressing one of the collective modes. Since the collective modes are orthogonal to each other, an arbitrary transformation performed on one of the modes will not affect the remaining modes.

In the first step, we adjust the driving lasers to propagate in the direction of  the clockwise mode~$a_{+}$. In this case, the dynamics of the system are described by the Hamiltonian (\ref{12.19}), and after a unitary transformation of the density matrix
\begin{eqnarray}
S_{\pm k}^{(2)}(\xi/\lambda)S_{\pm k}^{(1)}(-\lambda\xi)T\rho T^\dag
S_{\pm k}^{(1)}(\lambda \xi)S_{\pm k}^{(2)}(-\xi/\lambda) = \rho_{1} ,\label{12.40}
\end{eqnarray}
with the squeezing parameter
\begin{eqnarray}
\xi = \frac{1}{2\lambda}\ln\left(\frac{\beta_{u_1}+\beta_{s_2}}{\beta_{u_1}-\beta_{s_2}}\right) ,\label{12.41}
\end{eqnarray}
and with the laser Rabi frequencies and phases such that
\begin{eqnarray}
&&\beta_{u_2} = \lambda\beta_{u_1} ,\quad \beta_{s_1}=\lambda\beta_{s_2} ,\nonumber \\
&&\phi_{u_n} = \phi_{s_n}=0 ,\quad n=1,2 ,\label{12.42}
\end{eqnarray}
we find that the master equation of the transformed density matrix is of the form
\begin{equation}
\frac{d}{dt}\rho_{1} = -i[\tilde{H}_{1},\rho_{1}]+{\cal L}_{c} \rho_{1} ,\label{12.43}
\end{equation}
where
\begin{eqnarray}
\tilde{H}_{1} &=& \left[\beta_{u_1}\!\left(C_{0k}^{(1)}\!+\!\lambda C_{0k}^{(2)}\right)\!
+\!\beta_{s_2}\left(\lambda C_{0k}^{(1)\dag}\!+\!C_{0k}^{(2)\dag}\right)\right]\!a_+^\dag \nonumber\\
&+&\sqrt{\left(1\!+\!\lambda^2\right)\left(\beta_{u_1}^2-\beta_{s_2}^2\right)} C_{2k}^{(1)}a_-^\dag\!
+\!{\rm H.c.} \label{12.46}
\end{eqnarray}
We see that the choice of the laser parameters (\ref{12.42}) results in the collective mode $C_{2k}^{(1)}$, out of the four modes involved in the  squeezing operator (\ref{12.36}), being effectively coupled to the cavity modes, the other modes are decoupled. It may appear surprising that only one of the four collective modes is effectively coupled to the cavity mode. However, this is merely a consequence of the form of the effective interaction Hamiltonian~(\ref{12.19}). It has an obvious advantage that we can separately address each of the collective modes. One may also notice from Eq.~(\ref{12.46}) that apart form the collective mode $C_{2k}^{(1)}$, the modes $C_{0k}^{(n)}$ are also coupled to one of the cavity modes. Since the modes $C_{0k}^{(n)}$ are not involved in the multi-mode squeezed state~(\ref{12.35}) whose construction we are
interested in, we do not consider their evolution.

Fortunately, the master equation (\ref{12.43}) is of a similar form as the master equation~(\ref{12.26}). Therefore, we can follow the same arguments as before to conclude that the mode $C_{2k}^{(1)}$ will evolve in time and definitely after a sufficiently long time, $\sim 2/\kappa$, the
mode will be found in a stationary vacuum state. The other three modes will remain in undetermined states.

We now turn off the lasers propagating in the direction of the clockwise  mode, and perform the second step, in which we first adjust the driving lasers to propagate in the direction of  the
anti-clockwise mode~$a_{-}$.  In this case, we adjust the squeezing parameters such that
\begin{eqnarray}
&&\beta_{u_1} = \lambda\beta_{u_2} ,\quad \beta_{s_2}=\lambda\beta_{s_1} ,\nonumber \\
&&\phi_{u_n} = \phi_{s_n}=0 ,\quad n=1,2 ,\label{12.47}
\end{eqnarray}
to obtain the same value for the squeezing parameter now given by
\begin{eqnarray}
\xi = \frac{1}{2\lambda}\ln\left(\frac{\beta_{u_2}+\beta_{s_1}}{\beta_{u_2}-\beta_{s_1}}\right) .\label{12.48}
\end{eqnarray}
For this choice of the parameters, and bearing in mind that the dynamics of the system are now governed by the Hamiltonian~(\ref{12.20}), we find that after a unitary transformation, Eq.~(\ref{12.40}), the master equation of the transformed density matrix takes the form
\begin{equation}
\frac{d}{dt}\rho_{2} = -i[\tilde{H}_{2},\rho_{2}]+{\cal L}_{c} \rho_{2} ,\label{12.49}
\end{equation}
with
\begin{eqnarray}
\tilde{H}_{2} &=& \left[\beta_{u_2}\!\left(C_{0k}^{(1)}\!+\!\lambda C_{0k}^{(2)}\right)\!
+\!\beta_{s_1}\left(\lambda C_{0k}^{(1)\dag}\!+\!C_{0k}^{(2)\dag}\right)\right]\!a_{-}^\dag\nonumber\\
&+&\sqrt{\left(1+\lambda^2\right)\left(\beta_{u_2}^2-\beta_{s_1}^2\right)}\,
C_{-2k}^{(2)}a_{+}^\dag +{\rm H.c.} \label{12.50}
\end{eqnarray}
We see that the choice of the laser parameters, Eq.~(\ref{12.47}), has resulted in the collective mode $C_{-2k}^{(2)}$ to be effectively coupled to the cavity modes, with the other modes decoupled. The dynamics of the mode are determined by the master equation~(\ref{12.49}), which is of the similar form as Eq.~(\ref{12.26}). Thus, we can use the same arguments as in the previous step, and conclude that after a sufficiently long time, $\sim 2/\kappa$, also the mode
$C_{-2k}^{(2)}$ will be found in a stationary vacuum state.

In the next step, we change the direction of propagation of the driving lasers back to the direction of  the clockwise mode~$a_{+}$. Similarly as in the above two steps, we first perform the unitary
transformation of the density matrix, Eq.~(\ref{12.40}), with the laser parameters chosen such that
\begin{eqnarray}
&&\beta_{u_1} = \lambda\beta_{u_2} ,\quad \beta_{s_2}=\lambda\beta_{s_1} ,\nonumber \\
&&\phi_{u_1} = \phi_{s_1}=0 ,\quad \phi_{u_2} = \phi_{s_2} = \pi ,\label{12.51}
\end{eqnarray}
and
\begin{eqnarray}
\xi = \frac{\lambda}{2}\ln\left(\frac{\beta_{u_2}+\beta_{s_1}}{\beta_{u_2}-\beta_{s_1}}\right) .\label{12.52}
\end{eqnarray}
With this choice of the laser parameters (\ref{12.51}), we find that now the collective mode $C_{2k}^{(2)}$ is the only mode coupled to the cavity modes and the dynamics of the mode follows the same pattern as the modes considered in the above two steps. Thus, we may conclude that after a sufficiently long time, also the collective mode $C_{2k}^{(2)}$ will be found in a stationary vacuum state.

Finally, in the fourth step, we prepare the remaining mode~$C_{-2k}^{(1)}$ in the vacuum state. To do this, we again change the direction of the driving lasers to propagate in the direction of the anti-clockwise mode and choose the laser parameters such that
\begin{eqnarray}
&&\beta_{u_2} = \lambda\beta_{u_1} ,\quad \beta_{s_1}=\lambda\beta_{s_2} ,\nonumber \\
&&\phi_{u_1} = \phi_{s_1} = \pi ,\quad \phi_{u_2} = \phi_{s_2} = 0 ,\label{12.53}
\end{eqnarray}
and
\begin{eqnarray}
\xi = \frac{\lambda}{2}\ln\left(\frac{\beta_{u_1}+\beta_{s_2}}{\beta_{u_1}-\beta_{s_2}}\right) .\label{12.54}
\end{eqnarray}
Following the same procedure as in the previous three steps, one can easily show that the choice of the laser parameters, Eq.~(\ref{12.53}), results in a master equation for the transformed
density operator determined by the interaction Hamiltonian involving the collective mode $C_{-2k}^{(1)}$. Thus, as a result of the damping of the cavity modes, the collective mode will evolve towards a vacuum state.

In this way, the final state of the transformed system is a four-mode vacuum state determined by the density operator
\begin{equation}
\tilde{\rho}(\tau\sim 8/\kappa)=|\tilde{\Phi}\rangle\langle\tilde{\Phi} | ,\label{12.55}
\end{equation}
where
\begin{eqnarray}
|\tilde{\Phi}\rangle &=&   S_{\pm k}^{(1)}(-\lambda \xi)S_{\pm k}^{(2)}(\xi/\lambda)T|\Psi\rangle \nonumber\\
&=&  |0_{C_{2k}^{(1)}},0_{C_{-2k}^{(1)}},0_{C_{2k}^{(2)}},0_{C_{-2k}^{(2)}}\rangle \label{12.56}
\end{eqnarray}
represents the vacuum state of the transformed system and the ket $|\Psi\rangle$  represents the final stationary state of the system. 

Note that the density matrix (\ref{12.55}) cannot be factorized into the product of the density matrices of the individual modes. Thus, we may conclude that the atomic ensemble, after the interaction with the sequences of the laser pulses is prepared in four-mode entangled states.

Finally it may be mentioned that the procedures which construct entangled states by using only simple unitary transformations offer further interesting prospects for the study of multi-mode entanglement and creation of continuous variable entangled cluster states. Examples of cluster states and a procedure of their creation are discussed in the next section.

\section{Cluster states of atomic ensembles}

\noindent As the last issue of this review, we wish to make a further theoretical application of the bosonic continuous variable model, introduced in the preceding section, to discuss a practical scheme for creation of cluster states of separate atomic ensembles. Cluster states have been recognized as suitable for one-way quantum computation and possess a high persistency and robustness of their entanglement with regard to decoherence effects~\cite{br01,db04}.
\begin{figure}[hbp]
\includegraphics[width=7cm,keepaspectratio,clip]{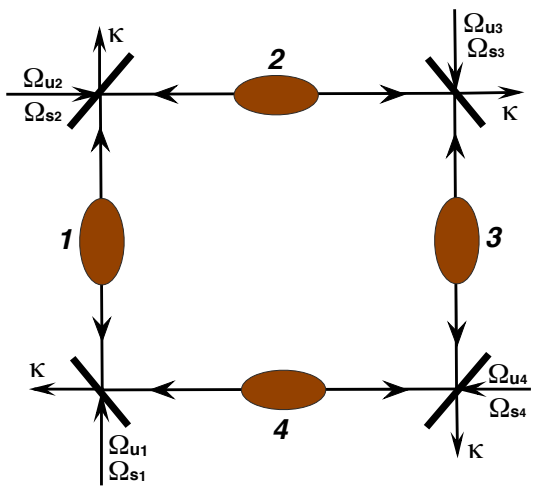}
\caption{Configuration of a ring cavity and four atomic ensembles for the preparation of entangled continuous variable cluster states.  External pulse lasers of the Rabi frequencies $\Omega_{un}$ and $\Omega_{sn}$ couple to only one (propagating) mode of the cavity and interact dispersively with the atoms.}
\label{figc31}
\end{figure}
There have been many proposals for creation of cluster states that 
involve squeezed light sources, a network of beam splitters and a measurement by the homodyne 
detection~\cite{ml06,mf07,lw07,st07}. Essentially all these proposals are based on linear optics schemes and 
measurements on the system. Apart from the schemes based on linear optics, a significant interest is now in 
the possibility of creating continuous variable cluster states of distinct atomic ensembles.

We present here the results of the study of a practical scheme for creation of continuous variable entangled 
cluster states of four separate atomic ensembles located inside a high-$Q$ ring cavity. Specifically, we employ 
the method of Li~\cite{li06} to demonstrate how to prepare the so-called continuous variable linear, square, 
and T-type cluster states of the atomic ensembles using the sequence of laser pulses and the cavity dissipation. 

We shall consider a system consisting of four atomic ensembles located inside a high-finesse ring cavity. 
The cavity is composed of four mirrors that create two modes, called propagating and counter-propagating modes, 
to which the atomic ensembles are equally coupled. External pulse lasers that are used to drive the atomic 
ensembles couple to only a single propagating mode, as it is illustrated in Fig.~\ref{figc31}. We assume that 
the atoms are homogeneously distributed inside the ensembles. In this case only the forward scattering occurs 
that allows us to neglect the coupling of the atoms to the counter-propagating mode and work in the single mode 
approximation. In other words, the cavity mode propagates with the laser fields. 
The effective Hamiltonian of the system, Eq.~(\ref{12.19}), can then be simplified to the form
\begin{eqnarray}
H_{e} = \sum\limits_{n=1}^{4}\left[ \left(\beta_{un} {\rm e}^{-i\phi_{un}}C_{n}\!+\!\beta_{sn}{\rm e}^{-i\phi_{sn}}C_{n}^{\dag}\right)a_+^\dag\!+\!{\rm H.c.}\right] ,\nonumber\\
\label{13.1}
\end{eqnarray}
where $C_{n}\equiv C_{0k}^{(n)}$.

The condition of the coupling of the atomic ensembles to only a single mode of the cavity can be easily fulfilled in trapped room-temperature atomic ensembles where fast atomic oscillations over the interaction time lead to a collectively enhanced coupling of the atoms to a single mode that is practically collinear with the laser fields~\cite{dcz}.

\begin{figure}[hbp]
\includegraphics[width=4cm,keepaspectratio,clip]{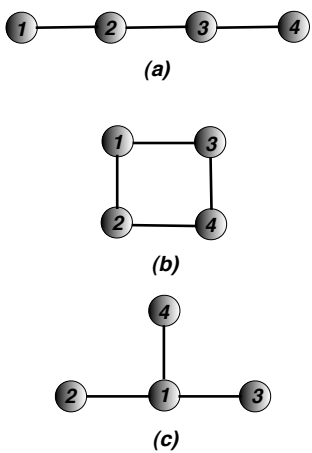}
\caption{Examples of continuous variable four-mode cluster states: (a) linear cluster state, (b) square cluster state, and (c) T-shape cluster state.}
\label{figc32}
\end{figure}

We now proceed to discuss the detailed procedure of the preparation of continuous variable four-mode cluster states of atomic ensembles located inside a single-mode ring cavity. In particular, we shall show how to deterministically prepare a linear cluster state, a square cluster state, and a T-shape cluster state, as shown in Fig.~\ref{figc32}.

We begin with a brief explanation of how one could distinguish that a given state belongs to the class of cluster states. Simply, a given state is quantified as a cluster state if the quadrature correlations are such that in the limit of infinite squeezing, the state becomes zero eigenstate of a set of quadrature combinations~\cite{yu08}
\begin{eqnarray}
\left(\hat{p}_{a} -\sum_{b\in N_{a}}\hat{x}_{b}\right) \rightarrow 0 ,\label{13.2}
\end{eqnarray}
where $\hat{x}$ and $\hat{p}$ are the position and momentum operators (quadratures) of a mode $a$, and the modes $b$ are the nearest neighbors $N_{a}$ of the mode $a$. In what follows, we
quantify a given state as a cluster state by evaluating the variances of linear combinations of the momentum and position operators of the involved field modes. If the variances vanish in the limit of the infinite squeezing then, according to the above definition, a given state is an example of  cluster state.

\subsection{Preparation of a linear cluster state}

\noindent Let us first discuss in greater detail the problem of the preparation of a four-mode linear cluster state involving four separate atomic ensembles, as shown in Fig.~\ref{figc32}(a). The essential idea is as follows. We begin by making an unitary  transformation 
\begin{eqnarray}
d_{L_n}=T_1 C_{n}T_1^\dag ,\quad  n=1,2,3,4 \label{13.3}
\end{eqnarray}
that transfers the $C_{n}$ operators into linear combinations
\begin{eqnarray}
d_{1} &=& -\frac{1}{\sqrt{2}}(iC_1+C_2) ,\nonumber\\
d_{2} &=& -\frac{1}{\sqrt{10}}(iC_1-C_2-2iC_3-2C_4) ,\nonumber\\
d_{3} &=& -\frac{1}{\sqrt{2}}(C_3+iC_4) ,\nonumber\\
d_{4} &=& -\frac{1}{\sqrt{10}}(2C_1+2iC_2+C_3-iC_4) .\label{13.4}
\end{eqnarray}
Since the operators $d_{n}$ commutate with each other, the combined modes are orthogonal to each other. In this case, the system
consists of completely decoupled modes, which will allow us to prepare each mode separately in a desired state. In other words, an arbitrary transformation performed on the operators of a given mode will not affect the remaining modes.

Next, we prepare the field modes in a desired state by using laser pulses of equal length and suitably chosen magnitudes of the Rabi frequencies and phases. More concretely, for a cluster state, all the modes $d_{n}$ should be prepared in a squeezed vacuum state. We employ the fact that the modes can be separately prepared in the squeezed vacuum state. Afterwards, if the variances of specific combinations of the quadrature components of the field operators vanish in the limit of the infinite squeezing, the desired state is a linear continuous variable cluster state.

In the first step of the preparation, we send a set of laser pulses driving the the atomic ensembles 1 and 2 only, the lasers driving the atomic ensembles 3 and 4 are turned off. We choose the Rabi frequencies and phases of the turned on lasers as
\begin{eqnarray}
\Omega_{un} &=& \frac{\Omega_{sn}}{r}=\sqrt{2}\Omega ,\quad n=1,2 , \nonumber\\
\phi_{u1}&=& \frac{3}{2}\pi ,\ \phi_{s1}=\frac{1}{2}\pi ,\ \phi_{u2} = \phi_{s2}=\pi .\label{13.5}
\end{eqnarray}

After the first step of the preparation, the mode $d_{1}$ is left in a state that is described by the density operator
\begin{eqnarray}
\rho_1=T_1\rho T_1^\dag ,\label{13.7}
\end{eqnarray}
which obeys a master equation
\begin{equation}
\frac{d}{dt} \rho _1 = -i\beta\left[\left(a^\dagger d_{1}+ra^\dagger
d_{1}^\dagger\right)+{\rm H.c.}, \rho _1\right] + L_a \rho _1 ,\label{13.8}
\end{equation}
where $\beta \equiv 2\beta_{u(1,2)}=2\beta_{s(1,2)}$, and $r\in (0,1)$. 

If we now perform the single-mode squeezing transformation for the mode $d_{1}$ as
\begin{eqnarray}
\tilde{\rho}_1=S^\dag_1(\xi)\rho_1S_1(\xi) ,\label{13.9}
\end{eqnarray}
with the single-mode squeezing operator~\cite{fd2}
\begin{eqnarray}
S_n(\xi)\!=\!\exp\!\left[\frac{\xi}{2}\left( d_{n}^2-d_{n}^{\dag2}\right)\right] ,\label{13.10}
\end{eqnarray}
where $\xi = \tanh^{-1}(r)$, we find that the master equation~(\ref{13.8}) becomes
\begin{eqnarray}
\frac{d}{dt}\tilde{\rho}_1=\beta\sqrt{1-r^2}\left[ad_{1}^\dag+a^\dag
d_{1},\tilde{\rho}_1\right]+L_a\tilde{\rho_1} .\label{13.11}
\end{eqnarray}
It can be shown from Eq.~(\ref{13.11}) that in the steady state, the cavity mode will be in the vacuum state and the mode $d_{1}$ will be in a squeezed vacuum state. This is easy to understand, the master equation (\ref{13.11}) represents two coupled modes with the cavity mode linearly damped with the rate $\kappa$. Since the remaining modes $d_{2}$, $d_{3}$, and~$d_{4}$ are decoupled from the mode $d_{1}$, they remain in an undetermined state determined by the density operator~$\rho_{d_{2}d_{3}d_{4}}(\tau)$.

In order to estimate the required time to reach the steady state, we calculate eigenvalues of Eq.~(\ref{13.11})
\begin{eqnarray}
\eta_\pm = -\frac{\kappa}{2}\pm\left[\left(\frac{\kappa}{2}\right)^2-\beta^2\left(1-r^2\right)\right]^\frac{1}{2} ,\label{13.12}
\end{eqnarray}
from which we observe that as long as $\beta\sqrt{1-r^2}>\kappa/2$, the time for the system  to reach its steady state is of order of $\sim 2/\kappa$. Therefore, the system will definitely
evolve into the steady state, provided the interaction time is sufficient long, for example, $\tau=4/\kappa$. The time $\tau=4/\kappa$ determines the time scale in our protocol for the preparation of the cluster states.

By taking the inverse unitary transformation, it follows that in the steady state, the total system is in a state determined by the density operator
\begin{eqnarray}
\rho_1\!(\tau)\!=\!S_1(\xi)|0_a,\!0_{d_{L_1}}\!\rangle
\langle0_a,\!0_{d_{L_1}}\!|S_1^\dag(\xi)\!\otimes\!\rho_{d_{L_2}\!d_{L_3}\!d_{L_4}}\!(\tau) .\label{13.13}
\end{eqnarray}
Briefly summarize what we have obtained after the first step of the preparation. The application of
suitably chosen laser pulses leaves the mode $d_{1}$ prepared in the single-mode squeezed vacuum state, with the cavity field found in the vacuum state, and the remaining combined modes~$d_{2}$, $d_{3}$, and~$d_{4}$ left in the states related to their initial~states.

In the second step, we turn off the first series of driving lasers and sequentially send another series of laser pulses with different parameters to preparation of the combined bosonic mode $d_{2}$ in the single-mode squeezed vacuum state so that a similar linearly mixing interaction between the cavity mode and another combined bosonic modes arises. As before, for the first series of pulses, a single-mode squeezed vacuum state for this combined bosonic mode can be prepared due to the cavity dissipation. The Rabi frequencies of the second series of the laser pulses, which are turned on during the time of $t\in[\tau,2\tau)$ for the preparation of the combined bosonic mode $d_{2}$ in the single-mode squeezed vacuum state $S_2(\xi)|0_{d_{2}}\rangle$, are
\begin{eqnarray}
\Omega_{un} &=& \frac{\Omega_{sn}}{r}=\frac{2}{\sqrt{10}}\Omega ,\quad n = 1,2 , \nonumber\\
\Omega_{um} &=& \frac{\Omega_{sm}}{r}=\frac{4}{\sqrt{10}}\Omega ,\quad m = 3,4 ,\label{13.14}
\end{eqnarray}
and the phases of the driving lasers
\begin{eqnarray}
\phi_{u1} &=& \frac{3}{2}\pi ,\ \phi_{s1}=\frac{1}{2}\pi ,\ \phi_{u3} = \frac{1}{2}\pi ,\ \phi_{s3}=\frac{3}{2}\pi ,\nonumber\\
\phi_{u2} &=& \phi_{s2}=\phi_{u4}=\phi_{s4} = 0 .\label{13.15}
\end{eqnarray}
The specific choice of the Rabi frequencies and the phases ensures the mode $d_{2}$ to be prepared in the state described by the density operator $\rho_{1}$ satisfying the same master 
as~Eq.~(\ref{13.11}) but with $d_{1}$ replaced by $d_{2}$.

In the third step, which is performed during the time of $t\in[2\tau,3\tau)$, the combined bosonic mode $d_{3}$ is being prepared in the single-mode squeezed vacuum state~$S_3(\xi)|0_{d_{3}}\rangle$. We turn off lasers driving ensembles 1 and 2, and the laser driving the ensembles 3 and 4 have Rabi frequencies and phases
\begin{eqnarray}
\Omega_{um} &=& \frac{\Omega_{sm}}{r} = \sqrt{2}\Omega ,\quad m=3,4 ,\nonumber\\
\phi_{u3} &=& \frac{3}{2}\pi ,\ \phi_{s3}=\frac{1}{2}\pi ,\
\phi_{u4} = \phi_{s4} = \pi .\label{13.17}
\end{eqnarray}

The fourth series, laser pulses are turned on during the time of $t\in[3\tau,4\tau)$. In this final series, the combined bosonic mode $d_{4}$ is prepared in the single-mode squeezed vacuum state $S_4(\xi)|0_{d_{4}}\rangle$. The laser pulses required to achieve this are of the Rabi frequencies
\begin{eqnarray}
\Omega_{un} &=& \frac{\Omega_{sn}}{r}=\frac{4}{\sqrt{10}}\Omega ,\quad n=1,2 ,\nonumber\\
\Omega_{um} &=& \frac{\Omega_{sm}}{r}=\frac{2}{\sqrt{10}}\Omega ,\quad m=3,4 ,\label{13.18}
\end{eqnarray}
and phases
\begin{eqnarray}
\phi_{u2} &=& \frac{1}{2}\pi ,\ \phi_{s2}=\frac{3}{2}\pi ,\ \phi_{u4} = \frac{3}{2}\pi ,\ \phi_{s4} = \frac{1}{2}\pi ,\nonumber\\
\phi_{um} &=& \phi_{sm}=0 ,\quad m=1,3 .\label{13.19}
\end{eqnarray}
Therefore after enough long time $4\tau$, the system evolves into the state determined by the density operator
\begin{equation}
\rho_1(4\tau)= |\Phi_L\rangle\langle \Phi_L| \otimes
 |0_a\rangle\langle0_a| ,\label{13.20}
\end{equation}
where
\begin{eqnarray}
|\Phi_{L}\rangle &=& T_1|\Psi_L\rangle =
S_1(\xi)|0_{d_{1}}\rangle
\otimes S_2(\xi)|0_{d_{2}}\rangle \nonumber \\
&& \otimes  S_3(\xi)|0_{d_{3}}\rangle \otimes  S_4(\xi)|0_{d_{4}}\rangle .\label{13.21}
\end{eqnarray}
This means that the transformed state is in the form of a quadripartite squeezed vacuum state. Once all the combined modes $d_{n}$ are prepared in the corresponding single-mode
squeezed vacuum state, the explicit form of the state created can be obtained by reversing the unitary transformation $T_1$. This leads to a pure linear continuous variable quadripartite state of the form
\begin{eqnarray}
|\Psi_L\rangle&=&\exp\left\{-\frac{\xi}{10}[C_1^2-C_2^2-C_3^2+C_4^2\right. \nonumber\\
&-&\left.  4(C_1+iC_2)(C_3-iC_4)\right. \nonumber\\
&-&\left. 8i(C_{1}C_2+C_3C_4)\right] -{\rm H.c.}\} |\{0\}\rangle  , \label{13.22}
\end{eqnarray}
where
\begin{eqnarray}
 |\{0\}\rangle = |0_{c_{1}},0_{c_{2}},0_{c_{3}},0_{c_{4}}\rangle .\label{13.23}
\end{eqnarray}

The question remains as to whether the state $|\Psi_L\rangle$ is an example of the linear continuous variable cluster state. To examine this, we introduce the quadrature amplitude and phase components of the four modes
\begin{eqnarray}
q_n = (C_{n}+C_{n}^\dag)/\sqrt{2} ,\ p_n = -i(C_n -C_n^\dag)/\sqrt{2} ,\label{13.24}
\end{eqnarray}
from which we easily find that the variances of the linear combinations of the components, evaluated according to the definition (\ref{13.1}), are
\begin{eqnarray}
V(p_1-q_2)&=& {\rm e}^{-2\xi} ,\ 
V(p_2-q_1-q_3) = \frac{3}{2}{\rm e}^{-2\xi} ,\nonumber\\
V(p_3-q_2-q_4)&=& \frac{3}{2}{\rm e}^{-2\xi} ,\ 
V(p_4-q_3) = {\rm e}^{-2\xi} ,\label{13.25}
\end{eqnarray}
where $V(X)=\langle X^2\rangle-\langle X\rangle^2$. Clearly, all the four variances tend to zero in the limit of infinite squeezing, $\xi\rightarrow\infty$. We therefore conclude that the state
$|\Psi_L\rangle$ is a four-mode linear continuous variable cluster state.

We may summarize that by an appropriate driving the four atomic ensembles, the four combined bosonic modes will be ultimately prepared in four single-mode squeezed vacuum states. It can be done in four steps by appropriately choosing the laser parameters such as phases and intensities. Then, by applying an inverse unitary transformation, the pure linear continuous variable  entangled state $|\Psi_L\rangle$ for the four atomic ensembles can be obtained.

\subsection{Preparation of a square cluster state}

\noindent We now turn to the calculation of the realization of an entangled continuous variable square cluster state $|\Psi_S\rangle$. 
We follow exactly the same route as we employed in the preparation of the continuous variable linear cluster~state.

A square continuous variable cluster state may be obtained by performing a unitary transformation $f_{n}=T_2C_nT_2^\dagger$ to obtain superposition modes determined by the following operators
\begin{eqnarray}
f_{1}&=&-\frac{1}{\sqrt{10}}(iC_1+iC_2+2C_3+2C_4) ,\nonumber\\
f_{2}&=&-\frac{i}{\sqrt{2}}(C_1-C_2) ,\nonumber\\
f_{3}&=&-\frac{1}{\sqrt{10}}(2C_1+2C_2+iC_3+iC_4) ,\nonumber\\
f_{4}&=&-\frac{i}{\sqrt{2}}(C_3-C_4) .\label{13.26}
\end{eqnarray}
Similar to the preparation of the linear continuous variable cluster state, we send series of laser pulses with different 
appropriate parameters, the phases and amplitudes. Each series serves to prepare the combined bosonic modes $f_{n}$ in a squeezed vacuum state. 

\vspace{3mm} \noindent{\footnotesize{Table 1}\quad Rabi frequencies of the laser pulses for the preparation of a square cluster state\\}
\vspace*{2mm}
\footnotesize \doublerulesep 0.4pt \tabcolsep 4pt
\begin{tabular}{ccccccccc}
 \hline
$t$  &   $\Omega_{u1}$ & $\Omega_{s1}$ & $\Omega_{u2}$  & $\Omega_{s2}$ & $\Omega_{u3}$ & $\Omega_{s3}$ & $\Omega_{u4}$ & $\Omega_{s4}$ \\\hline\hline $[0,\tau)$ & $\frac{2\Omega}{\sqrt{10}}$ &  $\frac{2r\Omega}{\sqrt{10}}$ &  $\frac{2\Omega}{\sqrt{10}}$ & $\frac{2r\Omega}{\sqrt{10}}$ & $\frac{4\Omega}{\sqrt{10}}$ & $\frac{4\Omega}{\sqrt{10}}$ & $\frac{4\Omega}{\sqrt{10}}$ & $\frac{4\Omega}{\sqrt{10}}$ \\
$[\tau, 2\tau)$& $\sqrt{2}\Omega$ & $\sqrt{2}r\Omega$ & $\sqrt{2}\Omega$ & $\sqrt{2}r\Omega$ & $0$ & $0$ & $0$ & $0$ \\
$[2\tau, 3\tau)$& $\frac{4\Omega}{\sqrt{10}}$ & $\frac{4r\Omega}{\sqrt{10}}$ & $\frac{4\Omega}{\sqrt{10}}$ & $\frac{4r\Omega}{\sqrt{10}}$ & $\frac{2\Omega}{\sqrt{10}}$ & $\frac{2\Omega}{\sqrt{10}}$ & $\frac{2\Omega}{\sqrt{10}}$ & $\frac{2\Omega}{\sqrt{10}}$\\
$[3\tau, 4\tau)$& $0$ & $0$ & $0$ & $0$ & $\sqrt{2}\Omega$ & $\sqrt{2}\Omega$ & $\sqrt{2}\Omega$ & $\sqrt{2}\Omega$ \\\hline
\end{tabular}
\normalsize
\vspace*{2mm}

The required Rabi frequencies and phases of the laser fields to create the square cluster state are listed in Tables 1 and 2. 
Different sets of laser pulses are chosen for different periods of time. Throughout this sequence of laser pulses each individual field
mode is incoherently damped by the cavity losses with the rate $\kappa$.

\vspace{3mm} \noindent{\footnotesize{Table 2}\quad Phases of the laser pulses for the preparation of a square cluster state\\}
\vspace*{2mm}
\footnotesize \doublerulesep 0.4pt \tabcolsep 6pt
\begin{tabular}{ccccccccc}
 \hline
$t$  &   $\phi_{u1}$ & $\phi_{s1}$ & $\phi_{u2}$  & $\phi_{s2}$ & $\phi_{u3}$ & $\phi_{s3}$ & $\phi_{u4}$ & $\phi_{s4}$ \\\hline\hline $[0,\tau)$ & $\frac{3\pi}{2}$ &  $\frac{\pi}{2}$ &  $\frac{3\pi}{2}$ & $\frac{\pi}{2}$ & $\pi$ & $\pi$ & $\pi$ & $\pi$ \\
$[\tau, 2\tau)$& $\frac{3\pi}{2}$ & $\frac{\pi}{2}$ & $\frac{\pi}{2}$ & $\frac{3\pi}{2}$ & $0$ & $0$ & $0$ & $0$ \\
$[2\tau, 3\tau)$& $\frac{3\pi}{2}$ & $\pi$ & $\frac{3\pi}{2}$ & $\pi$ & $\frac{3\pi}{2}$ & $\frac{\pi}{2}$ & $\frac{3\pi}{2}$ & $\frac{\pi}{2}$\\
$[3\tau, 4\tau)$& $0$ & $0$ & $0$ & $0$ & $\frac{3\pi}{2}$ & $\frac{\pi}{2}$ & $\frac{\pi}{2}$ & $\frac{3\pi}{2}$ \\\hline
\end{tabular}
\normalsize
\vspace*{2mm}

After the above sequence of the laser pulses, the system is left in a state described by the density operator
\begin{equation}
\rho_1(4\tau)=T_2|\Psi_S\rangle\langle \Psi_S|T_2^\dag \otimes
 |0_a\rangle\langle0_a| .\label{13.35}
\end{equation}
Inverting the transformation $T_2$, we find that the system is in the state
\begin{eqnarray}
|\Psi_S\rangle &=& \exp\left\{-\frac{\xi}{10}[C_1^2+C_2^2+C_3^2+C_4^2\right. \nonumber\\ 
&-&\left. 4i(C_1\!+\!C_2)(C_3\!+\!C_4)\right. \nonumber\\
&-&\left. 8(C_1C_2 + C_3C_4)]\!-\!{\rm H.c.}\right\}\!|\{0\}\rangle .\label{13.36}
\end{eqnarray}

Finally, we calculate the variances in the sum and difference operators  following the general procedure for quantifying the cluster states and find that the variances for the square type state
illustrated in Fig.~\ref{figc31}(b) are given~by
\begin{eqnarray}
V(p_1-q_3-q_4) &=& V(p_2-q_3-q_4) \nonumber\\
&=& V(p_3-q_1-q_2)=  V(p_4-q_1-q_2)\nonumber\\
& =& \frac{3}{2}{\rm e}^{-2\xi} .\label{13.37}
\end{eqnarray}
In the limit $\xi\rightarrow\infty$, it is easy to see that the variances tend to zero. Thus, according to the criterion (\ref{13.2}), the state $|\Psi_S\rangle$ is an example of a four-mode square cluster state. We conclude that the sequential application of the laser pulses, the continuous variable entangled quadripartite square cluster state~$|\Psi_S\rangle$ is unconditionally produced.

\subsection{Preparation of a T-shape cluster state}

\noindent The final example illustrates the procedure to prepare a~T-shape entangled cluster state of the form
\begin{eqnarray}
|\psi_T\rangle&=&\exp\left\{\frac{1}{4}\xi [C_1^2-C_2^2-C_3^2-C_4^2\right. \nonumber \\
&+&\left. 2(C_2C_3+C_2C_4+C_3C_4)\right. \nonumber\\
&+&\left. 2iC_1(C_2+C_3+C_4)]\!-\!{\rm H.c.}\}|\{0\right\}\rangle .\label{13.38}
\end{eqnarray}

To arrive at the cluster state (\ref{13.38}), we first perform a unitary transformation $h_{n}=T_3C_{n}T_3^\dagger$ that transfers the bosonic operators $C_{n}$ into superposition modes
\begin{eqnarray}
h_{1}&=&\frac{\sqrt{3}}{2}\left[iC_1-\frac{1}{3}(C_2+C_3+C_4)\right] ,\nonumber\\
h_{2}&=&\frac{\sqrt{6}}{3}\left[C_2-\frac{1}{2}(C_3+C_4)\right] ,\nonumber\\
h_{3}&=&\frac{1}{\sqrt{2}}\left(C_3-C_4\right) ,\nonumber\\
h_{4}&=&\frac{1}{2}\left(iC_1+C_2+C_3+C_4\right) .\label{13.39}
\end{eqnarray}
We will show that the transformation $T_3$ leads to the field  modes
that can be prepared in a T-shape cluster state.

Following the similar four-step procedure as in the above two examples, we first show that by a suitable choice of the Rabi 
frequencies and phases of the laser fields, one can prepare each individual mode $h_{n}$ in a squeezed vacuum state.

To achieve this, we choose a sequence of the laser pulses of the Rabi frequencies and phases as listed in the following 
Tables 3 and 4. Of course, throughout this sequence of laser pulses each field mode is damped to a steady-state by the cavity losses
with the rate $\kappa$.

\vspace{3mm} \noindent{\footnotesize{Table 3}\quad Rabi frequencies of the laser pulses for the creation of a~T-shape cluster state\\}
\vspace*{2mm}
\footnotesize \doublerulesep 0.4pt \tabcolsep 3.5pt
\begin{tabular}{ccccccccc}
 \hline
$t$  &   $\Omega_{u1}$ & $\Omega_{s1}$ & $\Omega_{u2}$  & $\Omega_{s2}$ & $\Omega_{u3}$ & $\Omega_{s3}$ & $\Omega_{u4}$ & $\Omega_{s4}$ \\\hline\hline $[0,\tau)$ & $\sqrt{3}\Omega$ &  $\sqrt{3}\Omega$ &  $\frac{\Omega}{\sqrt{3}}$ & $\frac{r\Omega}{\sqrt{3}}$ & $\frac{\Omega}{\sqrt{3}}$ & $\frac{r\Omega}{\sqrt{3}}$ & $\frac{\Omega}{\sqrt{3}}$ & $\frac{r\Omega}{\sqrt{3}}$ \\
$[\tau, 2\tau)$& $0$ & $0$ & $\frac{2\sqrt{6}\Omega}{3}$ & $\frac{2\sqrt{6}\Omega}{3}$ & $\frac{\sqrt{6}\Omega}{3}$ & $\frac{\sqrt{6}r\Omega}{3}$ & $\frac{\sqrt{6}\Omega}{3}$ & $\frac{\sqrt{6}r\Omega}{3}$ \\
$[2\tau, 3\tau)$& $0$ & $0$ & $0$ & $0$ & $\sqrt{2}\Omega$ & $\sqrt{2}r\Omega$ & $\sqrt{2}\Omega$ & $\sqrt{2}r\Omega$\\
$[3\tau, 4\tau)$& $\Omega$ & $r\Omega$ & $\Omega$ & $r\Omega$ & $\Omega$ & $r\Omega$ & $\Omega$ & $r\Omega$ \\\hline
\end{tabular}
\normalsize
\vspace*{2mm}

\vspace{3mm} \noindent{\footnotesize{Table 4}\quad Phases of the laser pulses for the creation of a~T-shape cluster state\\}
\vspace*{2mm}
\footnotesize \doublerulesep 0.4pt \tabcolsep 6pt
\begin{tabular}{ccccccccc}
 \hline
$t$  &   $\phi_{u1}$ & $\phi_{s1}$ & $\phi_{u2}$  & $\phi_{s2}$ & $\phi_{u3}$ & $\phi_{s3}$ & $\phi_{u4}$ & $\phi_{s4}$ \\\hline\hline $[0,\tau)$ & $\frac{\pi}{2}$ &  $\frac{3\pi}{2}$ &  $\pi$ & $\pi$ & $\pi$ & $\pi$ & $\pi$ & $\pi$ \\
$[\tau, 2\tau)$& $0$ & $0$ & $0$ & $0$ & $\pi$ & $\pi$ & $\pi$ & $\pi$ \\
$[2\tau, 3\tau)$& $0$ & $0$ & $0$ & $0$ & $0$ & $0$ & $\pi$ & $\pi$\\
$[3\tau, 4\tau)$& $\frac{\pi}{2}$ & $\frac{3\pi}{2}$ & $0$ & $0$ & $0$ & $0$ & $0$ & $0$ \\\hline
\end{tabular}
\normalsize
\vspace*{2mm}

After the above sequence of the laser pulses, the system is left in a state described by the density operator
\begin{equation}
\rho_1(4\tau)=T_3|\Psi_T\rangle\langle \Psi_T|T_3^\dag \otimes
 |0_a\rangle\langle0_a| .\label{13.46}
\end{equation}
Inverting the transformation $T_3$, we find that the system is in the pure state (\ref{13.38}).

The only thing left is to determine as to whether the state $|\Psi_T\rangle$ belongs  to the class of cluster states. It is done by calculating the variances of the sum and difference operators
from which we find that for the state $|\Psi_T\rangle$, the variances are given by
\begin{eqnarray}
V(p_2-q_1) &=& V(p_3-q_1) \nonumber\\
&=& V(p_4-q_1) = {\rm e}^{-2\xi} ,\nonumber\\
V(p_1-q_2-q_3-q_4) &=& 2{\rm e}^{-2\xi} .\label{13.47}
\end{eqnarray}
Clearly, in the limit $\xi\rightarrow\infty$, the variances tend to zero. Hence, we conclude that the state $|\Psi_T\rangle$ is an example of a four-mode continuous variable T-shape cluster~state.

In summary of this section, we have discussed a practical scheme for the preparation of entangled continuous variable cluster states of effective bosonic modes realized in four physically separated atomic ensembles interacting collectively with a single-mode optical 
cavity and driving laser fields. We have demonstrated robustness of the scheme on three examples of the so-called continuous variable linear, square and T-type cluster states. These examples show that it should be straightforward in principle to create different types of cluster states with only few laser pulses.

\section{Summary}

\noindent We have shown that the current research on entanglement and disentanglement of multi-atom systems covers a wide class of situations ranging from isolated atoms located inside high-$Q$ cavities to atoms coupled to a common multi-mode electromagnetic field. The results show unexpected behaviors of entanglement, such as entanglement sudden death, spontaneous revival and sudden birth of entanglement, triggered entanglement evolution by imperfection and steered entanglement transfer. We have explored the role of the irreversible process of spontaneous emission in creation of entanglement and in disentanglement of two atoms. We have shown that spontaneous emission does not necessary lead to disentanglement of an initial entangled atoms. Under some circumstances, this irreversible process can entangle already disentangled atoms. Finally, we have presented models for creation of pure continuous variable  entangled and cluster states between macroscopic atomic ensembles.

\section*{Acknowledgements}
 I acknowledge useful collaborations and discussions on work presented in this review with Ryszard Tana\'s, Gao-xiang Li, Margaret Reid, Levente Horvath, Jun Jing, Zhi-Guo L\"u, Sonny Natali and Stanley Chan.

\end{document}